\documentclass[usenatbib]{mn2e}
\usepackage{graphicx}
\usepackage{ifthen}
\usepackage{url}
\usepackage{amsmath}
\usepackage{bm}
\usepackage{lscape}
\usepackage{rotating}
\usepackage[graphicx]{realboxes}
\usepackage{supertabular}
\usepackage{pdfpages}
\usepackage{footnote}
\usepackage{hyperref}
\makesavenoteenv{tabular}
\makesavenoteenv{table}
\hypersetup{
  colorlinks   = true,    % Colours links instead of ugly boxes
  urlcolor     = blue,    % Colour for external hyperlinks
  linkcolor    = blue,    % Colour of internal links
  citecolor    = blue      % Colour of citations
}

\def\ltsima{$\; \buildrel < \over \sim \;$}
\def\lta{\lower.5ex\hbox{\ltsima}}
\def\gtsima{$\; \buildrel > \over \sim \;$}
\def\simgt{\lower.5ex\hbox{\gtsima}}
\def\kms{{\rm\,km \  s^{-1}}}
\def\mas{{\rm\,mas}}
\def\masyr{{\rm\,mas/yr}}
\def\kpc{{\rm\,kpc}}

\def\msun{{\rm\,M_\odot}}

\def\pc{{\rm\,pc}}

\def\mss{{\rm M}_\odot \rm pc^{-2}}

\def\s{\ifmmode \widetilde \else \~\fi}
\def\={\overline}

\def\spose#1{\hbox to 0pt{#1\hss}}

\def\eg{{e.g.,\ }}

\def\lta{\mathrel{\spose{\lower 3pt\hbox{$\mathchar"218$}}
     \raise 2.0pt\hbox{$\mathchar"13C$}}}
\def\gta{\mathrel{\spose{\lower 3pt\hbox{$\mathchar"218$}}
     \raise 2.0pt\hbox{$\mathchar"13E$}}}
\def\Dt{\spose{\raise 1.5ex\hbox{\hskip3pt$\mathchar"201$}}}    % upper case
\def\dt{\spose{\raise 1.0ex\hbox{\hskip2pt$\mathchar"201$}}}    % lower case

\def\dotsfill{\leaders\hbox to 1em{\hss.\hss}\hfill}

\def\Gyr{{\rm\,Gyr}}
\def\FeH{{\rm[Fe/H]}}
\def\MH{{\rm[M/H]}}

\loadboldmathitalic 
\title[Tracing the formation of the MW through UMPs]{Tracing the formation of the Milky Way through ultra metal-poor stars} 
\author[F. Sestito et al.] {Federico Sestito$^{1,2}$, Nicolas Longeard$^{1}$, Nicolas F. Martin$^{1,3}$, Else Starkenburg$^{2}$,
\newauthor Morgan Fouesneau$^{3}$, Jonay I. Gonz\'alez Hern\'andez$^{4,5}$, Anke Arentsen$^{2}$,  Rodrigo Ibata$^{1}$, 
\newauthor  David S. Aguado$^{6}$, Raymond G. Carlberg$^{7}$, Pascale Jablonka$^{8,9}$,
\newauthor Julio F. Navarro$^{10}$, Eline Tolstoy$^{11}$, Kim A. Venn$^{10}$\\
$^{1}$ Observatoire astronomique de Strasbourg, CNRS, UMR 7550, F-67000 Strasbourg, France
\\
$^{2}$ Leibniz Institute for Astrophysics Potsdam (AIP), An der Sternwarte 16, D-14482 Potsdam, Germany\\
$^{3}$ Max-Planck-Institut f\"ur Astronomie, K\"onigstuhl 17, D-69117, Heidelberg, Germany\\
$^{4}$ Instituto de Astrofísica de Canarias, V\'ia L\'actea, 38205 La Laguna, Tenerife, Spain\\
 $^{5}$ Universidad de La Laguna, Departamento de Astrof\'isica, 38206 La Laguna, Tenerife, Spain\\
 $^{6}$ Institute of Astronomy, University of Cambridge, Madingley Road, Cambridge CB3 0HA, UK\\
 $^{7}$ Department of Astronomy and Astrophysics, University of Toronto, Toronto, ON M5S 3H4, Canada\\
 $^{8}$ GEPI, Observatoire de Paris, Universit\'e PSL, CNRS, Place Jules Janssen, F-92190 Meudon, France\\
 $^{9}$ Institute of Physics, Laboratoire d'astrophysique, \'Ecole Polytechnique F\'ed\'erale de Lausanne (EPFL), Observatoire, CH-1290 Versoix, Switzerland\\
 $^{10}$ Department of Physics and Astronomy, University of Victoria, PO Box 3055, STN CSC, Victoria BC V8W 3P6, Canada\\
 $^{11}$ Kapteyn Astronomical Institute, University of Groningen, Landleven 12, NL-9747AD Groningen, the Netherlands}

\date{\today}
\begin{document} 
\maketitle 
\begin{abstract} 
We use Gaia DR2 astrometric and photometric data, published radial velocities and
MESA models to infer distances, orbits, surface gravities, and effective temperatures 
for all ultra metal-poor stars ($\FeH<-4.0$ dex) available in the literature. Assuming that these stars are old ($>11\Gyr$) and that they are expected to belong to the Milky Way halo, we find that these 42 stars (18 dwarf stars and 24 giants or
sub-giants) are currently within $\sim20\kpc$ of the Sun and that they map a wide variety of orbits. A large fraction of those stars
remains confined to the inner parts of the halo and was likely formed or
accreted early on in the history of the Milky Way, while others have
larger apocentres ($>30\kpc$), hinting at later accretion from dwarf galaxies. Of particular
interest, we find evidence that a significant fraction of all known
UMP stars ($\sim26$\%) are on prograde orbits confined within $3\kpc$ of the Milky Way
plane ($J_z < 100 \kms \kpc$). One intriguing interpretation is that these stars belonged to the massive building block(s) of the proto-Milky Way that formed the backbone of the Milky Way disc. Alternatively,  they might have formed in the early disc and have been dynamically heated, or have been brought into the Milky Way by one or more accretion events whose orbit was dragged into the plane by dynamical friction before disruption. The combination of the exquisite Gaia DR2 data and surveys of the very metal-poor sky opens an exciting era in which we can trace the very early formation of the Milky Way.
\end{abstract}
 

\section{Introduction}

Ultra metal-poor (UMP) stars, defined to have \FeH\footnote{$\FeH =\log(N_{Fe}/N_{H})_{\star} - \log(N_{Fe}/N_{H})_{\odot}$, with N$_{X}$= the number density of element $X$} $<-4$ dex
\citep{Beers05}, are extremely rare objects located mainly in the
Milky Way (MW) halo. Because they are ultra metal-poor, also relative to
their neighbourhood, it is assumed that they formed from relative
pristine gas shortly after the Big Bang \citep[e.g.,][]{Freeman02}. As such,
they belong to the earliest generations of stars formed in the
Universe \citep{Karlsson13}. Because they are old, observable UMPs must be low-mass
stars, however the minimum metallicity at which low-mass stars can form is
still an open question \citep[see][and references therein]{Greif15}. The search for, and study of, stars with the lowest metallicities are therefore important topics to answer questions on the masses of the first generation of stars and the universality of the initial mass function (IMF), as well as on the early formation stages of galaxies and the first supernovae \citep[e.g.,][and references therein]{Frebel15}. Careful studies over many decades have allowed us to build up a catalogue of 42 UMP stars throughout the Galaxy. Many of these stars were discovered in survey programs that were or are dedicated to finding metal-poor stars using some special pre-selection through prism techniques \citep[e.g., the HK and HES surveys;][]{Beers85,Christlieb02} or narrow-band photometry \citep[such as for instance the SkyMapper and Pristine survey programmes;][]{Wolf18,Starkenburg17a}. Others were discovered in blind but very large spectroscopic surveys such as SDSS/SEGUE/BOSS \citep{York00,Yanny09,Eisenstein11} or LAMOST \citep{Cui12}.

From the analysis of cosmological simulations, predictions can be made for the present-day distribution of such stars in MW-like galaxies. Since these predictions have been shown to be influenced by the physics implemented in these simulations, we can use the present-day distribution to constrain the physical  processes of early star formation. For instance, a comparison between the simulations of \citet{Starkenburg17b} and \citet{ElBadry18} indicates a clear sensitivity of the present-day distribution on the conditions applied for star formation and the modelling of the ISM.

In an effort to refine the comparison with models and unveil the phase-space properties of these rare stars, we combine the exquisite Gaia DR2 astrometry and photometry \citep{2018arXiv180409365G} with models of UMP stars (MESA isochrones and luminosity functions; \citealt{2011ApJS..192....3P,2016ApJS..222....8D,2016ApJ...823..102C}, \url{waps.cfa.harvard.edu/MIST}) to infer the distance, stellar properties, and orbits of all 42 known UMP stars.

This paper is organised as follows: Section~\ref{DATA} explains how we put our sample together while Section~\ref{METHOD} presents our statistical framework to infer the distance, effective temperature, surface gravity, and orbit of each star in the sample using the Gaia DR2 information (parallax, proper motion, and  $G$, $BP$, and $RP$ photometry). The results for the full sample are presented in Section~\ref{RESULTS} and we discuss the implications of the derived orbits in Section~\ref{DISCUSSIONS} before concluding in Section~\ref{CONCLUSIONS}. We refer readers who are interested in the results for individual stars to Appendix~\ref{indi}, in which each star is discussed separately.

\section{Data}
\label{DATA}
We compile the list of all known ultra metal-poor ($\FeH<-4.0$ dex), hyper metal-poor ($\FeH<-5.0$ dex), and mega metal-poor ($\FeH<-6.0$ dex) stars from the literature building from the JINA catalogue \citep{2018ApJS..238...36A}, supplemented by all relevant discoveries. The literature properties for these stars are listed in Table~\ref{literaturetable}. We crossmatch this list with the Gaia DR2 catalogue\footnote{\url{https://gea.esac.esa.int/archive/}} \citep{2018arXiv180409365G} in order to obtain the stars' photometric and astrometric information. This is listed in Table~\ref{gaiatable}.

Some stars were studied in more than one literary source, with different methods
involving 1D or 3D models and  considering the stellar atmosphere at  Local Thermodynamic Equilibrium (LTE) or non-Local Thermodynamic Equilibrium (non-LTE), leading to dissimilar results
on metallicity and stellar parameters. In this paper, when multiple
results are available, we report in Table~\ref{literaturetable}
preferentially results including 3D stellar atmosphere and/or
involving non-Local Thermodynamic Equilibrium (non-LTE) modelling. If all
results are in 1D LTE, we favour the most recent results. 

When the UMP stars are recognised to be in binary systems and the orbital parameters are known (see Table~\ref{literaturetable}), the reported radial velocity is the systemic value that is corrected for the binary orbital motion around the centre of mass.

Assuming that all stars in our sample are distant, we consider that all the extinction is in the foreground. Therefore, all stars are de-reddened using the \citet{1998ApJ...500..525S} extinction map as listed in Table~\ref{literaturetable} and the \citet{2008A&A...482..883M} coefficients for the Gaia filters based on \citet{Evans18}, i.e.

\begin{eqnarray}
G_0 & = & G-2.664 E(B-V),\\
BP_0 & = & BP-3.311 E(B-V),\\
RP_0 & = & RP-2.021 E(B-V).
\end{eqnarray} 
Extinction values remain small in most cases (Table~\ref{literaturetable}).

We assume that the distance between the Sun and the Galactic centre is 8.0 \kpc, that the Local Standard of Rest circular velocity is $V_c=239 \kms$, and that the peculiar motion of the Sun is ($U_0=11.10 \kms, V_0+V_c=251.24\kms, W_0=7.25\kms$) as described in \citet{Schoenrich10}.

\section{Inferring the properties of stars in the UMP sample}\label{METHOD}
\subsection{Distance inference}
It is ill advised to calculate the distance to a star by simply inverting the parallax measurement \citep{1538-3873-127-956-994}, especially for large relative measurement uncertainties (\eg $\delta\varpi/\varpi>0.2$) and negative parallaxes. Therefore, we infer the probability distribution function (PDF) of the heliocentric distance to a star by combining its photometric and astrometric data with a sensible MW stellar density prior. Following Bayes' rule \citep{Sharma17}, the posterior probability of having a star at a certain distance given its observables $\bm{\Theta}$ (\eg photometry, metallicity, parallax) and a model $\mathcal{M}$ is characterised by its likelihood $\mathcal{L}(\bm{\Theta} | \mathcal{M})$ and the prior $\mathcal{P}(\mathcal{M})$. The likelihood gives the probability of the set of observables $\bm{\Theta}$ given model $\mathcal{M}$, whereas the prior represents the knowledge of the model used for the representation of a phenomenon. With these notations,

\begin{equation}
\mathcal{P}(\mathcal{M} |\bm{\Theta} ) \propto \mathcal{L}(\bm{\Theta} |\mathcal{M}) \mathcal{P}(\mathcal{M}).
\label{Bayes}\end{equation}

In this work, the model parameters are $\mathcal{M}=\{\mu=5\log(r)-5,A\}$, with $\mu$ the distance modulus of the star, $r$ the distance to the star, and $A$ its age. The observables $\bm{\Theta}$ can be split into the Gaia photometric observables $\bm{\Theta}_\mathrm{phot}=\{G_0,BP_0,RP_0,\delta_G,\delta_{BP},\delta_{RP}\}$ and the Gaia astrometric (parallax) observables $\bm{\Theta}_\mathrm{astrom}=\{\varpi,\delta_\varpi\}$, with $\delta x$ the uncertainty associated with measurement $x$. Assuming that the photometric and astrometric information on the star are independent, Equation~(\ref{Bayes}) becomes

\begin{equation}
\mathcal{P}(\mathcal{M} |\bm{\Theta} ) \propto \mathcal{L}_\mathrm{phot}(\bm{\Theta}_\mathrm{phot} |\mathcal{M}) \mathcal{L}_\mathrm{astrom}(\bm{\Theta}_{\varpi} |\mathcal{M}) \mathcal{P}(\mathcal{M}).
\label{Bayes2}\end{equation}

\subsubsection{$\mathcal{L}_\mathrm{phot}(\bm{\Theta}_\mathrm{phot} |\mathcal{M})$}\label{photo}
In order to determine the photometric likelihood of a given star for a chosen $\mu$ and $A$, we rely on the isochrone models from the MESA/MIST library \citep{2011ApJS..192....3P,2016ApJS..222....8D,2016ApJ...823..102C}, as they are the only set of publicly available isochrones that reach the lowest metallicity ($\FeH=-4.0$ dex) and is therefore the most appropriate for our study. 

Any isochrone, $\mathcal{I}$, of a given age, $A$, associated with a luminosity function\footnote{This associated luminosity function, $\Phi$, assumes a Salpeter IMF \citep{1955ApJ...121..161S}. The choice of the IMF is not very sensitive for the type of stars we analyse.} $\Phi(M_G|A)$, predicts the density distribution triplet of absolute magnitudes $p(M_G,M_{BP},M_{RP}|\mathcal{I},\Phi)$ in the Gaia photometric bands. After computing the likelihood $p(\bm{\Theta}_\mathrm{phot}|M_G,M_{BP},M_{RP},\mu)$, of these predictions shifted to a distance modulus $\mu$, against the observed photometric properties of the star, $\mathcal{L}_\mathrm{phot}$ results from the marginalization along that isochrone:

\begin{equation}
\begin{split}
\mathcal{L}_\mathrm{phot}(\bm{\Theta}_\mathrm{phot} | \mu,A,\Phi) & \\
 = & \int_\mathcal{I} p(\bm{\Theta}_\mathrm{phot}|M_G,M_{BP},M_{RP},\mu)\\
&\times p(M_G,M_{BP},M_{RP}|\mathcal{I},\Phi)p(\mathcal{I}|A)d\mathcal{I},
\label{likelihood_pho2}
\end{split}
\end{equation}

\noindent with

\begin{equation}
\begin{split}
&p(\bm{\Theta}_\mathrm{phot}|M_G,M_{BP},M_{RP},\mu) \\
& = \mathcal{N}(G_0|M_G+\mu,\delta_G^2+0.01^2)\\
&\times\mathcal{N}((BP-RP)_0|M_{BP}-M_{RP},\delta_{BP}^2+\delta_{RP}^2+2\times0.01^2)
\label{p_CMD}
\end{split}
\end{equation}

\noindent and $\mathcal{N}(x|m,s^2)$ the value of a Gaussian function of mean $m$ and variance $s^2$ taken on $x$. In Equation~(\ref{p_CMD}), a systematic uncertainty of 0.01 mag is added to the photometric uncertainties in each band to represent the uncertainties on the models.

For most stars, we expect to find two peaks in $\mathcal{L}_\mathrm{phot}(\bm{\Theta}_\mathrm{phot} |\mathcal{M})$, corresponding to the dwarf and giant solutions but stars close to the main sequence turnoff naturally yield a PDF with a single peak.

\subsubsection{$\mathcal{L}_\mathrm{astrom}(\bm{\Theta}_{\varpi} |\mathcal{M})$}
Gaia DR2 provides us with a parallax $ \varpi $ and its uncertainty $ \delta_\varpi$, which is instrumental in breaking the dwarf/giant distance degeneracy for most stars. The astrometric likelihood is trivially defined as

\begin{equation}
\mathcal{L}_\mathrm{astrom}(\varpi | \delta_{\varpi},  r) =  \frac{1}{\sqrt{2\pi} \delta_{\varpi}} \exp\left(-\frac{1}{2}\left( \frac{\varpi-\varpi_0- r^{-1}}{\delta_\varpi} \right)^2\right).
\label{gaia}
\end{equation}

\noindent Here, $\varpi_0=-0.029\mas$ is the parallax zero-point offset measured by \citet{2018A&A...616A...2L}.

Even in cases for which the parallax is small and the associated uncertainties are large, the Gaia data are often informative enough to rule out a nearby (dwarf) solution. 

\subsubsection{$\mathcal{P}(\mathcal{M}$)}
\emph{Prior on the distance and position $(r|\ell,b)$ ---} The prior on the distance and position to the star folds in our knowledge of the distribution of UMP stars around the MW. Since we expect those stars to be among the oldest stars of the MW and (likely) accreted, we first assume a halo profile. In particular, we use the RR Lyrae density power-law profile inferred by \citet{0004-637X-859-1-31}, $\rho(r) \propto r^{-3.4}$, since RR Lyrae stars are also expected to be old halo tracers.

From this stellar density profile, the  probability density to have a star at distance $r$ from the Sun along the line of sight described by Galactic coordinates $(\ell,b)$ is

\begin{equation}
\mathcal{P}_{\mathrm{H}}(r|\ell,b) =  \rho_0 r^2 \left(  \frac{D_\mathrm{GC}(r|\ell,b)}{r_0} \right)^{-3.4}.
\label{haloprior}\end{equation}

\noindent In this equation, $D_\mathrm{GC}(r|\ell,b)$ is the distance of the star to the Galactic centre, while $\rho_0$ and $r_0$ are reference values for the density and the scale length of the halo. For this work, the specific values of $\rho_0$ and $r_0$ will not affect the result because they will be simplified during the normalisation of the posterior PDF.

Anticipating the results described in Section~\ref{RESULTS}, we find that, even when using a pure halo prior, $\sim 26\%$ of our sample remains confined to the MW plane and the distance inference for a small number of stars yields unrealistic (unbound) orbits. Hence we repeat the analysis described with a mixture of a thick disc and a halo prior to investigate if, and how, the choice of the prior affects our results. This alternative MW prior is defined as

\begin{equation}
\mathcal{P}_{\mathrm{DH}}(r|\ell,b) =\eta \mathcal{P}_{\mathrm{D,norm}}(r|\ell,b) +(1-\eta) \mathcal{P}_{\mathrm{H,norm}}(r|\ell,b),
\label{diskhaloprior}\end{equation}

\noindent with $\eta=1/2$ the mixture coefficient, $ \mathcal{P}_{\mathrm{H,norm}}(r|\ell,b)$ the normalised halo prior expressed in Equation~(\ref{haloprior}), and $\mathcal{P}_{\mathrm{D,norm}}(r|\ell,b)$ the normalised thick disc prior defined by \citet{Binney}:

\begin{equation}
\mathcal{P}_{\mathrm{D}}(r|\ell,b) = \frac{r^2 \Sigma_{\mathrm{T}}}{2z_{\mathrm{T}}}  \exp\left( -\frac{D_\mathrm{GC}(r,\ell,b)}{D_{\mathrm{T}}}   -  \frac{|z|}{z_{\mathrm{T}}} \right),
\label{diskprior}\end{equation}

\noindent with $\Sigma_{\mathrm{T}} = 268.648 \mss$ the disc surface density, $D_{\mathrm{T}}= 2 \kpc$ the radial scale length for the density and  $z_{\mathrm{T}}=0.9 \kpc$ the vertical scale length  \citep{BlandHawthorn16}.

\emph{Prior on the age $A$, $\mathcal{P}(A)$  ---} There is no well defined age constraint for UMP stars but they are usually assumed to be very old \citep{Starkenburg17b}. Hence we assume that all the stars studied here were formed at least 11.2 Gyr ago ($\log(A/\textrm{yr})=10.05$). Beyond this age, we assume a uniform prior on $\log(A)$ until 14.1 Gyr ($\log(A/\textrm{yr})=10.15$), which is the maximum value of the isochrone grid.

Finally, $\mathcal{P}(\mathcal{M}) = \mathcal{P}(r|\ell,b)\mathcal{P}(A)$.

\subsubsection{Posterior PDF on distance $r$}

So far, $\mathcal{M}=\{\mu,A\}$ but we aim to infer the PDF on the distance modulus (or the distance) to the star alone. In order to do so, we simply marginalise over the age:

\begin{equation}
P(r= 10^{(\mu+5)/5} |\bm{\Theta}) = \int \mathcal{P}(\mathcal{M} |\bm{\Theta})dA,
\end{equation}

\noindent assuming $\mu \geq 0$ mag ($r \geq 10 \pc$).

\subsection{Effective temperature and surface gravity inference}\label{TEMPGRA}
For each point of the theoretical isochrones $\mathcal{I}(A,\mu)$ corresponds a value of the surface gravity, $\log(g)$, and a value of the effective temperature, $T_\mathrm{eff}$. Marginalising the likelihood and prior over distance modulus and age instead of over the isochrone as in Equation~(\ref{likelihood_pho2}), we can find the posterior probability as a function of $\log(g)$ and $T_\mathrm{eff}$. In detail,

\begin{eqnarray}
&\mathcal{P}(\log(g),T_\mathrm{eff}| \bm{\Theta})= \iint \mathrm{P}(\bm{\Theta} | \log(g),T_\mathrm{eff}, \mathcal{I}(A),\mu) \nonumber\\
& \times \Phi(M(\log(g),T_\mathrm{eff}, A))  \mathcal{P}(r,\ell,b)   \mathcal{L}_\mathrm{astrom}(\varpi | r(\mu), \delta_{\varpi}) \,d\mbox{A}\,d\mu .  \nonumber\\
&
\label{logT}\end{eqnarray}

\begin{figure*}
\hspace*{-1cm}\includegraphics[scale=.6]{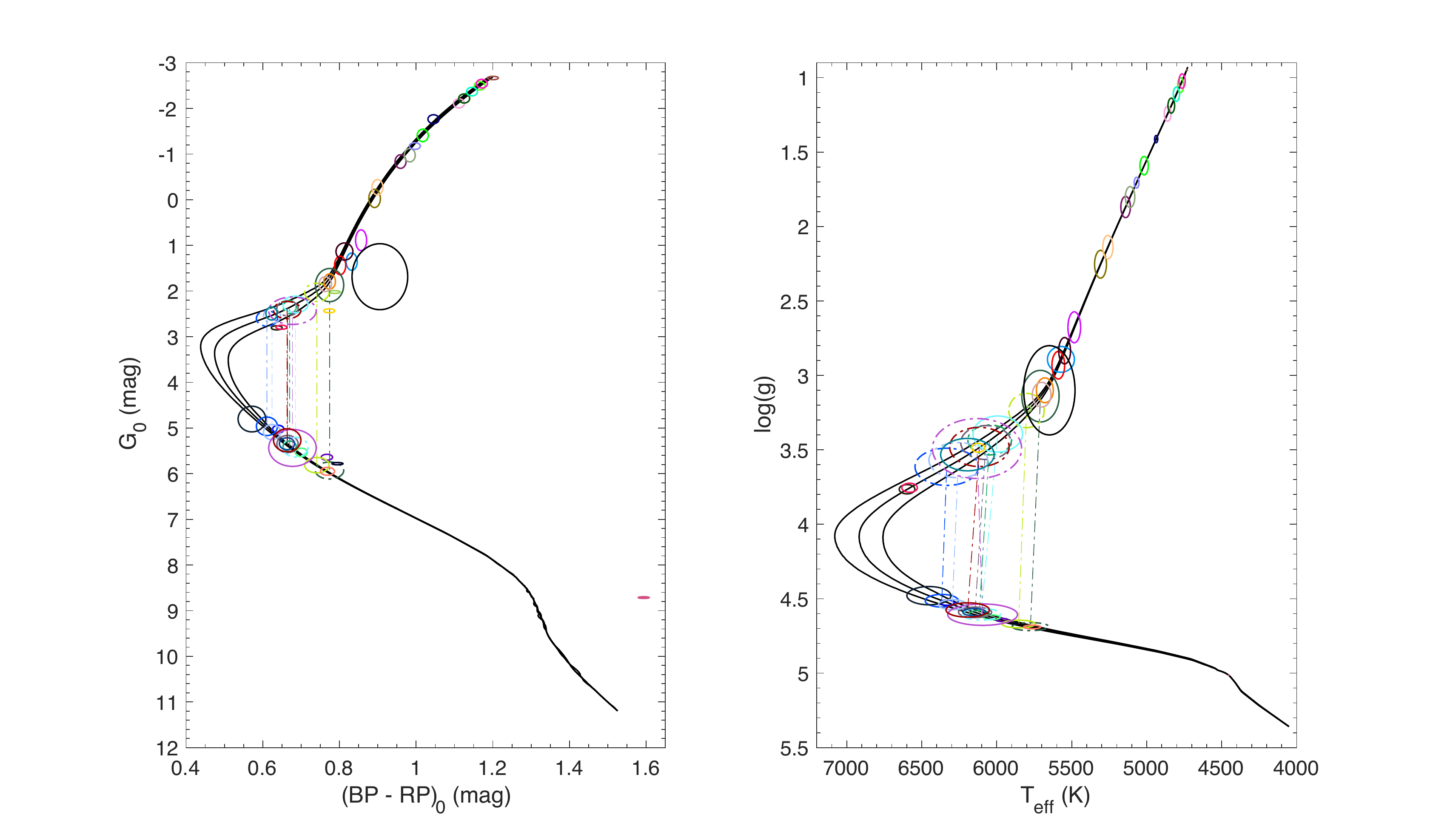}
\caption{Position of the sample stars in the CMD (left) and the $\log(g)$ vs. $T_\mathrm{eff}$ plane (right). The ellipses represent the position of the stars within 1 sigma and the black lines correspond to the three isochrones with $\log(A/\mathrm{yr}) = 10.05,10.10,10.15$ and metallicity $\FeH = -4$ dex. If the dwarf-giant degeneracy is not broken, the two possible solutions are represented and connected by a dot-dashed line of the same colour code. Each colour represents a star and the colour-code is the same as the colour-code for the markers in Figure~\ref{actiondiagram} and the panel's titles in Figures~\ref{HE 0020-1741} ~-~\ref{HE 2323-0256}. Solutions with integrated probability ($\int_{d-3\sigma}^{d+3\sigma} P(r) dr$) lower than $5\%$ are not shown and solutions with integrated probability in the range $[5\%, 50\%]$ are shown with dot-dashed ellipses.}
\label{CMD}
\end{figure*}

\subsection{Orbital inference}\label{ORBITS}
Gaia DR2 provides proper motions in right ascension and declination with their associated uncertainties and covariance. Combining this with the distance inferred through our analysis, we can calculate the velocity vector PDF $P(\bm{v}) =P(v_r,v_{\alpha},v_{\delta})$ for all 42 stars in our UMPs sample. This PDF, in turn, allows us to determine the properties of the orbit of the stars for a given choice of Galactic potential. We rely on the \texttt{galpy}\footnote{\url{http://github.com/jobovy/galpy}} package \citep{2015ApJS..216...29B} and choose their \textit{MWPotential14}, which is a MW gravitational potential composed of a power-law, exponentially cut-off bulge, a Miyamoto Nagai Potential disc, and a \citet{NavarroFrenkWhite97} dark matter halo. A more massive halo is chosen for this analysis, with a mass of $1.2 \cdot 10^{12} \msun$ compatible with the value from \citet{BlandHawthorn16} (vs.~$0.8 \cdot 10^{12} \msun$ for the halo used in \textit{MWPotential14}).

For each star, we perform a thousand random drawings from the position, distance, radial velocity, and proper motion PDFs. In the case of the two components of the proper motion ($\mu_{\alpha},\mu_{\delta}$), we consider their correlation given by the coefficients in Gaia DR2, drawing randomly these two parameters according to a multivariate gaussian function that takes into account the correlation. The possible correlation between coordinates and proper motions is not taken into account because it does not affect our result. For each drawing, we integrate this starting phase-space position backwards and forwards for 2 Gyr and extract the apocentre, $r_\mathrm{apo}$, pericentre, $r_\mathrm{peri}$, eccentricity, $\epsilon$, energy $E$, the angular momentum $L$ of the resulting orbit (note that in this frame of reference, $L_z>0$ means a prograde orbit), and the action-angle vector ($J_r$, $J_{\phi}=L_z$, $J_z$, where the units are in $\kms\kpc$). 

\section{Results}\label{RESULTS}
Tables~\ref{distancetable} and \ref{orbittable} summarise the results of the analysis and list the inferred stellar and orbital properties for all stars, respectively. In cases for which the (distance) PDF is double-peaked, we report the two solutions, along with their fractional probability.

Figure~\ref{CMD} shows the colour-magnitude diagram (CMD) and the temperature-surface gravity diagram for our UMP sample, plotted with three isochrones that cover the age range we considered ($\log(A/\mathrm{yr}) = 10.05,10.10,10.15$). For stars for which the dwarf/giant degeneracy is not broken, we show both solutions connected by a dot-dashed line, where the least probable solution is marked with a dot-dashed ellipse. Only results using a MW halo prior are shown here. As we can see, from the CMD plot (left panel of Figure~\ref{CMD}), the method overall works well, except for the HE~0330+0148 ($(BP-RP)_0 \approx1.6$ mag) that lays outside the colour range of the available set of isochrones. This special case is discussed in more detail in section~\ref{HE 0330+0148Sec}. The distances and stellar parameters lead to the conclusion that 18 stars ($\sim43\%$) are in the main sequence phase, and the other 24 are in the subgiant/giant phase ($\sim57\%$). This is of course a result of the observing strategies of the multiple surveys that led to the discovery of these stars.

For all 42 stars in our sample, we show the results of our analysis in Figures~\ref{HE 0020-1741} to~\ref{HE 2323-0256}. In all figures, the top-left panel shows the distance likelihood functions and posterior PDFs, the top-middle panel presents the log(g) PDF, while the top-right panel shows the effective temperature PDF. The orbit of the star in Galactic cartesian coordinates is presented in the bottom panels of the figures.

In the subsections of  Appendix~\ref{indi}, we discuss in detail the results for every star in the sample sorted by right ascension. Specifically, we focus on the inferred distances, stellar parameters, and orbits using a MW halo prior and, when it yields different results, we also discuss the use of the disc$+$halo prior.  A global comparison between the inferred stellar parameters form our work and the values from literature is described in Appendix~\ref{comp} and shown in the two panels of Figure~\ref{comparison}.

We did a comparison between the distances inferred in this work and the ones inferred by \citet{Bailer18}. These authors use a posterior probability composed by the astrometric likelihood shown in Equation~(\ref{gaia}) and a MW prior that is based on a Gaia-observed Galaxy distribution function accurately describing the overall distribution of all MW stars. This is naturally  more biased to higher densities in the thin disc and thus results in closer distances for most of the stars.

 \citet{Frebel18} compiled a list of 29 UMP stars inferring orbital parameters starting from the MW prior described in  \citet{Bailer18} but fixing the length-scale parameter to $L=0.5$. As both the initial assumptions and the focus of the analysis given in  \citet{Frebel18}  significantly differ from the approach taken in this work, we refrain from a further qualitative comparison.
 
\begin{figure*}
\hspace*{-1.3cm}\includegraphics[scale=.73]{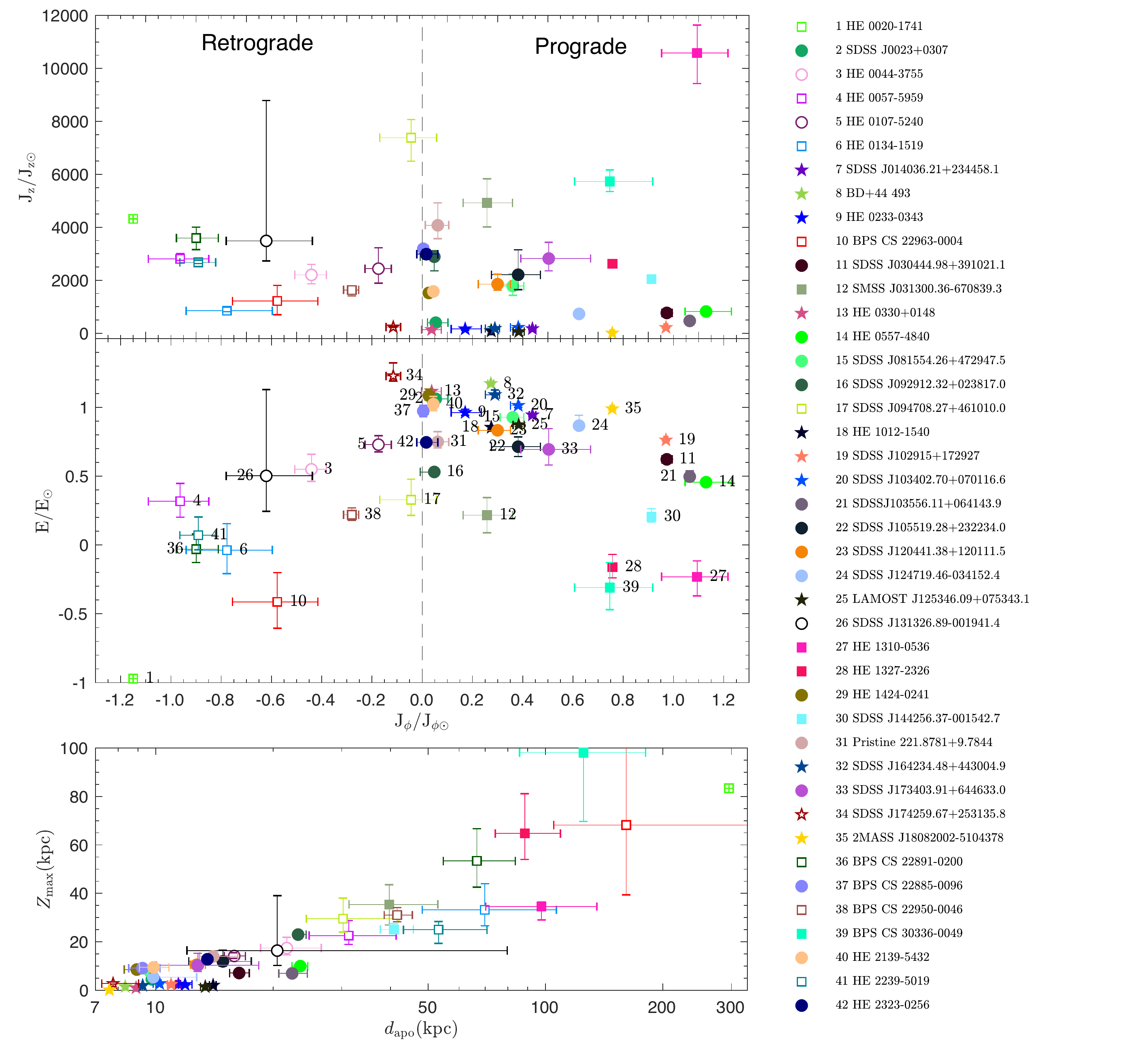}
\caption{Position of the sample stars in the rotational action $J_{\phi}$ ($=L_z$) and vertical action $J_z$ space (top panel), in the energy and rotational action space, and in the maximum height vs. apocentre of the stars' orbits (bottom panel). The rotational and vertical action and the Energy are scaled  by the Sun values respectively $J_{\phi\odot} = 2009.92 \kms\kpc$,  $J_{z\odot} = 0.35 \kms\kpc$ and $E_{\odot} = -64943.61$~km$^2$~s$^{-2}$. Stars with our ``MW planar'' sample that are confined close to the MW plane are marked with a star symbols, while ``inner halo'' and ``outer halo'' stars are represented by circles and squares, respectively. Retrograde stars, which are located on the left side of the top and central panels ($J_{\phi}<0\kms\kpc$) are denoted with empty marker, while prograde stars are shown with a filled marked. The colour-coding is the same as in Figure~\ref{CMD} and as the title of Figures~\ref{HE 0020-1741} ~-~\ref{HE 2323-0256} and helps to differentiate the stars. The full legend is provided on the side of this Figure. The number associated to each star also corresponds to the number of the subsection in the Appendix~\ref{indi} in which the individual results are discussed.} 
\label{actiondiagram} 
\end{figure*}

\section{Discussions}\label{DISCUSSIONS}
Our combined analysis of the Gaia DR2 astrometry and photometry with stellar population models for low metallicity stars allows us to infer the stellar parameters and orbital properties of the 42 known UMP stars. We derive well constrained properties for most stars and, in particular, we are now in a position to unravel the possible origin of the heterogeneous sample of UMP stars found to date.

\subsection{Insights on the orbits of UMP stars}
Apart from 2 ambiguous cases, we can classify the orbits of the UMP stars within three loosely defined categories:
\begin{itemize}
\item 19 ``inner halo'' stars, arbitrarily defined as having  apocentres smaller than $30\kpc$
\item 12 ``outer halo'' stars with apocentre larger than $30\kpc$
\item strikingly, 11 stars that have ``MW plane'' orbits, by which we mean that they stay confined close to the MW plane ($|Z|<3.0 \kpc$).
\end{itemize}
Figure~\ref{actiondiagram} attempts to show these different kind of orbits, displaying on the top panel the vertical component of the action-angle $J_z$ versus the rotational component $J_{\phi}$ ($=L_z$) for all the UMP in our sample. In this space, the stars confined to the MW plane (denoted by a star marker) are constrained to the lower part of the diagram, while the halo stars have larger $J_z$. Stars that have a prograde motion have $J_{\phi} >0$ and stars with retrograde orbits lie in the $J_{\phi}<0$ part of the diagram. We note how the Caffau star  (SDSS~J102915+172927) and 2MASS~J18082002-5104378 occupy a special place in this plane and they are the only stars on a quasi-circular orbit at large $J_{\phi}$ and low $J_z$.

It is appealing to assign a tentative origin to stars in these three categories. The ``inner halo'' stars could well be stars accreted onto the MW during its youth, when its mass was smaller and, therefore, its potential well less deep than it is now. At that time, more energetic orbits would have been unbound and left the MW in formation. ``Outer halo'' orbits tend to have very radial orbits in this sample (likely a consequence of the window function imparted by the various surveys that discovered these UMP stars; see below), which makes it easier to identify them. It is tempting to see those as being brought in through the accretion of faint dwarf galaxies onto the MW throughout the hierarchical formation of its halo. Although no UMP has been found in MW satellite dwarf galaxies yet, we know of many extremely metal-poor stars in these systems, down to [Fe/H]$ = -4$ \citep[e.g.,][]{Tafelmeyer10} and UMP stars are expected to be present as well \citep{Salvadori15}. We note that, among the two ``halo" categories, there is a distinct preference for prograde over retrograde orbits.

The 11 ``MW plane'' orbits are much more unexpected:
\begin{itemize}
\item 8 stars (SDSS~J014036.21+234458.1, BD+44~493, HE~0233-0343, HE~0330+0148, HE~1012-1540, SDSS~J103402.70+070116.6, LAMOST~J125346.09+075343, SDSS~J164234.48+443004.9) share similar rosette orbits within a wide range of angular momentum along the $z$ axis ($83 \lta L_z \lta 885\kms \kpc$). These stars orbit close to the plane, but not on circular orbits.
\item SDSS~J102915+172927 and 2MASS~J18082002-5104378 (Figures~\ref{SDSS J102915+172927} and~\ref{2MASS J18082002-5104378}), are on  almost circular orbits close to the solar radius.
\item SDSS~J174259.67+253135.8 (Figure~\ref{SDSS J174259.67+253135.8}) is retrograde  and more likely on an ``inner halo'' orbit that remains close to the MW plane.
\end{itemize}
\noindent The first ten of those stars, excluding SDSS~J174259.67+253135.8, all have positive $L_z$ and thus a prograde orbit, which is unlikely to be a random occurrence ($<1\%$ chance). It is worth noting that it is very unlikely the selection functions that led to the discovery of the UMP stars biased the sample for/against prograde orbit. The origin of those stars is puzzling but we can venture three different hypothesis for their presence in the sample, all of which must account for the fact that this significant fraction of UMP stars, which are expected to be very old, appears to know where the plane of the MW is located, even though the MW plane was unlikely to be in place when they formed.

\emph{Scenario 1:} The first obvious scenario is that these stars formed in the MW disc itself after the \textsc{HI} disc settled. In this fashion, the stars were born with a quasi-circular orbit and then the presence of a dynamical heating mechanism is mandatory to increase the eccentricity and the height from the plane as a function of time. We find that all the prograde ``MW plane'' stars and few catalogued as inner halo stars that are confined within  $Z_{\mathrm{max}}<15\kpc$ and $d_{\mathrm{apo}}<25 \kpc$ (see Figure~\ref{actiondiagram}) overlap in the parameters space ($Z_{\mathrm{max}}$, $d_{\mathrm{apo}}$, $L_z$, $E$) with a population of known stars at higher metallicity that \citet{Haywood18} hypothesise to be born in the thick disc and then dynamical heated by the interaction between the disc and a merging satellite. However, the question is whether in a relatively well-mixed \textsc{HI} disc it is possible to form stars so completely devoid of metals.

\emph{Scenario 2:} The second scenario is that these stars were brought into the MW by the accretion of a massive satellite dwarf galaxy. Cosmological simulations have shown that merger events are expected to sometimes be aligned with the disc. As a result, significant stellar populations currently in the disc might actually be merger debris \citep{Gomez17}. Alternatively, \citet{Scannapieco11}, show that 5--20\% of disc stars in their simulated MW-like disc galaxies where not formed in situ but, instead, accreted early from now disrupted satellites on co-planar orbits. Additionally, it is well known that the accretion of a massive system onto the MW will see its orbit align with the plane of the MW via dynamical friction, as shown by \citet{Abadi03} or \citet{Penarrubia02}. From these authors' simulations, one would expect orbits to become such that they would end up with larger eccentricities than the satellite's orbit at the start of the merging process and also aligned with the disc by dynamical friction and tidal interactions, which is compatible with our orbital inference for the remarkable UMP stars. If such an accretion took place in the MW's past, it could have brought with it a significant fraction of the UMP stars discovered in the solar neighbourhood. The accretion of the so-called Gaia-Enceladus satellite in the Milky Way's past \citep{Belokurov18,Haywood18,2018Natur.563...85H} could be an obvious culprit, however Gaia-Enceladus was discovered via the mainly halo-like and retrograde orbit of its stars whereas the vast majority of the stars we find here are on prograde orbits. In fact, there is no evidence of a particular overdensity of stars in the top-left region of the $J_z$ vs. $J_\phi$ of Figure~\ref{actiondiagram} where Gaia-Enceladus stars are expected to be found. It would therefore be necessary to summon the presence of another massive or several less massive accretion events onto the MW if this scenario is valid.

\emph{Scenario 3:} Finally, the third scenario that could explain the presence of this significant fraction of UMP stars that remain confined to the plane of the MW would be one in which these stars originally belonged to one or more of the building blocks of the proto-MW, as it was assembling into the MW that we know today. Fully cosmological simulations confirm that stars that are at the present time deeply embedded in our Galaxy do not need to have their origin in the proto-Galaxy. \citet{ElBadry18} find in their cosmological simulations that, of all stars formed before $z = 5$ presently within $10 \kpc$ of the Galactic centre, less than half were already in the main progenitor at $z = 5$. Over half of these extremely old stars would thus make their way into the main Galaxy in later merging events and find themselves at $z = 5$ inside different building blocks that are up to $300 \kpc$ away from the main progenitor centre. In such a scenario, we can expect that whatever gas-rich blocks formed the backbone of the MW disc brought with it its own stars, including UMP stars. Yet, for such a significant number of UMP stars to align with the current MW plane, it is necessary to assume that the formation of the MW's disc involved a single massive event that imprinted the disc plane that is aligned with the orbit of its stars. The presence of many massive building blocks would have likely led to changes in the angular \textsc{Hi} disc alignment. Similarly, the MW cannot have suffered many massive accretions since high redshift or the disc would have changed its orientation \citep{Scannapieco09}. This would be in line with expectations that the MW has had an (unusually) quiet accretion history throughout its life \citep{Wyse01,Stewart08}.

\subsubsection{The Caffau star and 2MASS J18082002-5104378}
SDSS~J102915+172927 (see Figure~\ref{SDSS J102915+172927}), also known as ``the Caffau star" \citep{Caffau2011aa}, and 2MASS~J18082002-5104378 (see Figure~\ref{2MASS J18082002-5104378}) both have a disc-like prograde orbit but while the Caffau Star reaches a height of $~2.3 \kpc$ from the MW plane, the latter star is confined within $~0.166 \kpc$, confirming the results from \citet{Schlaufman18}. Both  stars represent  outliers inside the surprising sample of ``MW planar" stars that typically have more eccentric orbits. For these stars, scenario 3, as outlined above, might be an interesting possibility. A merging between the building blocks of the proto-MW could have brought in these UMP stars and their orbit circularised by dynamical friction.

\subsubsection{Coincidence with the Sagittarius stream}

\begin{figure*}
\includegraphics[scale=.7]{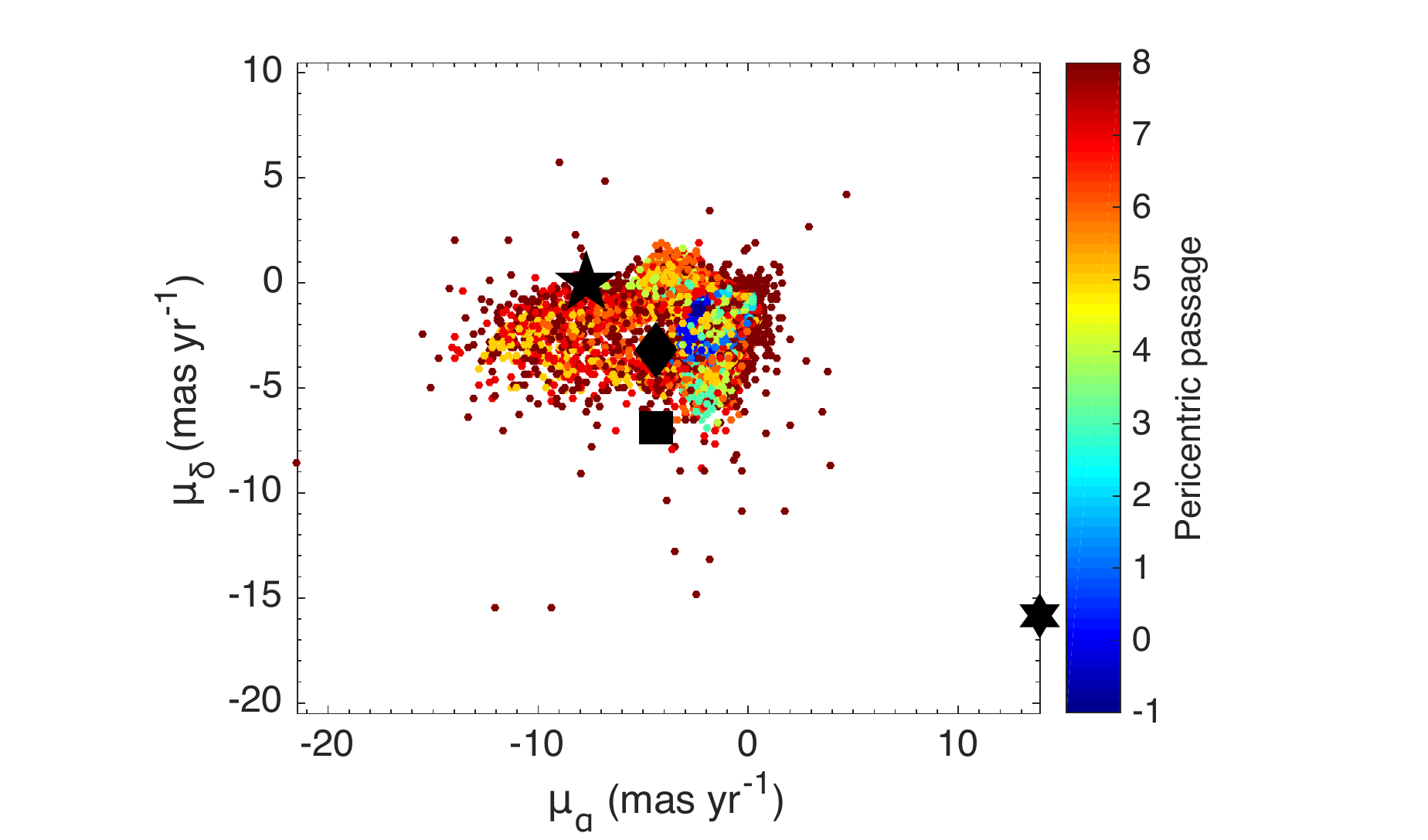}\\ 
\includegraphics[width=1.2\textwidth]{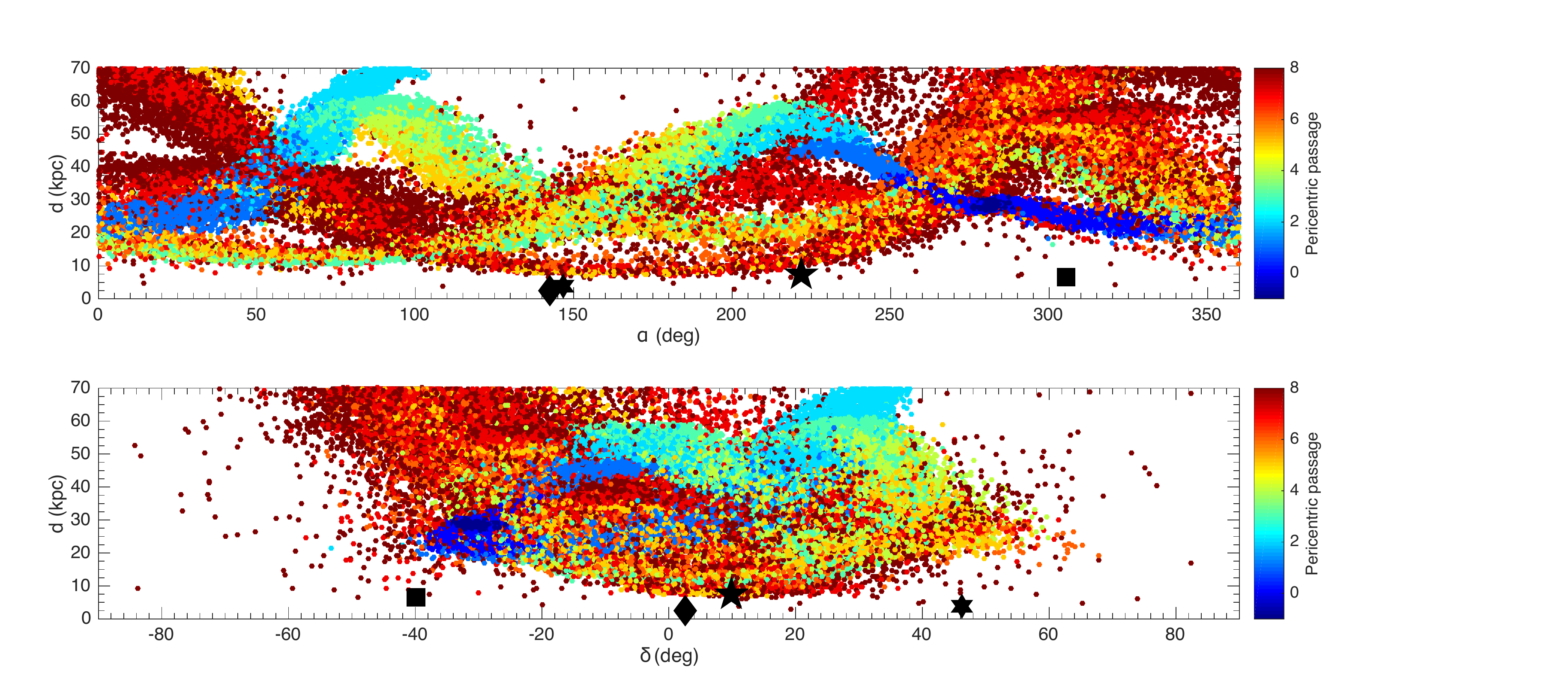}
\caption{Top: proper motion space for the particles of the LM10 simulation (dots), and SDSS~J092912.32+023817.0 (black diamond), SDSS~J094708.27+461010.0 (black hexagram), Pristine221.8781+9.7844 (black pentagram), and BPS~CS~22885-0096 (black square). The colour-code for the LM10 simulation indicates the pericentric passage on which the particle became unbound from Sgr. 
A pericentric passage value of $-1$ indicates debris which is still bound at the present day, while a value of $0$ indicates debris stripped on the most recent pericentric passage of Sgr, and a value above $1$ corresponds to successive pericentric passages.
Centre: heliocentric distance $d$ as a function of right ascension $\alpha$ for the LM10 simulation and the candidates. Bottom: heliocentric distance $d$ as a function of declination $\delta$ for the LM10 simulation and the candidates. The LM10 simulation is shown within $70 \kpc$ from the Sun for the centre and bottom panel.} 
\label{SgrStream} 
\end{figure*}

We note that four of the ``halo'' stars (SDSS~J092912.32+023817.0, SDSS~J094708.27+461010.0, Pristine221.8781+9.7844 and BPS~CS~22885-0096) have orbits that are almost perpendicular to the MW plane (see Figures~\ref{SDSS J092912.32+023817.0}, \ref{SDSS J094708.27+461010.0}, \ref{Pristine221.8781+9.7844}, and \ref{BPS CS 22885-0096}), coinciding with the plane of the stellar stream left by the Sagittarius (Sgr) dwarf galaxy as it as being tidally disrupted by the MW. We therefore investigate if these stars belong to the stream by comparing their proper motions and distances with the values provided by the N-body simulation of \citet{LM10}, hereafter LM10 (see Figure~\ref{SgrStream}). It is clear that SDSS~J094708.27+461010.0 has a proper motion that is incompatible with the simulation's particles. On the other hand, we find that SDSS~J092912.32+023817.0, Pristine221.8781+9.7844, and BPS~CS~22885-0096 have proper motions that are in broad agreement with those of the simulation. These stars could be compatible with the oldest wraps of the Sgr galaxy but we are nevertheless cautious in this assignment since only the young wraps of the stream were constrained well with observations in the \citet{LM10} model. Older wraps rely on the simulation's capability to trace the orbit back in the MW potential, that is itself poorly constrained and has likely changed over these timescales, and the true 6D phase-space location the older warps could therefore easily deviate significantly from the simulation's expectations.

\subsubsection{A connection between SDSS~J174259.67+253135.8 and $\omega$ Centauri?}

SDSS~J174259.67+253135.8 is the only star of the ``MW planar'' sample that has a retrograde motion and its orbital properties are, in fact, similar enough to those of the  $\omega$ Centauri ($\omega$Cen) stellar cluster to hint at a possible connection between the two. It should be noted, however, that the $L_z$ of $\omega$Cen's orbit is about twice that of this star. Nevertheless, given the dynamically active life that $\omega$Cen must have had in the commonly-held scenario that it is the nucleus of a dwarf galaxy accreted by the Milky Way long ago (\eg~\citealt{Zinnecker88,Mizutani03}), the similarity of the orbits is intriguing enough to warrant further inspection.

\subsection{Limits of the analysis and completeness}\label{SELECTION}
The heterogeneous UMP sample comes from multiple surveys conducted over the years, with their own, different window functions for the selection of the targets and it can thus by no means be called a complete or homogeneous sample. To reconstruct the full selection function of this sample is nearly impossible since it includes so many inherited window functions from various surveys and follow-up programs. As far as we can deduce, however, none of the
programs would have specifically selected stars on particular orbits. We therefore consider the clear preference of the UMP star population for orbits in the plane of the MW disc a strong result of this work but we caution the reader not to consider the ratio of ``inner halo,'' ``outer halo,'' and ``MW plane'' orbits as necessarily representative of the true ratios, which will require a more systematic survey to confirm.

We note that due to the different abundance patterns of these stars, $\FeH$ is not always a good tracer of the total metallicity $\MH$. However, not all stars in this sample are equally well-studied and therefore constraints on $\MH$ are inhomogeneous. This has led us to nevertheless choose a cut on $\FeH$ as this is the common quantity measured by all the cited authors.

Another limitation of this work comes from the isochrones we use, which are the most metal-poor
isochrones available in the literature at this time and have [Fe/H]$=-4$ dex with solar-scaled
$\alpha$-abundances. Beyond the fact that some stars in our sample are significantly more metal-poor than this, not all stars follow this
abundance pattern and as a result their total metal-content can
change, in turn affecting the colour of the isochrones. We estimate,
however, that this will be a small effect at these low metallicities,
as low-metallicity isochrones are relatively insensitive to small
variations in metallicity, and take this into account adding a
systematic uncertainty of 0.01 mag in quadrature to the model  (see
Section~\ref{photo}). This is unlikely to affect the final results on
the evolutionary phase and the typology of the orbits. A final potential limitation of this work stems from the possible binary of some of the studied stars. If, unbeknownst to us, a star is in fact a binary system whose component are in the same or a similar evolutionary phase, their photometry would not be representative of their true properties and our distant inference would be biased. Similarly a binary star would like have its velocity be affected, leading to flawed orbital parameters. For known binary stars, we nevertheless take these effects into account and our distance and orbital inference should not be severely affected by this binarity issue.

\subsection{Future outlook}
As described in \ref{SELECTION}, the current sample and analysis of
their dynamics is quite limited by an unknown and complicated
selection function. With proper motion, parallax, and the exquisite photometry from Gaia
DR2, we plan to apply the same bayesian framework described in
Section~\ref{METHOD} to all the EMP stars within the Pristine survey
\citep{Starkenburg17a} to investigate their stellar properties and
orbits. As the completeness and purity of this sample is very well
understood \citep{Youakim17} and this sample is much larger, this will open up more quantitative
avenues to explore the role of extremely metal-poor stars in the big
picture of the accretion history of the MW.

\section{Conclusions}
\label{CONCLUSIONS}
Combining the Gaia DR2 photometric and astrometric information in a statistical framework, we determine the posterior probability distribution function for the distance, the stellar parameters (temperature and surface gravity), and the orbital parameters of 42 UMPs (see Tables~\ref{distancetable} and \ref{orbittable}). Given that 11 of those stars remain confined close to the MW plane, we use both a pure halo prior and a combined disc$+$halo prior. Folding together distance posterior and orbital analysis we find that 18 stars are on the main sequence and the other 24 stars are in a more evolved phase (subgiant or giant).

Through the orbital analysis, we find that 11 stars are orbiting in the plane of disc, with maximum height above the disc within 3 kpc. We hypothesise that they could have once belonged to a massive building blocks of the proto-MW that formed the backbone of the MW disc, or that they were brought into the MW via a specific, massive hierarchical accretion event, or they might have formed in the early disc and have been dynamically heated. Another 31 stars are from both the ``inner halo'' (arbitrarily defined as having $r_\mathrm{apo}<30\kpc$) and were accreted early on in the history of the MW, or the ``outer halo'' hinting that they were accreted onto the Galaxy from now-defunct dwarf galaxies. Of these halo stars, SDSS~J092912.32+023817.0, Pristine221.8781+9.7844 and BPS~CS~22885-0096, could possibly be associated with the Sagittarius stream, although they would need to have been stripped during old pericentric passages of the dwarf galaxy. SDSS~J174259.67+253135.8 could also possibly be associated with $\omega$Cen  as its progenitor.

The work presented here provides distances, stellar parameters, and orbits for all known UMP stars and, hence, some of the oldest stars known. To understand their position and kinematics within the Galaxy it is very important to reconstruct the early formation of the MW and/or the hierarchical formation of some of its components. We foresee a statistical improvement of this first study with the arrival of homogeneous and large datasets of EMP stars, such as observed within the Pristine or SkyMapper surveys \citep{Starkenburg17a,Wolf18}. With these surveys, the window function and the selection criteria of the objects for which distances and orbits are derived will be much better known.

\section*{Acknowledgements}
We would like to thank Benoit Famaey, Misha Haywood and Paola Di Matteo for the insightful discussions and comments.

FS, NFM, NL, and RI gratefully acknowledge funding from CNRS/INSU through the Programme National Galaxies et Cosmologie and through the CNRS grant PICS07708. FS thanks the Initiative dExcellence IdEx from the University of Strasbourg and the Programme Doctoral International PDI for funding his PhD. This work has been published under the framework of the IdEx Unistra and benefits from a funding from the state managed by the French National Research Agency as part of the investments for the future program.
 ES and AA gratefully acknowledge funding by the Emmy Noether program from the Deutsche Forschungsgemeinschaft (DFG). JIGH acknowledges financial support from the Spanish Ministry project MINECO AYA2017-86389-P, and from the Spanish MINECO under the 2013 Ram\'on y Cajal program MINECO RYC-2013-14875. KAV thanks NSERC for research funding through the Discovery Grants program.

This research has made use of use of the SIMBAD database, operated at CDS, Strasbourg, France \citep{Simbad}.
This work has made use of data from the European Space Agency (ESA) mission
{\it Gaia} (\url{https://www.cosmos.esa.int/gaia}), processed by the {\it Gaia}
Data Processing and Analysis Consortium (DPAC,
\url{https://www.cosmos.esa.int/web/gaia/dpac/consortium}). Funding for the DPAC
has been provided by national institutions, in particular the institutions
participating in the {\it Gaia} Multilateral Agreement.

\begin{table*}
\centering
\caption{Physical parameters of the analysed UMPs found in literature. [Fe/H], [C/Fe], $v_r$, $T_{lit}$, log(g)$_{lit}$ are from the articles listed in the column References. $v_r$ and the binarity flag denoted with $a$ are from \citet{Arentsen18}, the $v_r$ values for binary systems denoted with $a$ are the systemic radial velocities corrected for the binary orbital motion. $v_r$ values for stars that are not known to be in a binary system and from the compilation of  \citet{Arentsen18} are calculated with a weighted average of all the $v_r$ measurements. E(B-V) is from \citet{1998ApJ...500..525S}. In case the star is in a binary system, the binarity flag is equal to Y, while stars labelled with N are not in a binary system or the binarity is not known.}
\resizebox{\textwidth}{!}{
\begin{tabular}{llllllllllllllll}
\hline
Identifier          & $\alpha_{J2000}$ & $\delta_{J2000}$ & [Fe/H]  & $\delta_{[Fe/H]}$ & [C/Fe]  & $\delta_{[C/Fe]}$ & $v_r$  & $\delta_{v_r}$ & $T_{lit}$ & $\delta_{T_{lit}}$ & log(g)$_{lit}$ & $\delta_{log(g)_{lit}}$ & E(B-V) & Binarity & References \\ 
& (deg) & (deg) & (dex)& (dex)&(dex)&(dex)& ($\kms$) & ($\kms$) & (K) & (K) & (dex) & (dex) & (mag) & & \\ \hline
HE 0020-1741 & 5.6869167 & $-17.4080944$ & $-4.05$ & $-$ & 1.4 & $-$ & 93.06 & 0.83 & 4630.0 & 150 & 0.95 & 0.3 & 0.021 & N & \citet{Placco16} \\ \hline
SDSS J0023+0307            & 5.80834363858 & 3.13284420892  & $<-6.6$ & $-$       & $<2.0$   &  $-$       & $-195.5$  & 1.0         & 6140 & 132       & 4.8       & 0.6            & 0.028         & N    & \citet{Aguado2018b}       \\ \hline
HE 0044-3755                  & 11.6508144643 & $-37.6593210379$ & $-4.19$ & $-$    & $-0.3$  & $-$       & 48.3   & 2.5         & 4800 & 100       & 1.5       & 0.1            & 0.010         & N    & \citet{Cayrel2004}       \\ \hline
HE 0057-5959               & 14.9749409617 & $-59.7249294278$ & $-4.08$ & $-$      & 0.86  & $-$      & 375.64a  & 1        & 5257 & $-$        & 2.65      & $-$           & 0.016         & N    & \citet{2007ApJ...670..774N}, \citet{2013ApJ...762...25N}        \\ \hline
HE 0107-5240               & 17.3714810637 & $-52.4095009821$ & $-5.5$ & 0.2       & 3.85  & $-$     & 46.0a   & 2.0        & 5100 & 150       & 2.2       & 0.3            & 0.011         & Ya    & \citet{2004ApJ...603..708C}       \\ \hline
HE 0134-1519               & 24.2724039774 & $-15.0729979538$ & $-4.0$  & 0.2       & 1.00   & 0.26      & 244   & 1        & 5500 & 100       & 3.2       & 0.3            & 0.016         & N   & \citet{2015ApJ...807..173H}       \\ \hline
SDSS J014036.21+234458.1            & 25.1509195676 & 23.7495011637  & $-4.0$  & 0.3       & 1.1  & 0.3       & $-197$a  & 1       & 5703 & 100       & 4.7      & 0.3            & 0.114         & Ya   & 
\citet{2013ApJ...762...26Y}       \\ \hline
BD+44 493                  & 36.7072451683 & 44.9629239592  & $-4.3$ & 0.2       & 1.2  & 0.2       & $-150.14$ & 0.63        & 5430 & 150       & 3.4       & 0.3            & 0.079         & N    & \citet{2013ApJ...773...33I}       \\ \hline
HE 0233-0343    & 39.1241380137 & $-3.50167460698$ & $-4.7$  & 0.2       & 3.48  & 0.24      & 64    & 1         & 6300 & 100       & 3.4       & 0.3            & 0.022         & N    & \citet{2015ApJ...807..173H}       \\ \hline
BPS CS 22963-0004    & 44.1940476203 & $-4.85483952327$ & $-4.09$ & 0.15      & 0.40   & 0.23      & 292.4   & 0.2         & 5060 & 42        & 2.15      & 0.16           & 0.045         & N    & \citet{Roederer2014}       \\ \hline
SDSS J030444.98+391021.1        & 46.1874375223 & 39.1725764233  & $-4.0$  & 0.2       & 0.7   & $-$     & 87   & 8         & 5859 & 13        & 5.0       & 0.5            & 0.111         & N    & \citet{Aguado2017b}       \\ \hline
SMSS J031300.36-670839.3   & 48.2515614545 & $-67.1442601577$ & $<-6.53$  & $-$     & 4.5   & 0.2       & 298.5a    & 0.5       & 5125 & $-$        & 2.3       & $-$           & 0.032         & N    & \citet{Keller2014aa}, \citet{Nordlander2017}       \\ \hline
HE 0330+0148                    & 53.158696449  & 1.96666957231  & $-4.0$ & 0.1       & 2.6   & $-$     & $-33.6$a   & 1       & 4100 & 200       & 5.2      & 0.1            & 0.094         & Y    & \citet{Plez}       \\ \hline
HE 0557-4840    & 89.6636087844 & $-48.6658029727$ & $-4.8$  & 0.2       & 1.65  & $-$      & 211.9  & 0.8        & 4900 & 100       & 2.2       & 0.3            & 0.037         & N    & \citet{2007ApJ...670..774N}       \\ \hline
SDSS J081554.26+472947.5            & 123.976115075 & 47.4965559814  & $<-5.8$  & $-$     & $>$5.0   & $-$    & $-95$   & 23        & 6215 & 82        & 4.7       & 0.5            & 0.063         & N    & \citet{Aguado2018a}       \\ \hline
SDSS J092912.32+023817.0            & 142.301366238 & 2.63806158906  & $-4.97$ & $-$      & $<3.91$  & $-$      & 388.3  & 10.4        & 5894 & $-$        & 3.7       & $-$           & 0.053         & Y    & \citet{2015AA...579A..28B}, \citet{2016AA...595L...6C}      \\ \hline
SDSS J094708.27+461010.0        & 146.784471294 & 46.1694746754  & $-4.1$  & 0.2       & 1.0   & 0.4       & $-5$    & 12        & 5858 & 73        & 5.0       & 0.5            & 0.013         & N   & \citet{Aguado2017}       \\ \hline
HE 1012-1540    & 153.722814524 & $-15.9314366402$ &$ -4.17$ & 0.16      & 2.2   & $-$      & 225.8a  & 0.5        & 5230 & 32        & 2.65      & 0.2            & 0.061         & N    & \citet{Roederer2014}       \\ \hline
SDSS J102915+172927        & 157.313121378 & 17.4910907404  & $-4.99$ & 0.06      & $<$0.7   & $-$      & $-35$   & 4         & 5850 & 100      & 4.0       & 0.2            & 0.023         & N    & \citet{Caffau2011aa}       \\ \hline
SDSS J103402.70+070116.6 & 158.511301205 & 7.02129528322 & $-4.01$ & 0.14 & $-$ & $-$ & 153 & 3 & 6270 & $-$ & $4.0$ & $-$ & 0.02 & N & \citet{Bonifacio2018}\\ \hline
SDSS J103556.11+064143.9            & 158.983818359 & 6.6955582264   & $<-5$ & $-$      & 3.08  & $-$       & $-45$  & 6         & 6262 & $-$        & 4       & $-$            & 0.024         & N   & \citet{2015AA...579A..28B}       \\ \hline
SDSS J105519.28+232234.0        & 163.830333515 & 23.3761158455  & $-4.00$  & 0.07      & $<$0.7   & $-$       & 62    & 4        & 6232 & 28        & 4.9       & 0.1           & 0.015         & N    & \citet{Aguado2017b}       \\ \hline
SDSS J120441.38+120111.5   & 181.172452065 & 12.019865284   & $-4.34$ & 0.05      & $<$1.45  & $-$       & 51   & 3        & 5917 & $-$        & 3      & $-$            & 0.024         & N   & \citet{2015ApJ...809..136P}       \\ \hline
SDSS J124719.46-034152.4        & 191.831114232 & $-3.69791795379$ & $-4.11$ & 0.18      & $<$1.61  & $-$       & 84   & 6         & 6332 & $-$         & 4      & $-$            & 0.022         & N    & \citet{2013AA...560A..71C}       \\ \hline
LAMOST J125346.09+075343.1 & 193.44189217  & 7.89526036289  & $-4.02$ & 0.06      & 1.59  & $-$       & 78.0    & 0.4         & 6030 & 135       & 3.65      & 0.16           & 0.025         & N    & \citet{2015PASJ...67...84L}       \\ \hline
SDSS J131326.89-001941.4   & 198.3620349838832 &  $-0.3281488686298$ &  $-4.7$ &0.2 &2.8 &0.3 & 268 & 4&  5525 & 106 &3.6 & 0.5 & 0.024 & Y & \citet{Allende15},\\ 
&&&&&&&&&&&&&&& \citet{FrebelJ1313}, \citet{Aguado2017b}\\ \hline
HE 1310-0536    & 198.379940261 & $-5.87014820763$ & $-4.2$  & 0.2       & 2.36  & 0.23      & 113.2   & 1.7         & 5000 & 100       & 1.9       & 0.3            & 0.037          & N    &       \citet{2015ApJ...807..173H}  \\ \hline
HE 1327-2326               & 202.524748159 & $-23.6971386187$ & $-5.96$ & $-$       & 3.78  &$-$      & 64.4a   & 1.3        & 6200 & 100      & 3.7       & 0.3            & 0.066         & N   & \citet{0004-637X-684-1-588}       \\ \hline
HE 1424-0241           & 216.668044499 & $-2.90763517546$ & $-4.05$ & $-$       & $<$0.63  & $-$       & 59.8    & 0.6         & 5260 & $-$      & 2.66      & $-$            & 0.055  & N    & \citet{2013ApJ...762...25N}, \citet{2008ApJ...672..320C}       \\ \hline
SDSS J144256.37-001542.7        & 220.734907425 & $-0.26188939275$ & $-4.09$ & 0.21      & $<$1.59  & $-$      & 225  & 9         & 5850 & $-$        & 4       &$-$          & 0.036         & N    & \citet{2013AA...560A..15C}       \\ \hline
Pristine221.8781+9.7844           & 221.878064787 & 9.78436859397  & $-4.66$ & 0.13      & $<$1.76  & $-$     & $-$149.0  & 0.5         & 5792 & 100       & 3.5       & 0.5            & 0.020         & N    & \citet{Pristine221}       \\ \hline
SDSS J164234.48+443004.9        & 250.643694345 & 44.5013644484  & $-4.0$ & 0.2       & 0.55  & 0.0       & $-136$  & 4         & 6280 & 150       & 5.0       & 0.3            & 0.011         & N    & \citet{Aguado2016}       \\ \hline
SDSS J173403.91+644633.0        & 263.516273652 & 64.7758235012  & $-4.3$  & 0.2       & 3.1   & 0.2       & $-258$ & 13        & 6183 & 78        & 5.0       & 0.5            & 0.028         & N    & \citet{Aguado2017}       \\ \hline
SDSS J174259.67+253135.8            & 265.748669215 & 25.526636261   & $-4.8$  & 0.07      & 3.6  & $0.2$       & $-221.93$ & 10.0        & 6345 & $-$         & 4       & $-$            & 0.055          & N    &\citet{2015AA...579A..28B}       \\ \hline
2MASS J18082002-5104378    & 272.083464041 & $-51.0771900644$ & $-4.07$ & 0.07      & $<0.5$   & $-$       & $16.54$     & $0.12$         & 5440 & 100      & 3.0       & 0.2            & 0.101         & Y    & \citet{2016AA...585L...5M}\\ 
&&&&&&&&&&&&&&& \citet{Schlaufman18}       \\ \hline
BPS CS 22891-0200    & 293.829490257 & $-61.7067706698$ & $-4.06$ & 0.15      & $-$   & $-$       & 131   & 10        & 4490 & 33        & 0.5       & 0.1           & 0.068         & N   & \citet{Roederer2014}        \\ \hline
BPS CS 22885-0096    & 305.213220651 & $-39.8917320574$ & $-4.21$ & 0.07      &$-$   & $-$   & $-248$  & 10        & 4580 & 34        & 0.75      & 0.15           & 0.048         & N    & \citet{Roederer2014}        \\ \hline
BPS CS 22950-0046    & 305.368323431 & $-13.2760006492$ & $-4.12$ & 0.14      & $-$  & $-$       & 111   & 10        & 4380 & 32        & 0.5       & 0.1           & 0.054         & N   & \citet{Roederer2014}        \\ \hline
BPS CS 30336-0049    & 311.348055352 & $-28.7099758468$ & $-4.04$ & 0.09      & $-0.28$ & 0.31      &$ -236.6$  & 0.8         & 4827 & 100       & 1.5      & 0.2            & 0.054         & N    & \citet{0004-637X-681-2-1524}      \\ \hline
HE 2139-5432               & 325.676864649 & $-54.3119357441$ & $-4.02$ &$-$      & $-$   & $-$      & 105a  & 3       & 5457 & 44        & 2.0       & 0.2            & 0.017         & Ya    & \citet{2013ApJ...762...25N}       \\ \hline
HE 2239-5019    & 340.611864594 &$-50.0669213083$ & $-4.2$  & 0.2       & $<$1.7   & $-$       & 368.7   & 0.5         & 6100 & 100       & 3.5       & 0.3            & 0.010         & N    & \citet{2015ApJ...807..173H}      \\ \hline
HE 2323-0256    & 351.62419731  & $-2.66612144628 $& $-4.38$ & 0.15      & $-$  & $-$       & $-125.8$a  & 0.3        & 4630 & 34        & 0.95      & 0.13           & 0.043         & N    &  \citet{Roederer2014}    \\ \hline
\end{tabular}
}
\label{literaturetable}
\end{table*}

%\pagebreak
%\clearpage

\begin{table*}
\centering
\caption{Gaia properties of the stars. Coordinates at J$2015.5$, the  dereddened G$_0$, BP$_0$ and RP$_0$ magnitudes, proper motion $\mu_{\alpha}$, $\mu_{\delta}$ and the parallax $\varpi$ for the analysed sample of UMPs \citep[\url{https://gea.esac.esa.int/archive/}]{2016AA...595A...1G,2018arXiv180409365G} are listed. G$_0$, BP$_0$ and RP$_0$ magnitudes are dereddened using the  \citet{1998ApJ...500..525S} extinction map.  The parallaxes $\varpi$ are not corrected for the offset $\varpi_0=0.029\mas$.}
\resizebox{\textwidth}{!}{
\begin{tabular}{llllllllllllllll}
\hline
Identifier          &     $\alpha_{J2015.5}$ & $\delta_{J2015.5}$ & Gaia id  & G$_0$   & $\delta_{G}$  & BP$_0$ & $\delta_{BP}$    & RP$_0$              & $\delta_{RP}$               & $\mu_{\alpha}$    & $\delta_{\mu_{\alpha}}$ & $\mu_{\delta}$    & $\delta_{\mu_{\delta}}$ & $\varpi$ & $\delta_{\varpi}$ \\ 
                           &       (deg)              &  (deg)             &              &          (mag)         &       (mag)                &     (mag)               &       (mag)                &            (mag)        &      (mag)       &  ($\masyr$)       &       ($\masyr$)      &    ($\masyr$)      &    ($\masyr$)          &   ($\mas$)       &     ($\mas$)      \\ \hline
HE 0020-1741 & 5.68699047782 & -17.40811466246 & 2367173119271988480 & 12.5609 & 0.00017 & 13.0699 & 0.00095 & 11.904 & 0.00055 & 14.424 & 0.064 & $-4.546$ & 0.043 & 0.1456 & 0.0384 \\ \hline
SDSS J0023+0307            & 5.80835977813   & 3.1327843082    & 2548541852945056896 & 17.5638 & 0.001   & 17.7947 & 0.0074   & 17.1246 & 0.0074   & 3.743   & 0.318       & $-13.912$  & 0.187        & 0.2697   & 0.1406          \\ \hline
HE 0044-3755                  & 11.65089731416  & $-37.65935345272$ & 5000753194373767424 & 11.6633 & 0.0003  & 12.1427 & 0.0009   & 11.0310  & 0.0009   & 15.234  & 0.061       & $-7.529$   & 0.041        & 0.2152   & 0.0344          \\ \hline
HE 0057-5959               & 14.97496136508  &$ -59.72497472878$ & 4903905598859396480 & 15.0507 & 0.0004  & 15.3857 & 0.0025   & 14.5292 & 0.0025   & 2.389   & 0.042       & $-10.522$  & 0.041        & 0.1982   & 0.0254          \\ \hline
HE 0107-5240               & 17.37149810186  & $-52.40951706252$ & 4927204800008334464 & 14.9334 & 0.0003  & 15.3232 & 0.0019   & 14.3638 & 0.0019   & 2.414   & 0.033       & $-3.735$   & 0.035        & 0.0789   & 0.0258          \\ \hline
HE 0134-1519               & 24.27251527664  & $-15.07304490506$ & 2453397508316944128 & 14.227  & 0.0003 & 14.5501 & 0.0022   & 13.7181 & 0.0022   & 24.961  & 0.056       &$ -10.905$  & 0.039        & 0.3454   & 0.0299          \\ \hline
SDSS J014036.21+234458.1            & 25.15092436121  & 23.74940873996  & 290930261314166528  & 15.0495 & 0.0006  & 15.3423 & 0.0034   & 14.575  & 0.0034   & 1.019   & 0.176       &$-21.466$  & 0.091        & 1.0482   & 0.0562          \\ \hline
BD+44 493                  & 36.70796538815  & 44.96278519908  & 341511064663637376  & 8.6424  & 0.0005 & 8.9634  & 0.0016   & 8.1758  & 0.0016   & 118.359 & 0.141       & $-32.229$  & 0.105        & 4.7595   & 0.0660           \\ \hline
HE 0233-0343    & 39.12435352835  & $-3.50172027632$  & 2495327693479473408 & 15.2126 & 0.0005  & 15.4433 & 0.0027   & 14.8029 & 0.0027   & 49.962  & 0.073       &$ -10.607 $ & 0.072        & 0.7925   & 0.0545          \\ \hline
BPS CS 22963-0004    & 44.1941414394   & $-4.85485100336$  & 5184426749232471808 & 14.6906 & 0.0005  & 14.9991 & 0.0024   & 14.1973 & 0.0024   & 21.712  & 0.058       & $-2.666 $  & 0.059        & 0.2220    & 0.0364          \\ \hline
SDSS J030444.98+391021.1        & 46.18743595787  & 39.17249343121  & 142874251765330944  & 17.0085 & 0.0019  & 17.3215 & 0.0088   & 16.5085 & 0.0088   &$ -0.282 $ & 0.336       & $-19.276$  & 0.241        & 0.0752   & 0.1929          \\ \hline
SMSS J031300.36-670839.3   & 48.25163934361  & $-67.14425547143$ & 4671418400651900544 & 14.4342 & 0.0003  & 14.8379 & 0.0018   & 13.8545 & 0.0018   & 7.027   & 0.032       & 1.088    & 0.03         & 0.0981   & 0.0162          \\ \hline
HE 0330+0148                    & 53.15953261866  & 1.96344241611   & 3265069670684495744 & 13.0859 &0.0004  & 13.8664 & 0.0032   & 12.2728 & 0.0032   & 194.093 & 0.453       & $-749.533$ & 0.499        & 12.7174  & 0.2106          \\ \hline
HE 0557-4840    & 89.66361346726  &$ -48.66579980934 $& 4794791782906532608 & 15.0976 & 0.0004 & 15.5156 & 0.0028   & 14.4984 & 0.0028   & 0.718   & 0.043       & 0.735    & 0.044        & 0.0389   & 0.0207          \\ \hline
SDSS J081554.26+472947.5            & 123.97602487635 & 47.49645166114  & 931227322991970560  & 16.5417 & 0.0006  & 16.8056 & 0.0057   & 16.1052 & 0.0057   & $-14.154$ & 0.135       & $-24.229$  & 0.09         & 0.4441   & 0.0837          \\ \hline
SDSS J092912.32+023817.0            & 142.30134736257 & 2.63804791153   & 3844818546870217728 & 17.8302 & 0.0023  & 18.136  & 0.0316   & 17.3618 & 0.0316   &$ -4.379$  & 0.342       & $-3.177$   & 0.364        & 0.1276   & 0.1872          \\ \hline
SDSS J094708.27+461010.0        & 146.78455769932 & 46.16940656739  & 821637654725909760  & 18.7343 & 0.0021  & 19.0195 & 0.0221   & 18.2783 & 0.0221   & 13.898  & 0.317       & $-15.819 $ & 0.332        & 0.1989   & 0.2299          \\ \hline
HE 1012-1540    & 153.7223563828  & $-15.93131552666$ & 3751852536639575808 & 13.7019 & 0.0004  & 14.0084 & 0.0033   & 13.2135 & 0.0033   & $-102.32$ & 0.046       & 28.13    & 0.04         & 2.5417   & 0.0280          \\ \hline
SDSS J102915+172927        & 157.31307233934 & 17.49107327845  & 3890626773968983296 & 16.4857 & 0.0013  & 16.7665 & 0.0062   & 15.9976 & 0.0062   & $-10.863 $& 0.146       &$ -4.056$  & 0.113        & 0.7337   & 0.0780           \\ \hline
SDSS J103402.70+070116.6 & 158.51126738928 & 7.02126631404 & 3862721340654330112& 17.1906 & 0.0018 & 17.4051 & 0.0227 & 16.7943 & 0.0063& $-7.795$ & 0.236 & $-6.728$ & 0.291 & 0.2874 & 0.1367 \\ \hline
SDSS J103556.11+064143.9            & 158.98383317025 & 6.69554785085   & 3862507691800855040 & 18.3472 & 0.0034  & 18.623  & 0.0197   & 17.9584 & 0.0197   & 3.416   & 0.403       &$ -2.41  $  & 0.369        & $-0.3912$  & 0.3163          \\ \hline
SDSS J105519.28+232234.0        & 163.83036912138 & 23.37606935407  & 3989873022818570240 & 17.5182 & 0.0025  & 17.7015 & 0.0317   & 17.1298 & 0.0317   & 7.591   & 0.291       &$ -10.798 $ & 0.324        & 0.5909   & 0.1821          \\ \hline
SDSS J120441.38+120111.5   & 181.17245380263 & 12.01984412118  & 3919025342543602176 & 16.027  & 0.0005  & 16.3239 & 0.0043   & 15.5497 & 0.0043   & 0.395   & 0.11        & $-4.915$   & 0.067        & 0.2454   & 0.0656          \\ \hline
SDSS J124719.46-034152.4        & 191.83107728926 & $-3.69791015204$  & 3681866216349964288 & 18.1908 & 0.0016  & 18.3958 & 0.0118   & 17.7716 & 0.0118   & $-8.562$  & 0.439       & 1.812    & 0.226        & 0.3075   & 0.2098          \\ \hline
LAMOST J125346.09+075343.1 & 193.44198364753 & 7.895007511     & 3733768078624022016 & 12.228  &0.0002& 12.4603 & 0.0011   & 11.8239 & 0.0011   & 21.045  & 0.082       & $-58.727 $ & 0.049        & 1.4053   & 0.0378          \\ \hline
SDSS J131326.89-001941.4 & 198.36201866349555 & $-0.32817714440715445$ & 3687441358777986688 & 16.3560 & 0.0010 & 16.7237 & 0.0058 & 15.8183 & 0.0710 & $-3.790$ & 0.160 & $-6.567$ & 0.078 & 0.2976 & 0.0972     \\ \hline
HE 1310-0536    & 198.37991838382 & $-5.8701554707$   & 3635533208672382592 & 14.0256 & 0.0004 & 14.5363 & 0.0021   & 13.3649 & 0.0021   &$ -5.054 $ & 0.053       & $-1.687$   & 0.042        & 0.0078   & 0.0342          \\ \hline
HE 1327-2326               & 202.52450119109 & $ -23.69694272263$ & 6194815228636688768 & 13.2115 & 0.0004  & 13.45   & 0.0019   & 12.8012 & 0.0019   &$ -52.524 $& 0.04        & 45.498   & 0.035        & 0.8879   & 0.0235          \\ \hline
HE 1424-0241           & 216.66802803117 & $-2.90764744641$  & 3643332182086977792 & 15.0437 & 0.0007 & 15.3934 & 0.0046   & 14.5017 & 0.0046   & $-3.82$   & 0.087       &$ -2.85 $   & 0.066        & 0.1152   & 0.0469          \\ \hline
SDSS J144256.37-001542.7        & 220.73490626598 & $-0.26186035888 $ & 3651420563283262208 & 17.5635 & 0.0023  & 17.8216 & 0.0277   & 17.1364 & 0.0277   & $-0.269 $ & 0.315       & 6.743    & 0.396        & $-0.3910$   & 0.2981          \\ \hline
Pristine221.8781+9.7844    & 221.87803086877 & 9.78436834556   & 1174522686140620672 & 16.1846 & 0.0009  & 16.4688 & 0.0053   & 15.706  & 0.0053   &$ -7.763$  & 0.110        & $-0.058$   & 0.116        & 0.1187   & 0.0940           \\ \hline
SDSS J164234.48+443004.9        & 250.643641407   & 44.50138608236  & 1405755062407483520 & 17.4658 & 0.0012  & 17.6987 & 0.0112   & 17.0356 & 0.0112   &$ -8.769$  & 0.149       & 5.025    & 0.244        & 0.3122   & 0.0906          \\ \hline
SDSS J173403.91+644633.0        & 263.51630029934 & 64.77581642801  & 1632736765377141632 & 19.1198 & 0.0038  & 19.3849 & 0.0465   & 18.7074 & 0.0465   & 2.638   & 0.44        & $-1.643$   & 0.553        & $-0.1052$  & 0.2702          \\ \hline
SDSS J174259.67+253135.8            & 265.74864014534 & 25.52658646063  & 4581822389265279232 & 18.5115 & 0.0022  & 18.7628 & 0.0248   & 18.0991 & 0.0248   &$ -6.093 $ & 0.248       & $-11.567$  & 0.292        & $-0.1628 $ & 0.1870           \\ \hline
2MASS J18082002-5104378    & 272.08342547713 & $-51.07724449784 $& 6702907209758894848 & 11.488  & 0.0003  & 11.7853 & 0.0024   & 11.0119 & 0.0024   & $-5.627 $ & 0.068       & $-12.643$  & 0.058        & 1.6775   & 0.0397          \\ \hline
BPS CS 22891-0200    & 293.82944462026 & $-61.70676742367$ & 6445220927325014016 & 13.4478 & 0.0003  & 13.9306 & 0.0017   & 12.8053 & 0.0017   &$ -5.024 $ & 0.053       & 0.754    & 0.036        & 0.1135   & 0.0342          \\ \hline
BPS CS 22885-0096    & 305.21319576813 & $-39.89176180812 $& 6692925538259931136 & 12.9385 & 0.0003  & 13.3482 & 0.0017   & 12.3514 & 0.0017   & $-4.434 $ & 0.038       & $-6.91$    & 0.028        & 0.1708   & 0.0247          \\ \hline
BPS CS 22950-0046    & 305.36833037469 & $-13.27600846442 $& 6876806419780834048 & 13.7403 & 0.0002  & 14.2631 & 0.0011   & 13.0627 & 0.0011   & 1.57    & 0.045       & $-1.815$   & 0.028        & 0.0587   & 0.0270           \\ \hline
BPS CS 30336-0049    & 311.34804708033 &$ -28.71001086007 $& 6795730493933072128 & 13.5803 & 0.0002  & 14.074  & 0.0013   & 12.9283 & 0.0013   & $-1.685 $ & 0.038       & $-8.132$   & 0.027        & 0.0418   & 0.0227          \\ \hline
HE 2139-5432               & 325.676883449   & $-54.31195504869$ & 6461736966363075200 & 14.9386 & 0.0003  & 15.2991 & 0.0017   & 14.4    & 0.0017   & 2.547   & 0.046       & $-4.484 $  & 0.041        & $-0.0067 $ & 0.0298          \\ \hline
HE 2239-5019    & 340.61191653735 &$ -50.06702317874$ & 6513870718215626112 & 15.6038 & 0.0007  & 15.8336 & 0.0034   & 15.2107 & 0.0034   & 7.744   & 0.054       &$ -23.66$   & 0.076        & 0.2200     & 0.0545          \\ \hline
HE 2323-0256    & 351.6242048175  & $-2.66612932812 $ & 2634585342263017984 & 13.9922 & 0.0004  & 14.4286 & 0.0031   & 13.3832 & 0.0031   & 1.742   & 0.062       & $-1.831$   & 0.048        & 0.0038   & 0.0359          \\ \hline
\end{tabular}
}

\label{gaiatable}
\end{table*}

%\pagebreak
%\clearpage

\begin{table*}
\centering
\caption{Inferred stellar parameters for the stars in the sample. Distances D, effective temperatures T$_{eff}$ and surface gravities log(g) obtained in this work for the UMPs sample. If a second peak in the PDF is present, an estimate of the subtended area around the two peaks within $\pm 3\sigma$ is shown (Area= $\int_{d_1-3\sigma}^{d_1+3\sigma} P(r) dr$). The column \textit{Prior} indicates the MW prior used for inferring the parameters (i.e. H means halo prior, D$+$H indicates the disc$+$halo prior). }
\begin{tabular}{lllllllll}
\hline
Identifier          & D  & $\delta_D$ & T$_{eff}$ &$\delta_{T_{eff}}$ & log(g) & $\delta_{log(g)}$ & Area & Prior  \\ 
                           & ($\kpc$)  & ($\kpc$)      & (K)    & (K)           & (dex)    & (dex)           &    &   \\ \hline
 HE 0020-1741 &10.3 &0.4 &   4774 & 20 &   1.05 & 0.05 & & H \\     
                         &10.3 &0.4 &   4774 & 20 &   1.05 & 0.05 & & D$+$H \\   \hline            
SDSS J0023+0307            & 2.710   & 0.139      & 6116 & 66          & 4.6      & 0.1             & 88\%  & H \\ 
                           & 11.03 & 0.73      & 6047& 146        & 3.4      & 0.1             & 12\%   & H \\  
                           & 2.693 & 0.136  & 6108  &  65 & 4.6 & 0.1 & 99.6\%  & D$+$H \\
                           &11.02 &  0.74 & 6050  & 154  & 3.4  & 0.1 & 0.4\% & D$+$H \\ \hline
HE 0044-3755       & 5.70  &  0.25     &  4852 & 22           & 1.2      & 0.1             &      & H \\
                                                     & 5.65& 0.26 & 4863  & 23  &  1.2& 0.1 &  & D$+$H \\ \hline
HE 0057-5959               & 6.80  & 0.71      & 5483 & 42          & 2.7      & 0.1             &      & H \\ 
                                      & 6.50& 0.72  &  5501 &  44 &2.7  & 0.1 &  & D$+$H \\ \hline
HE 0107-5240               & 14.3  & 1.0      & 5141 & 32          & 1.9      & 0.1             &      & H \\ 
                                   & 14.2& 1.0  & 5141 & 32 & 1.9 &0.1  &  & D$+$H \\ \hline
HE 0134-1519               & 3.75   & 0.33      & 5572 & 90          & 2.9      & 0.1             &      & H \\ 
                                   &3.61 & 0.30  &  5589 & 37  &2.9  & 0.1 &  & D$+$H \\ \hline
SDSS J014036.21+234458.1            & 0.762  & 0.022      & 5963 & 41          & 4.6      & 0.1             &      & H \\ 
                                   & 0.761& 0.022  & 5962  &  40 & 4.6 & 0.1 &  & D$+$H \\ \hline
BD+44 493                  & 0.211  & 0.003      & 5789 & 19          & 3.2      & 0.1             &      & H \\ 
                                   & 0.211& 0.003  & 5794  & 20  & 3.2 & 0.1 &  & D$+$H \\ \hline
HE 0233-0343    & 1.090   & 0.043      & 6331 & 47          & 4.5      & 0.1              &      & H \\ 
                                   &1.088 & 0.043  & 6327  & 47  & 4.5 & 0.1 &  & D$+$H \\ \hline
BPS CS 22963-0004    & 4.47  & 0.42      & 5589 & 42          & 2.9      & 0.1             &      & H \\ 
                                   &4.36 & 0.39  & 5601  & 43   & 3.0 &0.1  &  & D$+$H \\ \hline
SDSS J030444.98+391021.1        & 14.9 & 1.3      & 5547 & 39          & 2.8      & 0.1             &   99\%   & H \\ 
                           & 1.505  & 0.071      & 5649 & 68          & 4.7      & 0.1             &   1\%   & H \\ 
                              & 14.3&  2.5 &  5548 & 74  & 2.8 & 0.2 & 79\% & D$+$H \\
                           &1.503 & 0.071  &  5648 & 68  & 4.7 & 0.1  & 21\% & D$+$H \\ \hline
SMSS J031300.36-670839.3   & 12.0 & 0.8      & 5111 & 31          & 1.8      & 0.1             &      & H \\ 
                           &12.1 & 0.8  & 5111  & 32  & 1.8 &0.1 &  & D$+$H \\ \hline
HE 0330+0148                    & 0.075  & 0.001      & 4454 & 1           & 5.0      & 0.1            &      & H \\ 
                           & 0.075& 0.001  &  4460 &  1 & 5.0 & 0.1  &  & D$+$H \\ \hline
HE 0557-4840    & 20.0 & 1.3      & 5017 & 28          & 1.6      & 0.1             &      & H \\ 
                           & 20.0&1.3   &  5018 & 30 & 1.6 &0.1 &  & D$+$H \\ \hline
SDSS J081554.26+472947.5            & 1.591  & 0.067      & 6034 & 56          & 4.6      & 0.1              &      & H \\  
                                                     &1.588 &  0.066 & 6031  &  56 &  4.6& 0.1 &  & D$+$H \\ \hline
SDSS J092912.32+023817.0            & 15.6  & 2.6      & 5708 & 124         & 3.1      & 0.2             &  68\%    & H \\ 
                           & 2.398  & 0.205      & 5775 & 122         & 4.7      & 0.1             &   32\%   & H \\ 
                                                         &2.367 &  0.198 & 5756  &  120 & 4.7 & 0.1 & 95\% & D$+$H \\
                           &15.5 & 2.6  &5713   &  125 & 3.1 & 0.2 &5\%  & D$+$H \\ \hline
SDSS J094708.27+461010.0        & 3.84  & 0.30      & 5854 & 110         & 4.7      & 0.1             & 82\%      & H \\ 
                           & 21.9 & 2.0      & 5801 & 118         & 3.2      & 0.1             & 18\%    & H \\
                                                         & 3.76& 0.28  &  5823 &  55 & 4.7 & 0.1 &98\%  & D$+$H \\
                           & 21.9&2.0   & 5802  &  120 & 3.2 & 0.1 & 2\% & D$+$H \\ \hline
HE 1012-1540    & 0.384  & 0.004      & 5872 & 16          & 4.7      & 0.1              &      & H \\ 
                           & 0.384&0.004   &  5870 & 16  & 4.7 & 0.1 &  & D$+$H \\ \hline
SDSS J102915+172927        & 1.281  & 0.051      & 5764 & 57          & 4.7      & 0.1              &      & H \\ 
                           &1.278 & 0.050  & 5761  &  56 & 4.7 & 0.1 &  & D$+$H \\ \hline

\end{tabular} 

\label{distancetable}
\end{table*}

\begin{table*}
\contcaption{from previous page.}
\centering
\begin{tabular}{lllllllll}
\hline
Identifier          & D  & $\delta_D$ & T$_{eff}$ & $\delta_{T_{eff}}$ & log(g) & $\delta_{log(g)}$ & Area & Prior\\ 
                           & ($\kpc$)  & ($\kpc$)      & (K)    & (K)           & (dex)    & (dex)           &   &   \\ \hline
SDSS J103402.70+070116.6 & 2.79 & 0.26 & 6366 & 110 & 4.5 & 0.1 & 89\% & H \\
                                       & 8.28 & 0.64 & 6333 & 211 & 3.6 & 0.1 & 11\% & H \\ 
                                                                     & 2.75&  0.25 &  6330 & 110  & 4.5 &0.1  & 99.4\% & D$+$H \\
                           & 8.18&  0.65 & 6320  & 200  & 3.6 &  0.1& 0.6\% & D$+$H \\ \hline
SDSS J103556.11+064143.9            & 3.97  & 0.35      & 6144 & 110         & 4.6      & 0.1              &    67\%  & H \\ 
                           & 15.6 & 1.2      & 6072 & 168         & 3.5      & 0.1             &  33\%    & H \\ 
                                                         &3.88 & 0.32  & 6114  &  106 & 4.6 & 0.1  & 95.5\% & D$+$H \\
                           & 15.6& 1.2  &  6073 & 175  &  3.5& 0.1 & 0.5\% & D$+$H \\ \hline
SDSS J105519.28+232234.0        & 3.49   & 0.45      & 6452 & 147         & 4.5      & 0.1             &   96\%  & H \\ 
                           & 8.84  & 0.94       & 6581 & 248         & 3.8      & 0.2             &  4\%    & H \\ 
                                                        &3.30 & 0.39  & 6387  & 138  & 4.5 & 0.1 & 99.7\% & D$+$H \\
                           &8.79 & 0.99  & 6606  &  257 & 3.8 & 0.2 & 0.3\% & D$+$H \\ \hline
SDSS J120441.38+120111.5   & 7.03  & 0.54      & 5679 & 56          & 3.1      & 0.1             &      & H \\ 
                                                      &6.96 &0.53   &5686   & 59 & 3.1 &  0.1&  & D$+$H \\ \hline
 SDSS J124719.46-034152.4        & 4.17  & 0.32      & 6296 & 92          & 4.5      & 0.1             &   92\%   & H \\ 
                           & 13.5 & 1.0      & 6256 & 196         & 3.6      & 0.1             &    8\%  & H \\ 
                                                                                   & 4.09& 0.30  & 6273  &  90 & 4.5 & 0.1 & 99\% & D$+$H \\
                           & 13.4& 1.0  &  6263 & 205  & 3.6 & 0.1 & 1\% & D$+$H \\ \hline
 LAMOST J125346.09+075343.1 & 0.766  & 0.016      & 6598 & 52          & 3.8      & 0.1              &      & H \\ 
                            &0.766 & 0.016  & 6608  & 52  & 3.8 & 0.1 &  & D$+$H \\ \hline
SDSS J131326.89-001941.4 & 8.59 & 2.86 & 5649&171 & 3.1&0.3 & 99.96\% & H   \\
                                               &  1.765 &     0.248 & 6278&171 &4.5 & 0.1&  0.04\% & H \\    
                                               &  8.07 & 2.70  & 5687&185 & 3.1& 0.3& 96.85\% & D$+$H \\
                                               &1.707 & 0.227 &6237 &164 & 4.6& 0.1& 3.15\% & D$+$H \\ \hline
HE 1310-0536    & 20.6 & 0.9      & 4788 & 20    & 1.0      & 0.1     &      & H \\ 
                           &20.6 & 0.9  & 4764  & 21  & 1.0 & 0.1 &  & D$+$H \\ \hline
HE 1327-2326               & 1.212  & 0.024      & 6581 & 52          & 3.8      & 0.1              &      & H \\ 
                           &1.212 &  0.024 &   6591&  51 & 3.8 & 0.1 &  & D$+$H \\ \hline
HE 1424-0241           & 10.3 & 1.0      & 5308 & 40          & 2.3      & 0.1             &      & H \\ 
                           &10.3 &1.0   &5308   & 40  & 2.3 & 0.1 &  & D$+$H \\ \hline
SDSS J144256.37-001542.7        & 11.3 & 1.0       & 5993 & 165         & 3.4      & 0.1             &   87\%   & H \\ 
                           & 2.683  & 0.266      & 6104 & 128         & 4.6      & 0.1              &  13\%    & H \\ 
                                                                                   & 2.634& 0.249  & 6079  &  124 & 4.6 &  0.1& 84\% & D$+$H \\
                           &11.3 &  1.0 &  5998 & 172  & 3.4 &  0.1& 16\% & D$+$H \\ \hline
Pristine221.8781+9.7844           & 7.36  & 0.55      & 5700 & 63          & 3.1      & 0.1             &      & H \\ 
                           &7.28 & 0.52  & 5710  &  65 & 3.1 & 0.1 &  & D$+$H \\ \hline                     
                                         
 SDSS J164234.48+443004.9        & 2.66  & 0.16       & 6149 & 77          & 4.6      & 0.1              &  99\%    & H \\ 
                           & 10.2 & 0.7      & 6126 & 163         & 3.5      & 0.1             &   1\%   & H \\ 
                                     & 2.64&0.16   &  6140 & 76  & 4.6 & 0.1 & 99.95\% & D$+$H \\
                           & 10.1&  0.7 & 6148  & 172  & 3.5 &  0.1 & 0.05\% & D$+$H \\ \hline
SDSS J173403.91+644633.0        & 5.46  & 1.02      & 6094 & 233         & 4.6      & 0.1             & 86\%     & H \\ 
                           & 21.8 & 3.0      & 6131 & 297         & 3.5      & 0.2             &   14\%   & H \\ 
                                     & 5.05& 0.79  &  5992 & 208  & 4.6 & 0.1 &  97\%& D$+$H \\
                           &21.7 &  3.0 &  6134 & 302  & 3.5 & 0.2 & 3\% & D$+$H \\ \hline
SDSS J174259.67+253135.8            & 4.46  & 0.52      & 6194 & 145         & 4.6      & 0.1              &   63\%   & H \\ 
                           & 16.6  & 1.4      & 6115 & 198         & 3.5      & 0.1             &   37\%   & H \\ 
                                     &4.34 & 0.48  &6162   &  140 & 4.6 & 0.1 & 94\% & D$+$H \\
                           &16.5 & 1.4  & 6118  & 206  & 3.5 &  0.1& 6\% & D$+$H \\ \hline
 2MASS J18082002-5104378    & 0.647  & 0.012      & 6124 & 44          & 3.5      & 0.1              &      & H \\ 
                                                       &0.647 & 0.012  &  6133 &  44 & 3.5 & 0.1 &  & D$+$H \\ \hline
BPS CS 22891-0200    & 14.7 & 0.5        & 4789 & 2           & 1.2      & 0.1              &      & H \\ 
                           &13.6 & 0.6  & 4836  &  22 & 1.2 & 0.1 &  & D$+$H \\ \hline

\end{tabular}

\end{table*}

\begin{table*}
\contcaption{from previous page.}
\centering
\begin{tabular}{lllllllll}
\hline
Identifier          & D  & $\delta_D$ & T$_{eff}$ &$\delta_{T_{eff}}$ & log(g) & $\delta_{log(g)}$ & Area & Prior\\ 
                           & ($\kpc$)  & ($\kpc$)      & (K)    & (K)           & (dex)    & (dex)           &   &   \\ \hline                          
  BPS CS 22885-0096    & 6.65   & 0.22      & 5068 & 16          & 1.7      & 0.1              &      & H \\ 
                           &6.61 & 0.38  &  5070 & 27  & 1.7 & 0.1  &  & D$+$H \\ \hline                         
 BPS CS 22950-0046   & 19.1 & 0.3      & $<$4780  &       $-$     & $<$1.0      & $-$              &      & H \\ 
                           & 19.1& 0.3  & $<$4780  & $-$  & $<$1.0  & $-$ &  & D$+$H \\ \hline
BPS CS 30336-0049    & 15.5 & 0.7      & 4809 & 20          & 1.1      & 0.1              &      & H \\ 
                           & 15.5& 0.7  &  4802 & 21  & 1.1 & 0.1 &  & D$+$H \\ \hline                          
HE 2139-5432               & 11.0  & 0.9      & 5259 & 34          & 2.1      & 0.1             &      & H \\ 
                           & 11.0& 0.9  &  5259 &  34 & 2.1 & 0.1 &  & D$+$H \\ \hline
HE 2239-5019    & 4.19  & 0.28      & 6195 & 179         & 3.5      & 0.1             &      & H \\ 
                           &4.13 & 0.16  &  6411 &  100 &  3.6& 0.1 &  & D$+$H \\ \hline
HE 2323-0256    & 14.2 & 0.6      & 4937 & 22          & 1.4      & 0.1             &      & H \\ 
                           & 14.2&0.6   & 4937  & 22  & 1.4 & 0.1 &  & D$+$H \\ \hline
\end{tabular}
\end{table*}

   \begin{table*}
   \def\arraystretch{2.0}
    \centering
    \caption{Inferred orbital parameters of the stars in the sample. Position (X,Y,Z), the apocentre and pericentre distances in the galactocentric frame,  the velocity (U,V,W) in the heliocentric frame, the eccentricity $\epsilon= (r_\mathrm{apo}-r_\mathrm{peri})/(r_\mathrm{apo} + r_\mathrm{peri})$ of the orbit, the $z$-component of the angular momentum, the energy and the kind of orbit (IH = inner halo with $r_\mathrm{apo}<30 \kpc$, OH = outer halo with $r_\mathrm{apo}>30 \kpc$, P = close to the MW plane, S = possible Sgr stream member, $\omega$ = possible $\omega$Cen member) are listed. For the unbound orbits, all the orbital parameters and the kind of orbit are denoted by NB.}
    \resizebox{\textwidth}{!}{
    \begin{tabular}{llllllllllllllll}
    \hline
    Identifier  & X   & Y  & Z  & U & V  & W & Apo & Peri  & $\epsilon$ & L$_\mathrm{z}$ & E & Orbit \\ 
      & ($\kpc$) & ($\kpc$) & ($\kpc$) & ($\kms$) & ($\kms$) & ($\kms$) & ($\kpc$) & ($\kpc$) & & ($\kms\kpc$) & (km$^{2}$ s$^{-2}$) &   \\ \hline
    HE 0020-1741 & 7.909$^{+0.0}_{-0.0}$ & 1.84$^{+0.0}_{-0.0}$ & -8.846$^{+0.0}_{-0.0}$ & -428.7$^{+0.0}_{-0.0}$ & -446.4$^{+0.0}_{-0.0}$ & -192.2$^{+0.0}_{-0.0}$ & 295.8$^{+0.0}_{-0.0}$ & 12.0$^{+0.0}_{-0.0}$ & 0.92$^{+0.0}_{-0.0}$ & -2311.5$^{+0.0}_{-0.0}$ & 63046.1$^{+0.0}_{-0.0}$  &IH        \\ \hline 
    SDSS J0023+0307 & 8.456$^{+0.039}_{-0.026}$ & 1.311$^{+0.113}_{-0.075}$ & -2.375$^{+0.133}_{-0.199}$ & 76.8$^{+3.9}_{-5.2}$ & -251.2$^{+8.6}_{-12.9}$ & 69.2$^{+8.1}_{-5.4}$ & 9.8$^{+0.0}_{-0.0}$ & 0.6$^{+0.1}_{-0.1}$ & 0.88$^{+0.03}_{-0.04}$ & 108.2$^{+97.9}_{-65.3}$ & -68950.0$^{+0.0}_{-0.0}$    & IH   \\ 
    & NB & NB & NB & NB & NB & NB & NB & NB & NB & NB & NB     &NB     \\ \hline 
HE 0044-3755 & 7.353$^{+0.028}_{-0.036}$ & -0.824$^{+0.036}_{-0.045}$ & -5.61$^{+0.246}_{-0.308}$ & -235.9$^{+11.0}_{-12.9}$ & -397.1$^{+17.8}_{-22.0}$ & -18.1$^{+2.5}_{-2.9}$ & 21.8$^{+4.6}_{-3.0}$ & 4.5$^{+0.7}_{-0.7}$ & 0.66$^{+0.01}_{-0.01}$ & -885.7$^{+102.1}_{-134.0}$ & -35273.2$^{+6822.9}_{-5458.3}$          & IH \\ \hline 
HE 0057-5959 & 6.077$^{+0.187}_{-0.22}$ & -3.206$^{+0.312}_{-0.367}$ & -5.839$^{+0.568}_{-0.668}$ & 206.5$^{+10.9}_{-9.8}$ & -456.5$^{+26.5}_{-29.9}$ & -129.0$^{+20.5}_{-19.4}$ & 31.3$^{+10.2}_{-5.9}$ & 9.0$^{+0.4}_{-0.3}$ & 0.56$^{+0.07}_{-0.06}$ & -1947.5$^{+229.5}_{-256.5}$ & -20642.9$^{+9464.8}_{-6625.3}$          & OH  \\ \hline 
HE 0107-5240 & 5.255$^{+0.167}_{-0.191}$ & -5.497$^{+0.334}_{-0.382}$ & -12.877$^{+0.782}_{-0.894}$ & 12.5$^{+2.4}_{-2.0}$ & -294.3$^{+18.3}_{-19.5}$ & 77.2$^{+8.7}_{-7.7}$ & 15.9$^{+1.0}_{-0.9}$ & 3.2$^{+1.5}_{-1.0}$ & 0.66$^{+0.08}_{-0.1}$ & -354.3$^{+98.9}_{-105.5}$ & -46879.6$^{+3604.2}_{-3604.2}$          & IH \\ \hline 
HE 0134-1519 & 9.025$^{+0.103}_{-0.082}$ & 0.244$^{+0.025}_{-0.019}$ & -3.679$^{+0.291}_{-0.369}$ & -302.5$^{+18.8}_{-23.8}$ & -416.8$^{+34.1}_{-43.2}$ & -197.1$^{+3.7}_{-3.3}$ & 70.2$^{+49.3}_{-20.7}$ & 4.1$^{+5.9}_{-0.6}$ & 0.87$^{+0.02}_{-0.07}$ & -1555.1$^{+313.6}_{-425.6}$ & 2196.4$^{+16677.9}_{-10006.7}$          & OH \\ \hline 
SDSS J014036.21+234458.1 & 8.447$^{+0.023}_{-0.025}$ & 0.414$^{+0.021}_{-0.023}$ & -0.473$^{+0.027}_{-0.024}$ & 132.8$^{+3.6}_{-3.6}$ & -153.5$^{+3.6}_{-4.3}$ & 61.5$^{+4.2}_{-4.7}$ & 11.4$^{+0.2}_{-0.1}$ & 2.5$^{+0.1}_{-0.1}$ & 0.64$^{+0.02}_{-0.02}$ & 884.9$^{+29.2}_{-30.8}$ & -61399.9$^{+465.4}_{-413.7}$          & P \\ \hline 
BD+44 493 & 8.157$^{+0.005}_{-0.005}$ & 0.131$^{+0.004}_{-0.004}$ & -0.054$^{+0.002}_{-0.002}$ & 30.6$^{+2.4}_{-2.6}$ & -184.5$^{+2.7}_{-2.8}$ & 51.8$^{+0.4}_{-0.4}$ & 8.3$^{+0.0}_{-0.0}$ & 1.5$^{+0.2}_{-0.1}$ & 0.69$^{+0.01}_{-0.03}$ & 549.6$^{+21.8}_{-22.6}$ & -76078.3$^{+236.5}_{-236.5}$          & P \\ \hline 
HE 0233-0343 & 8.62$^{+0.04}_{-0.04}$ & 0.063$^{+0.004}_{-0.004}$ & -0.913$^{+0.061}_{-0.056}$ & -175.5$^{+9.0}_{-8.6}$ & -209.8$^{+13.8}_{-13.3}$ & 27.5$^{+5.0}_{-5.2}$ & 11.9$^{+0.5}_{-0.4}$ & 1.0$^{+0.3}_{-0.3}$ & 0.85$^{+0.04}_{-0.05}$ & 344.6$^{+119.0}_{-114.2}$ & -62501.0$^{+1475.7}_{-1248.6}$          & P  \\ \hline 
BPS CS 22963-0004 & 10.739$^{+0.268}_{-0.231}$ & -0.087$^{+0.007}_{-0.008}$ & -3.62$^{+0.304}_{-0.352}$ & -421.8$^{+20.9}_{-24.0}$ & -359.1$^{+29.5}_{-34.1}$ & -39.2$^{+19.1}_{-16.6}$ & 155.8$^{+183.4}_{-55.0}$ & 3.0$^{+8.7}_{-1.1}$ & 0.96$^{+0.0}_{-0.01}$ & -1134.0$^{+337.8}_{-391.2}$ & 25397.6$^{+15195.6}_{-11396.7}$   & OH  \\ \hline 
SDSS J030444.98+391021.1 & NB & NB & NB & NB & NB & NB & NB & NB & NB & NB & NB     & NB     \\ 
    & 9.24$^{+0.063}_{-0.059}$ & 0.735$^{+0.037}_{-0.035}$ & -0.435$^{+0.021}_{-0.022}$ & -77.0$^{+6.6}_{-7.0}$ & -33.8$^{+5.1}_{-5.8}$ & -139.9$^{+6.8}_{-6.1}$ & 16.5$^{+1.1}_{-1.0}$ & 7.9$^{+0.1}_{-0.1}$ & 0.35$^{+0.03}_{-0.03}$ & 1960.2$^{+38.6}_{-47.4}$ & -40260.6$^{+1798.9}_{-1574.0}$     & IH     \\ \hline 
SMSS J031300.36-670839.3 & 5.821$^{+0.134}_{-0.16}$ & -8.413$^{+0.516}_{-0.619}$ & -8.566$^{+0.526}_{-0.631}$ & -218.6$^{+18.4}_{-19.5}$ & -459.4$^{+16.2}_{-18.2}$ & -32.7$^{+12.9}_{-12.2}$ & 40.4$^{+14.7}_{-7.8}$ & 5.8$^{+1.5}_{-1.5}$ & 0.76$^{+0.02}_{-0.01}$ & 519.6$^{+227.0}_{-176.6}$ & -14038.7$^{+9814.5}_{-7633.5}$          & OH \\ \hline 
HE 0330+0148 & 8.055$^{+0.002}_{-0.002}$ & -0.003$^{+0.0}_{-0.0}$ & -0.049$^{+0.002}_{-0.002}$ & 106.0$^{+8.6}_{-9.1}$ & -240.9$^{+9.1}_{-11.1}$ & -71.1$^{+8.1}_{-8.1}$ & 9.0$^{+0.6}_{-0.0}$ & 0.5$^{+0.3}_{-0.2}$ & 0.89$^{+0.04}_{-0.06}$ & 83.2$^{+73.0}_{-89.8}$ & -72495.6$^{+717.6}_{-717.6}$          & P \\ \hline 
HE 0557-4840 & 12.281$^{+0.27}_{-0.254}$ & -17.142$^{+1.012}_{-1.075}$ & -9.58$^{+0.566}_{-0.601}$ & -110.0$^{+5.3}_{-5.9}$ & -203.0$^{+2.8}_{-3.1}$ & -32.5$^{+6.1}_{-4.4}$ & 23.4$^{+1.2}_{-1.1}$ & 8.2$^{+0.9}_{-0.7}$ & 0.48$^{+0.02}_{-0.03}$ & 2273.1$^{+186.3}_{-165.6}$ & -29646.0$^{+2020.2}_{-2272.7}$          & IH \\ \hline 
SDSS J081554.26+472947.5 & 9.332$^{+0.082}_{-0.091}$ & 0.185$^{+0.011}_{-0.013}$ & 0.892$^{+0.055}_{-0.061}$ & -12.2$^{+18.0}_{-18.0}$ & -176.1$^{+12.1}_{-9.9}$ & -156.2$^{+12.6}_{-15.2}$ & 9.7$^{+0.2}_{-0.2}$ & 5.0$^{+0.5}_{-0.5}$ & 0.32$^{+0.04}_{-0.03}$ & 705.3$^{+101.9}_{-92.2}$ & -59940.5$^{+2067.3}_{-1447.1}$          & IH  \\ \hline 
SDSS J092912.32+023817.0 & 9.225$^{+0.165}_{-0.064}$ & -1.502$^{+0.203}_{-0.078}$ & 1.401$^{+0.189}_{-0.073}$ & -218.9$^{+9.2}_{-5.2}$ & -275.2$^{+13.3}_{-7.8}$ & 178.3$^{+1.6}_{-2.4}$ & 23.5$^{+2.6}_{-1.4}$ & 2.7$^{+0.3}_{-0.1}$ & 0.79$^{+0.04}_{-0.02}$ & 91.4$^{+71.5}_{-48.8}$ & -33988.5$^{+3752.4}_{-2170.6}$          & IH/S \\  
  & 16.138$^{+0.072}_{-0.234}$ & -9.971$^{+0.088}_{-0.287}$ & 9.303$^{+0.082}_{-6.944}$ & -321.7$^{+9.5}_{-43.6}$ & -447.3$^{+4.7}_{-35.4}$ & -102.6$^{+1.1}_{-40.7}$ & 193.7$^{+17.6}_{-11.8}$ & 21.1$^{+0.1}_{-0.4}$ & 0.8$^{+0.02}_{-0.01}$ & -51.8$^{+197.4}_{-928.4}$ & 34372.9$^{+3808.7}_{-2625.2}$          & OH  \\ \hline 
SDSS J094708.27+461010.0 & 10.521$^{+0.216}_{-0.189}$ & 0.326$^{+0.028}_{-0.024}$ & 2.941$^{+0.251}_{-0.22}$ & 205.0$^{+18.6}_{-16.3}$ & -264.7$^{+20.6}_{-22.1}$ & 197.6$^{+22.6}_{-18.8}$ & 30.0$^{+11.4}_{-5.7}$ & 8.2$^{+0.8}_{-0.8}$ & 0.58$^{+0.07}_{-0.05}$ & -71.0$^{+198.5}_{-213.8}$ & -21786.8$^{+9921.8}_{-7717.0}$     & OH \\ 
    & NB & NB & NB & NB & NB & NB & NB & NB & NB & NB & NB          & NB \\ \hline 
HE 1012-1540 & 8.074$^{+0.002}_{-0.002}$ & -0.316$^{+0.008}_{-0.008}$ & 0.207$^{+0.005}_{-0.005}$ & -222.5$^{+4.3}_{-4.6}$ & -191.1$^{+0.4}_{-0.4}$ & 49.0$^{+1.9}_{-1.8}$ & 14.0$^{+0.3}_{-0.3}$ & 1.3$^{+0.1}_{-0.0}$ & 0.83$^{+0.0}_{-0.01}$ & 552.7$^{+3.7}_{-3.7}$ & -55562.1$^{+850.4}_{-850.4}$          & P \\ \hline 
SDSS J102915+172927 & 8.537$^{+0.038}_{-0.033}$ & -0.481$^{+0.03}_{-0.034}$ & 1.062$^{+0.075}_{-0.066}$ & -31.1$^{+3.4}_{-3.6}$ & -23.7$^{+2.7}_{-3.3}$ & -68.7$^{+3.9}_{-4.4}$ & 10.9$^{+0.3}_{-0.2}$ & 8.6$^{+0.0}_{-0.0}$ & 0.12$^{+0.01}_{-0.01}$ & 1952.3$^{+15.6}_{-19.6}$ & -49546.0$^{+552.3}_{-552.3}$          & P \\ \hline 
SDSS J103402.70+070116.6 & 8.917$^{+0.075}_{-0.094}$ & -1.482$^{+0.122}_{-0.153}$ & 2.209$^{+0.181}_{-0.227}$ & -97.0$^{+1.2}_{-3.4}$ & -178.5$^{+6.2}_{-8.9}$ & 35.4$^{+7.0}_{-11.0}$ & 10.2$^{+0.1}_{-0.2}$ & 2.3$^{+0.2}_{-0.3}$ & 0.63$^{+0.03}_{-0.04}$ & 775.0$^{+41.4}_{-55.4}$ & -65816.8$^{+49.8}_{-102.9}$          & P \\  
  & 10.714$^{+0.263}_{-0.02}$ & -4.385$^{+0.425}_{-0.033}$ & 6.535$^{+0.633}_{-0.241}$ & -188.8$^{+36.2}_{-1.5}$ & -366.7$^{+24.4}_{-0.1}$ & -126.1$^{+38.5}_{-0.8}$ & 24.3$^{+17.8}_{-0.0}$ & 9.3$^{+2.1}_{-0.1}$ & 0.46$^{+0.12}_{-0.0}$ & -426.0$^{+81.6}_{-27.7}$ & -27566.6$^{+17448.6}_{-82.0}$          & IH \\ \hline 
SDSSJ103556.11+064143.9 & 9.26$^{+0.121}_{-0.035}$ & -2.106$^{+0.202}_{-0.058}$ & 3.131$^{+0.3}_{-0.087}$ & 87.9$^{+12.0}_{-2.3}$ & 2.2$^{+4.7}_{-6.3}$ & -20.0$^{+4.8}_{-0.2}$ & 22.2$^{+1.0}_{-1.0}$ & 7.1$^{+0.2}_{-0.1}$ & 0.52$^{+0.03}_{-0.01}$ & 2137.3$^{+60.4}_{-50.3}$ & -32395.0$^{+1110.9}_{-1568.4}$          & IH \\  
  & 12.915$^{+0.102}_{-0.061}$ & -8.218$^{+0.171}_{-0.102}$ & 12.216$^{+0.254}_{-0.346}$ & 300.8$^{+42.6}_{-25.2}$ & -60.8$^{+43.7}_{-31.4}$ & 25.1$^{+11.4}_{-13.5}$ & 147.8$^{+25.5}_{-11.8}$ & 11.9$^{+0.4}_{-0.2}$ & 0.86$^{+0.02}_{-0.01}$ & -99.4$^{+873.1}_{-641.6}$ & 25522.1$^{+5791.8}_{-2555.6}$          & OH \\ \hline 
SDSS J105519.28+232234.0 & 9.313$^{+0.197}_{-0.173}$ & -0.922$^{+0.119}_{-0.136}$ & 3.204$^{+0.481}_{-0.421}$ & 155.4$^{+29.0}_{-25.8}$ & -150.7$^{+19.3}_{-19.3}$ & 90.2$^{+8.6}_{-5.4}$ & 15.2$^{+2.7}_{-2.7}$ & 4.9$^{+0.4}_{-0.2}$ & 0.52$^{+0.04}_{-0.05}$ & 790.4$^{+169.0}_{-236.6}$ & -45745.3$^{+4608.7}_{-4608.7}$          & IH \\ \hline 
SDSS J120441.38+120111.5 & 8.213$^{+0.021}_{-0.016}$ & -2.304$^{+0.168}_{-0.228}$ & 6.814$^{+0.678}_{-0.499}$ & 92.3$^{+9.6}_{-9.1}$ & -149.9$^{+10.1}_{-13.8}$ & 6.2$^{+4.3}_{-5.2}$ & 12.8$^{+0.7}_{-0.6}$ & 3.7$^{+0.3}_{-0.3}$ & 0.55$^{+0.04}_{-0.04}$ & 597.9$^{+110.1}_{-169.3}$ & -53841.3$^{+1714.2}_{-1142.8}$          & IH \\ \hline 
SDSS J124719.46-034152.4 & 6.874$^{+0.081}_{-0.108}$ & -1.879$^{+0.135}_{-0.18}$ & 3.581$^{+0.351}_{-0.263}$ & -141.4$^{+15.7}_{-19.6}$ & -102.9$^{+5.2}_{-10.4}$ & 88.2$^{+6.9}_{-5.6}$ & 9.9$^{+2.9}_{-0.0}$ & 5.0$^{+0.3}_{-0.2}$ & 0.41$^{+0.05}_{-0.05}$ & 1263.5$^{+31.8}_{-31.8}$ & -55819.2$^{+2448.5}_{-2448.5}$          & IH \\ 
    & NB & NB & NB & NB & NB & NB & NB & NB & NB & NB & NB          & NB \\ \hline 
LAMOST J125346.09+075343.1 & 7.856$^{+0.006}_{-0.005}$ & -0.208$^{+0.008}_{-0.008}$ & 0.723$^{+0.028}_{-0.027}$ & 188.3$^{+6.8}_{-6.8}$ & -148.3$^{+4.7}_{-4.9}$ & 2.6$^{+2.6}_{-2.7}$ & 13.4$^{+0.4}_{-0.3}$ & 1.9$^{+0.1}_{-0.2}$ & 0.75$^{+0.03}_{-0.02}$ & 766.8$^{+40.7}_{-42.5}$ & -56611.4$^{+914.8}_{-774.1}$          & P \\ \hline 
SDSS J131326.89-001941.4 & 4.472$^{+0.866}_{-1.126}$ & -3.561$^{+0.874}_{-1.136}$ & 9.327$^{+3.014}_{-2.318}$ & 92.1$^{+7.3}_{-6.7}$ & -443.7$^{+81.5}_{-124.4}$ & 99.5$^{+33.6}_{-43.7}$ & 20.4$^{+39.8}_{-8.5}$ & 8.5$^{+3.3}_{-1.6}$ & 0.41$^{+0.26}_{-0.13}$ & -1256.6$^{+383.5}_{-264.5}$ & -34000.2$^{+35769.3}_{-16692.3}$          & IH \\ \hline 
HE 1310-0536 & 0.259$^{+0.29}_{-0.29}$ & -8.32$^{+0.311}_{-0.311}$ & 17.184$^{+0.644}_{-0.644}$ & -284.9$^{+13.2}_{-12.5}$ & -448.9$^{+16.8}_{-15.2}$ & 47.0$^{+3.4}_{-3.2}$ & 99.7$^{+38.3}_{-26.0}$ & 19.0$^{+0.7}_{-0.7}$ & 0.68$^{+0.06}_{-0.07}$ & 2216.2$^{+238.5}_{-238.5}$ & 15227.9$^{+8052.8}_{-8052.8}$          &OH  \\ \hline 
HE 1327-2326 & 7.332$^{+0.028}_{-0.022}$ & -0.686$^{+0.028}_{-0.023}$ & 0.755$^{+0.025}_{-0.031}$ & -279.3$^{+12.7}_{-10.7}$ & -68.7$^{+1.4}_{-1.3}$ & 287.8$^{+8.5}_{-10.6}$ & 91.4$^{+18.5}_{-17.0}$ & 7.4$^{+0.0}_{-0.0}$ & 0.85$^{+0.02}_{-0.03}$ & 1522.3$^{+4.9}_{-5.6}$ & 11193.7$^{+5154.1}_{-6091.2}$          &  OH\\ \hline 
HE 1424-0241 & 1.915$^{+0.468}_{-0.54}$ & -1.738$^{+0.134}_{-0.154}$ & 8.108$^{+0.723}_{-0.627}$ & -26.0$^{+5.5}_{-7.0}$ & -233.7$^{+16.1}_{-22.8}$ & 44.9$^{+2.2}_{-2.1}$ & 8.8$^{+0.7}_{-0.6}$ & 0.9$^{+0.5}_{-0.3}$ & 0.81$^{+0.06}_{-0.09}$ & 58.6$^{+38.9}_{-23.9}$ & -71087.9$^{+2804.7}_{-1869.8}$          & IH \\ \hline     
      \end{tabular}
    }
     \label{orbittable}
    \end{table*}

 \begin{table*}
   \def\arraystretch{2.0}
   \contcaption{from previous page.}
    \centering
    \resizebox{\textwidth}{!}{
    \begin{tabular}{llllllllllllllll}
    \hline
    Identifier  & X   & Y  & Z  & U & V  & W & Apo& Peri  & $\epsilon$ & L$_\mathrm{z}$ & E   & Orbit  \\ 
      & (kpc) & (kpc) & (kpc) & ($\kms$) & ($\kms$) & ($\kms$) & (kpc) & (kpc) & & ($\kms\kpc$) & (km$^{2}$ s$^{-2}$)    &  \\ \hline
SDSS J144256.37-001542.7 & NB & NB & NB & NB & NB & NB & NB & NB & NB & NB & NB          & NB \\ 
    & 6.366$^{+0.089}_{-0.221}$ & -0.233$^{+0.013}_{-0.032}$ & 2.033$^{+0.283}_{-0.113}$ & 93.2$^{+6.6}_{-7.6}$ & 42.0$^{+8.0}_{-6.4}$ & 217.1$^{+9.0}_{-9.0}$ & 39.1$^{+5.2}_{-2.6}$ & 6.7$^{+0.1}_{-0.1}$ & 0.71$^{+0.03}_{-0.02}$ & 1832.0$^{+27.4}_{-18.2}$ & -14290.8$^{+3324.2}_{-1662.1}$          & OH  \\ \hline 
Pristine 221.8781+9.7844 & 3.941$^{+0.29}_{-0.38}$ & 0.432$^{+0.041}_{-0.031}$ & 6.4$^{+0.602}_{-0.46}$ & -249.6$^{+12.4}_{-18.2}$ & -194.4$^{+12.6}_{-19.9}$ & -6.7$^{+12.7}_{-8.7}$ & 14.1$^{+2.3}_{-1.1}$ & 4.9$^{+0.7}_{-0.5}$ & 0.49$^{+0.01}_{-0.01}$ & 123.6$^{+83.8}_{-119.7}$ & -47818.2$^{+5527.2}_{-3684.8}$          & IH/S \\ \hline 
SDSS J164234.48+443004.9 & 7.283$^{+0.048}_{-0.072}$ & 1.869$^{+0.196}_{-0.13}$ & 1.742$^{+0.182}_{-0.121}$ & -124.6$^{+6.1}_{-9.1}$ & -143.0$^{+4.5}_{-6.1}$ & -6.7$^{+8.7}_{-5.8}$ & 9.2$^{+0.0}_{-0.0}$ & 1.6$^{+0.2}_{-0.2}$ & 0.72$^{+0.03}_{-0.04}$ & 539.8$^{+82.0}_{-41.0}$ & -70959.7$^{+2586.2}_{-0.0}$          & P  \\ \hline 
SDSS J173403.91+644633.0 & 8.374$^{+0.131}_{-0.058}$ & 4.823$^{+1.691}_{-0.752}$ & 3.089$^{+1.083}_{-0.481}$ & 66.2$^{+25.0}_{-16.7}$ & -174.4$^{+20.0}_{-15.0}$ & -200.0$^{+16.2}_{-23.1}$ & 13.0$^{+5.5}_{-2.7}$ & 7.7$^{+1.8}_{-1.3}$ & 0.33$^{+0.07}_{-0.06}$ & 980.4$^{+414.3}_{-207.1}$ & -43731.4$^{+9796.3}_{-7347.2}$          & IH \\ 
    & NB & NB & NB & NB & NB & NB & NB & NB & NB & NB & NB          & NB  \\ \hline 
SDSS J174259.67+253135.8 & 5.293$^{+0.294}_{-0.336}$ & 3.185$^{+0.399}_{-0.349}$ & 2.0$^{+0.25}_{-0.219}$ & 74.6$^{+26.6}_{-21.6}$ & -348.1$^{+23.9}_{-27.3}$ & -58.7$^{+8.8}_{-7.6}$ & 8.0$^{+1.3}_{-0.7}$ & 1.0$^{+0.2}_{-0.2}$ & 0.77$^{+0.06}_{-0.04}$ & -227.0$^{+71.9}_{-61.6}$ & -78634.3$^{+5992.5}_{-3424.3}$          & IH/$\omega$/P \\ 
    & NB & NB & NB & NB & NB & NB & NB & NB & NB & NB & NB          & NB  \\ \hline 
    2MASS J18082002-5104378  & 7.40$^{+0.02}_{-0.03}$  & $-0.189^{+0.007}_{-0.008}$ & $-0.163^{+0.006}_{-0.007}$& $2.19^{+0.53}_{-0.53}$ & $-45.4^{+1.6}_{-1.7}$ & $-5.2^{+0.2}_{-0.2}$ & $7.6^{+0.1}_{-0.1}$ & $6.3^{+0.1}_{-0.1}$ & $0.091^{+0.006}_{-0.005}$ & $1520.0^{+17.0}_{-18.4}$ & $-64227.3^{+509.0}_{-509.0}$& P \\ \hline
BPS CS 22891-0200 & -2.803$^{+0.469}_{-0.469}$ & -5.036$^{+0.219}_{-0.219}$ & -6.553$^{+0.285}_{-0.285}$ & 255.7$^{+11.1}_{-11.1}$ & -93.4$^{+4.6}_{-4.9}$ & 222.1$^{+14.0}_{-12.5}$ & 64.0$^{+18.9}_{-11.1}$ & 7.4$^{+0.6}_{-0.6}$ & 0.8$^{+0.03}_{-0.02}$ & -1788.2$^{+160.3}_{-171.0}$ & 1304.9$^{+7251.3}_{-6445.6}$          & OH \\ \hline 
BPS CS 22885-0096 & 2.413$^{+0.302}_{-0.302}$ & 0.123$^{+0.007}_{-0.007}$ & -3.701$^{+0.2}_{-0.2}$ & -145.0$^{+8.6}_{-8.1}$ & -241.8$^{+12.1}_{-12.9}$ & 223.8$^{+7.2}_{-7.2}$ & 9.3$^{+0.6}_{-1.3}$ & 3.6$^{+0.3}_{-0.3}$ & 0.44$^{+0.06}_{-0.06}$ & 5.2$^{+36.0}_{-29.7}$ & -63004.7$^{+2672.9}_{-2672.9}$          & IH/S \\ \hline 
BPS CS 22950-0046 & -6.594$^{+0.353}_{-0.243}$ & 8.665$^{+0.144}_{-0.21}$ & -8.233$^{+0.199}_{-0.137}$ & 59.4$^{+8.2}_{-7.7}$ & -70.6$^{+5.8}_{-5.5}$ & -222.9$^{+6.8}_{-5.6}$ & 41.7$^{+3.6}_{-3.2}$ & 2.5$^{+0.4}_{-0.4}$ & 0.89$^{+0.02}_{-0.02}$ & -572.2$^{+56.5}_{-56.5}$ & -14149.5$^{+2904.8}_{-2582.0}$          & OH \\ \hline 
BPS CS 30336-0049 & -4.037$^{+0.507}_{-0.475}$ & 3.41$^{+0.135}_{-0.144}$ & -9.176$^{+0.386}_{-0.362}$ & -31.8$^{+6.6}_{-6.6}$ & -643.6$^{+25.2}_{-23.6}$ & 119.8$^{+1.9}_{-2.4}$ & 122.7$^{+51.1}_{-41.4}$ & 8.1$^{+0.8}_{-0.8}$ & 0.88$^{+0.03}_{-0.04}$ & 1489.5$^{+305.3}_{-305.3}$ & 19080.8$^{+11833.4}_{-11833.4}$          & OH \\ \hline 
HE 2139-5432 & 0.746$^{+0.54}_{-0.607}$ & -2.466$^{+0.183}_{-0.206}$ & -8.023$^{+0.597}_{-0.672}$ & -48.5$^{+11.6}_{-12.3}$ & -264.7$^{+18.4}_{-20.7}$ & -113.8$^{+8.5}_{-8.5}$ & 9.8$^{+1.1}_{-0.9}$ & 1.1$^{+0.5}_{-0.3}$ & 0.79$^{+0.05}_{-0.06}$ & 87.2$^{+37.9}_{-27.1}$ & -66571.3$^{+4078.3}_{-3625.1}$          & IH \\ \hline 
HE 2239-5019 & 5.857$^{+0.137}_{-0.142}$ & -0.731$^{+0.047}_{-0.049}$ & -3.406$^{+0.217}_{-0.226}$ & 125.2$^{+4.6}_{-4.6}$ & -540.5$^{+29.3}_{-30.5}$ & -248.0$^{+3.9}_{-3.7}$ & 52.9$^{+16.6}_{-10.4}$ & 6.8$^{+0.0}_{-0.0}$ & 0.77$^{+0.05}_{-0.05}$ & -1792.6$^{+141.3}_{-141.3}$ & -4551.9$^{+7794.6}_{-7145.0}$          & OH \\ \hline 
HE 2323-0256 & 6.687$^{+0.076}_{-0.053}$ & 7.11$^{+0.29}_{-0.411}$ & -11.698$^{+0.674}_{-0.476}$ & -53.7$^{+4.4}_{-5.3}$ & -199.4$^{+8.4}_{-6.1}$ & 20.4$^{+5.7}_{-4.2}$ & 15.4$^{+0.5}_{-0.6}$ & 2.8$^{+0.2}_{-0.2}$ & 0.68$^{+0.03}_{-0.03}$ & 44.7$^{+94.9}_{-78.1}$ & -48598.2$^{+1025.7}_{-1172.2}$          &IH  \\ \hline 

    \end{tabular}
    }
    \end{table*}

\newcommand{\mnras}{MNRAS}
\newcommand{\pasa}{PASA}
\newcommand{\nat}{Nature}
\newcommand{\araa}{ARAA}
\newcommand{\aj}{AJ}
\newcommand{\apj}{ApJ}
\newcommand{\apjl}{ApJL}
\newcommand{\apjs}{ApJSupp}
\newcommand{\aap}{A\&A}
\newcommand{\aaps}{A\&ASupp}
\newcommand{\pasp}{PASP}
\newcommand{\pasj}{PASJ}

\bibliography{biblio_article}
\bibliographystyle{mn2e}

\appendix
\section{Individual results}\label{indi}

\subsection{HE 0020-1741}
\begin{figure*}
\includegraphics[scale=.4]{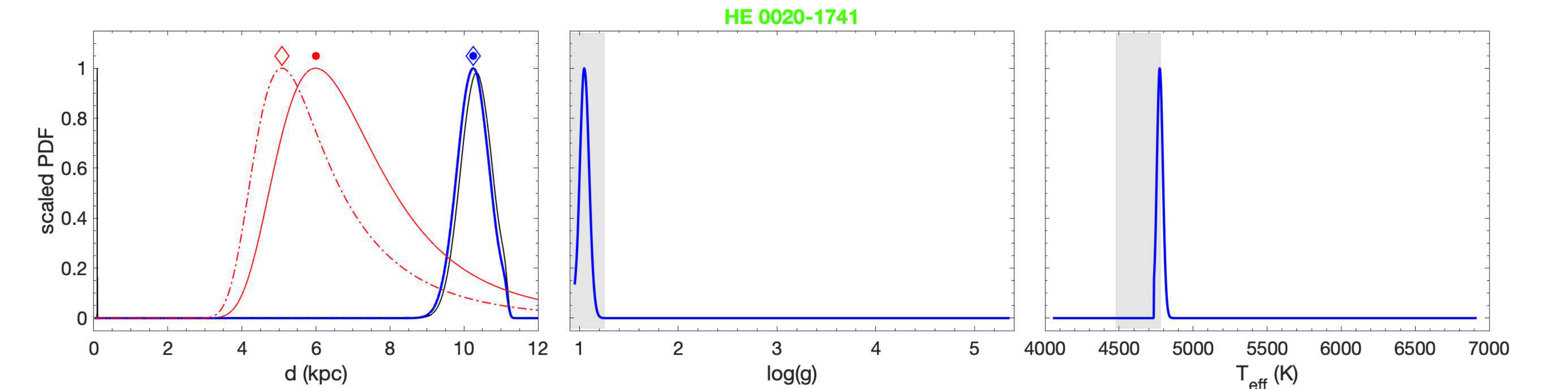}\\ 
\includegraphics[scale=.4]{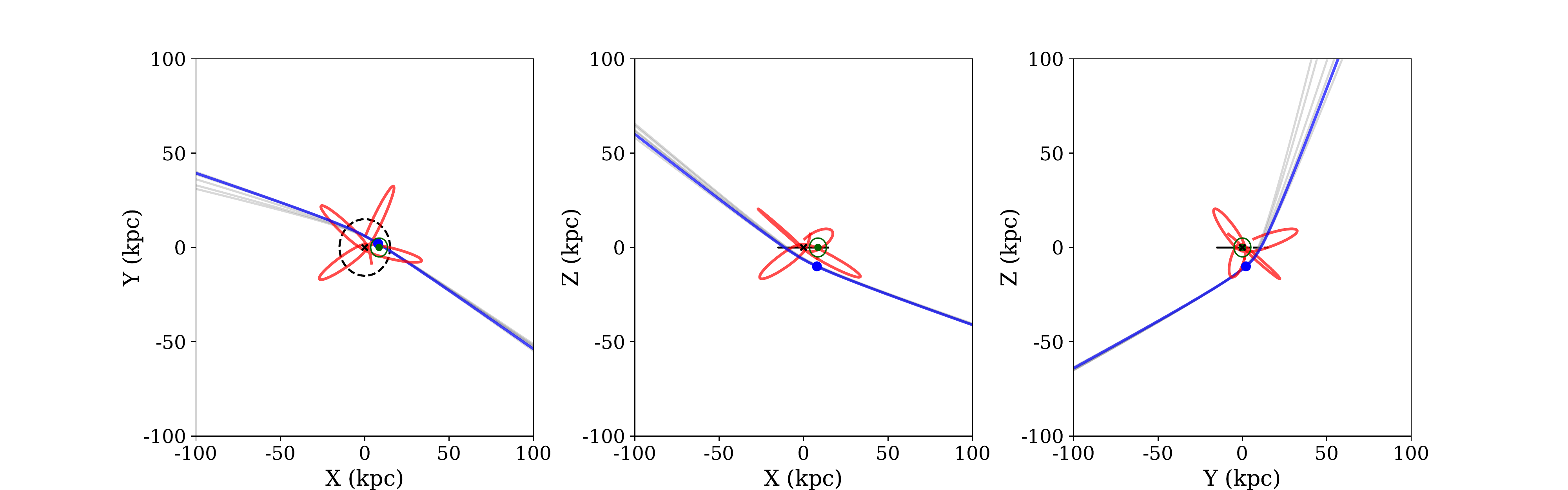} 
\caption{Left top: posterior probability (blue solid line and blue dot-dashed line respectively using a halo and a disc$+$halo prior), photometric likelihood  (black line), and the product between the astrometric likelihood and MW prior (red solid line and red dot-dashed line respectively using a halo and a disc$+$halo prior)  as a function of distance for HE 0020-1741.  The coloured dots and the diamonds represent the position of the maxima of their same colour distribution respectively using a halo and a disc$+$halo prior.   Center top: posterior probability as a function of log(g) (blue solid line for MW halo prior and blue dash-dot line for disc$+$halo prior). The gray box represents the surface gravity from literature within $1\sigma$. Right top: posterior probability as a function of T$_{eff}$ (blue solid line for MW halo prior and blue dot-dashed line for disc$+$halo prior). The gray box represents the effective temperature from literature within $1\sigma$. The PDFs are rescaled to 1. Bottom panels: Blue and red lines are, respectively, the  projected orbits of  HE 0020-1741 for the most probable distance from PDF and for the distance from Gaia astrometric only inference in the plane YX (left), ZX (center) and ZY (right).   The Galactic plane within $15\kpc$ (black line) and the Sun (green dot) are shown. Gray orbits represent randomisations around the values of position, distance, radial velocity and proper motions.} 
\label{HE 0020-1741} 
\end{figure*}

Figure~\ref{HE 0020-1741}  shows our results for HE 0020-1741, an ultra metal-poor star studied by \citet{Placco16}. Even if the parallax from Gaia DR2 has a large uncertainty ($\varpi = 0.1456 \pm 0.0384 \mas$; red solid line for the MW halo prior and a red dot-dashed  line for the disc$+$halo prior in the top-left pane), it is enough to break the dwarf/giant degeneracy obtained from the photometric solution (black line). The final PDFs are shown, using the MW halo prior and the disc$+$halo prior, respectively as the solid blue and the dot-dashed blue curves in that panel and, in both cases, the final scenario is a giant located at $10.3\pm0.4 \kpc$. The stellar parameters we infer are in agreement with the values from the literature. In the lower panels of Figure~\ref{HE 0020-1741}, both the orbits calculated from the inferred distances from the PDF and Gaia astrometry only are shown, respectively marked by the blue and the red lines. The orbital parameters relative to the distance PDF represent an unbound orbit, while the Gaia astrometric distance leads to a more benign orbit that remains in the inner part of the MW halo.

\subsection{SDSS J0023+0307}
\begin{figure*}
\includegraphics[scale=.4]{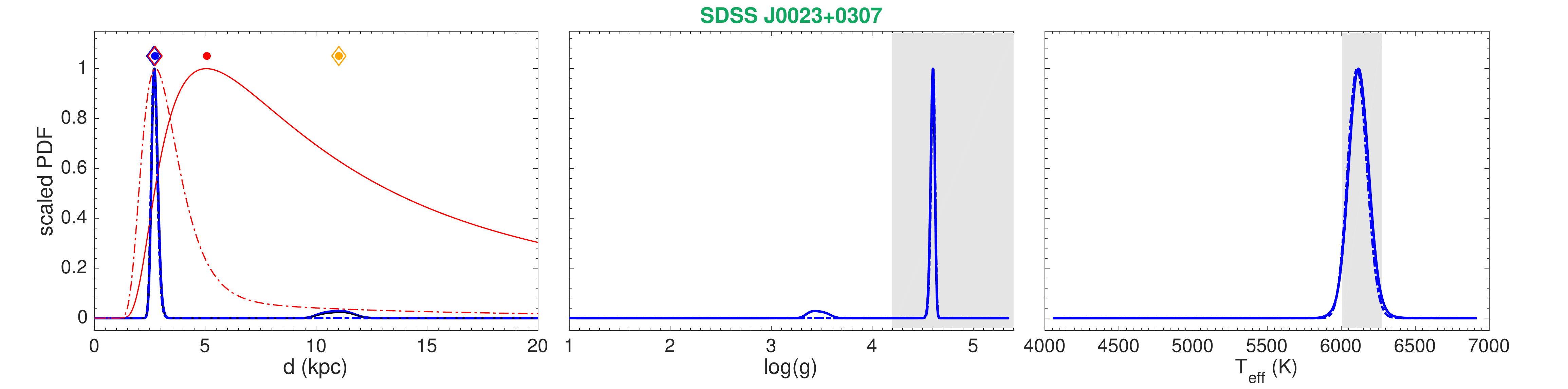}\\ 
\includegraphics[scale=.4]{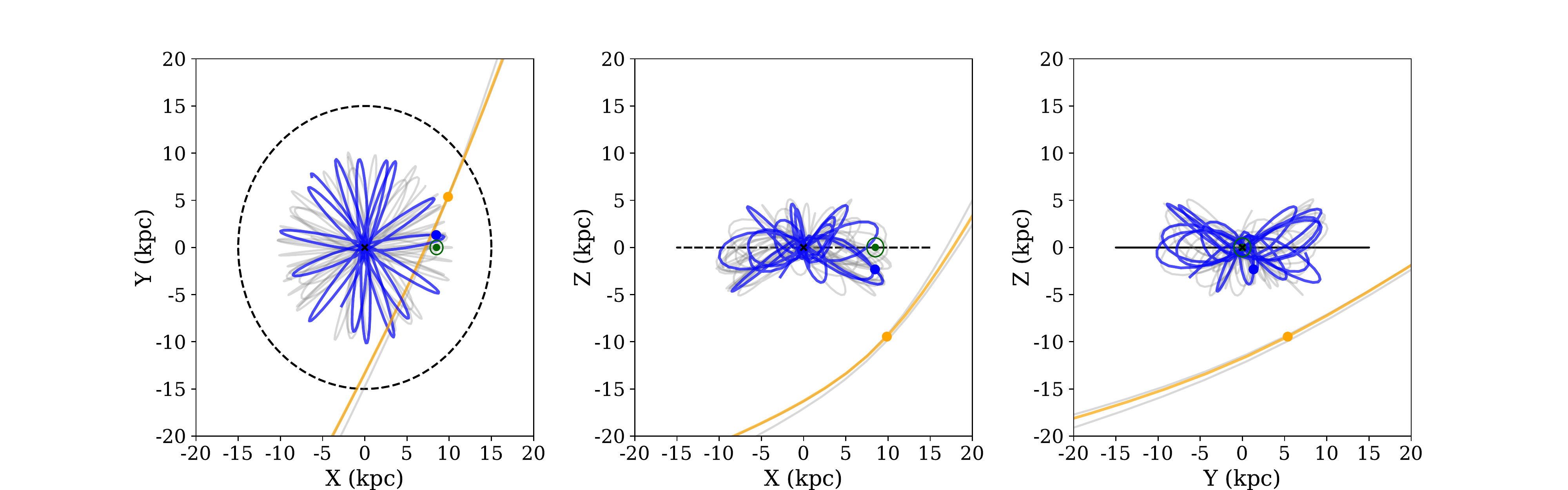} 
\caption{Left top: posterior probability (blue solid line and blue dot-dashed line respectively using a halo and a disc$+$halo prior), photometric likelihood  (black line), and the product between the astrometric likelihood and MW prior (red solid line and red dot-dashed line respectively using a halo and a disc$+$halo prior)  as a function of distance for SDSS J0023+0307.  The coloured dots and the diamonds represent the position of the maxima of their same colour distribution respectively using a halo and a disc$+$halo prior.   Center top: posterior probability as a function of log(g) (blue solid line for MW halo prior and blue dash-dot line for disc$+$halo prior). The gray box represents the surface gravity from literature within $1\sigma$. Right top: posterior probability as a function of T$_{eff}$ (blue solid line for MW halo prior and blue dot-dashed line for disc$+$halo prior). The gray box represents the temperature from literature within $1\sigma$. The PDFs are rescaled to 1. Bottom panels: Blue and orange lines are, respectively, the  projected orbits of  SDSS J0023+0307 for the most probable distance and for the second peak in the distance posterior in the plane YX (left), ZX (center) and ZY (right).   The Galactic plane within $15\kpc$ (black line) and the Sun (green dot) are shown. Gray orbits represent randomisations around the values of position, distance, radial velocity and proper motions.} 
\label{SDSS J0023+0307} 
\end{figure*}

Figure~\ref{SDSS J0023+0307} summarises our results for SDSS J0023+0307, which is a mega metal-poor star found by \citet{Aguado2018b}. The Gaia parallax is not very informative ($\varpi=0.2697\pm0.1406\mas$; red solid line for the MW halo prior and a red dot-dashed  line for the disc$+$halo prior in the top-left panel) and cannot break the dwarf/giant degeneracy inherent to the photometric solution (black line). It is nevertheless entirely compatible with that inference. The final PDFs are shown, using the MW halo prior and the disc$+$halo prior, respectively as the solid blue and the dot-dashed blue curves in that panel and, in both cases, yields a more likely dwarf solution at $2.71 \pm 0.14\kpc$ along with a less likely sub-giant solution at $11.03 \pm 0.73\kpc$. The stellar parameters we infer for the most likely dwarf solution are entirely compatible with the literature values. Combined with the exquisite Gaia proper motions, the two distance solutions yield drastically different orbits. The sub-giant distance peak implies an unbound orbit that is shown in orange, while the (more likely) dwarf solution produces a more benign orbit that remains within the inner MW (shown in blue), supporting the distance of the latter solution as the valid one. While eccentric, this orbit surprisingly remains confined close to the MW plane ($|Z|<5.0\kpc$)

\subsection{HE 0044-3755}
\begin{figure*}
\includegraphics[scale=.4]{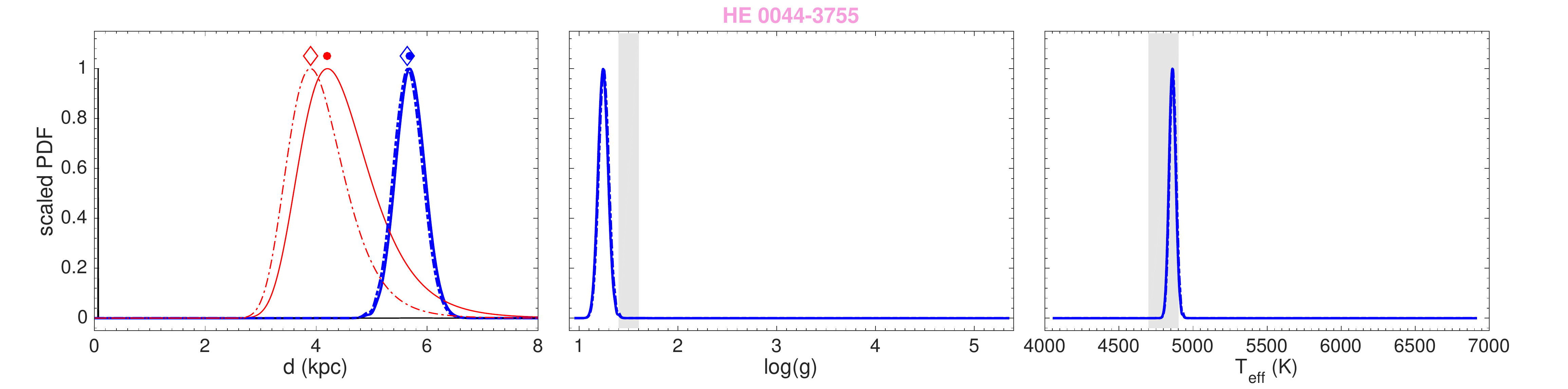}\\ 
\includegraphics[scale=.4]{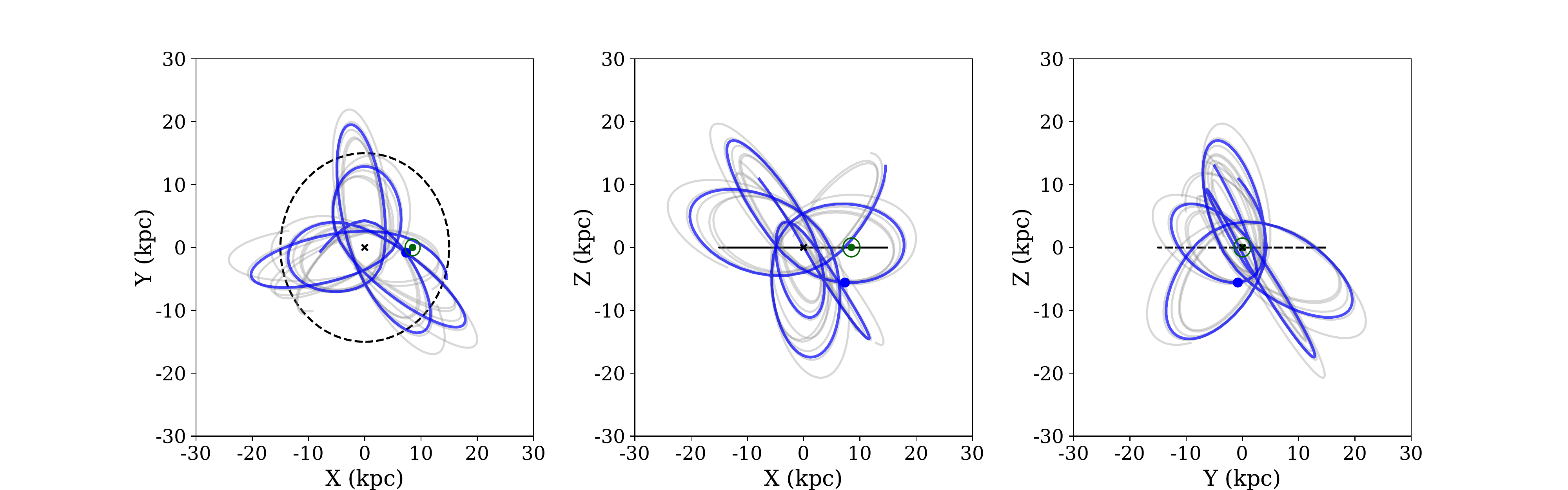} 
\caption{Same as Figure~\ref{SDSS J0023+0307}, but for HE 0044-3755.} 
\label{HE 0044-3755} 
\end{figure*}

HE 0044-3755 is an ultra metal-poor star studied by \citet{Cayrel2004} and our results for this star are shown in Figure~\ref{HE 0044-3755}. The distance PDF constrains the distance to $5.70\pm0.25 \kpc$. This result leads to a giant solution that is compatible with the values in the literature. The orbit of this star is typical of a halo star.

\subsection{HE 0057-5959}
\begin{figure*}
\includegraphics[scale=.4]{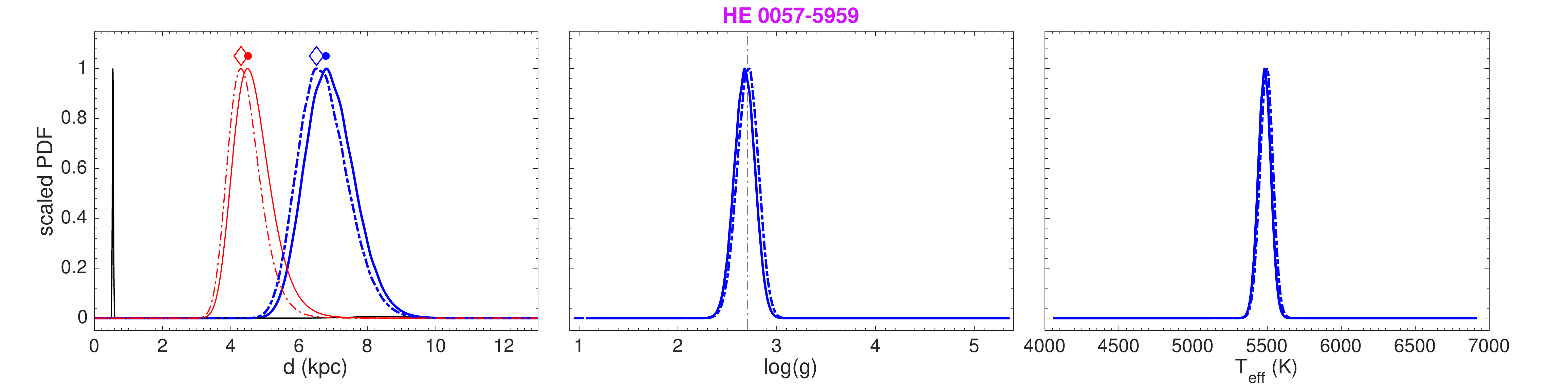}\\ 
\includegraphics[scale=.4]{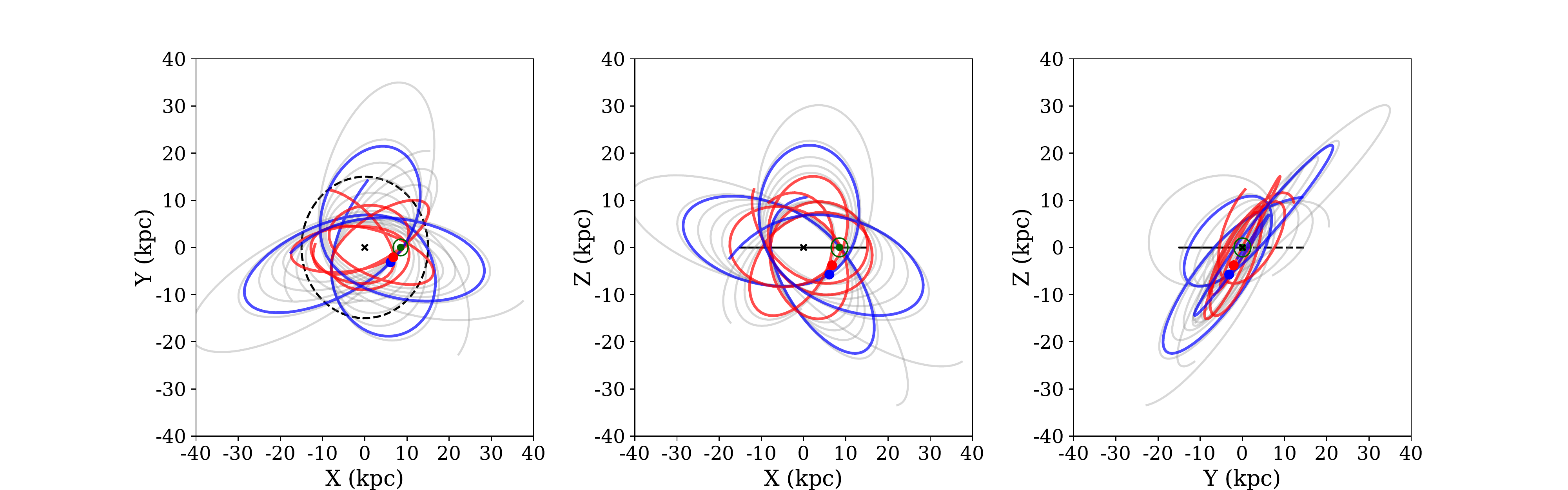} 
\caption{Same as Figure~\ref{SDSS J0023+0307}, but for HE 0057-5959. For this star, the orbit inferred from the product between the astrometric likelihood and MW halo prior is shown with the red line.} 
\label{HE 0057-5959} 
\end{figure*}

Our results for HE 0057-5959 are shown in Figure~\ref{HE 0057-5959}, taking the literature values from  \citet{2013ApJ...762...25N}. From the distance PDF, we see a disagreement between the photometric and the astrometric likelihoods, which we cannot trace to any obvious source, but the astrophysical parameter inference is compatible with the literature values. For this giant, we show the orbit inferred both from our full astrometric and photometric analysis (blue orbit) and when using only the Gaia astrometry with the MW prior (red orbit). In both cases, HE~0057-5959 remains in the inner region of the MW halo (apocentre $<30\kpc$).

\subsection{HE 0107-5240}
\begin{figure*}
\includegraphics[scale=.4]{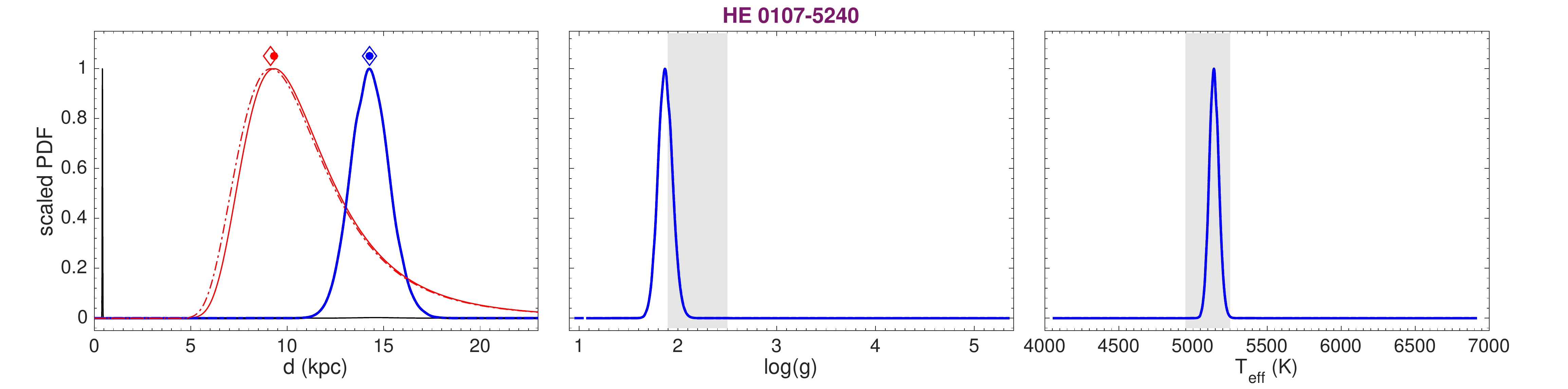}\\ 
\includegraphics[scale=.4]{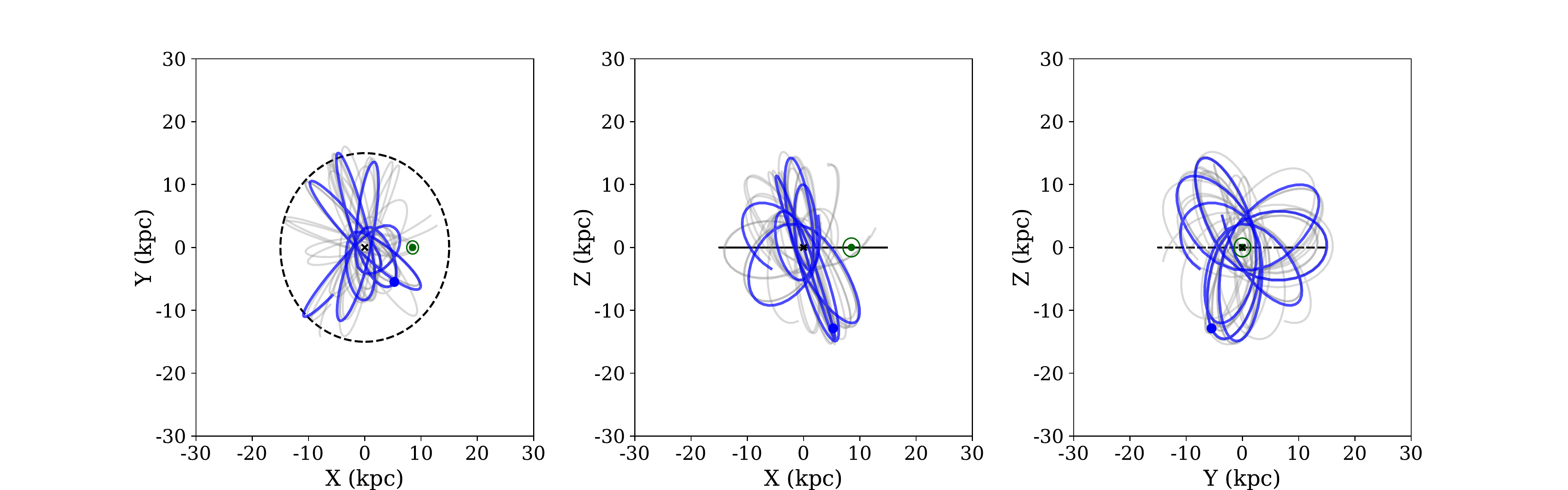} 
\caption{Same as Figure~\ref{SDSS J0023+0307}, but for HE 0107-5240.} 
\label{HE 0107-5240} 
\end{figure*}

HE~0107-5240 is likely a binary system \citep{Arentsen18} discovered by radial velocity variation. Its spectrum does not present double lines indicating that the light is not polluted by the secondary. It is a hyper metal-poor star analysed by \citet{2004ApJ...603..708C}. Our results are shown in Figure~\ref{HE 0107-5240} and we infer a distance of $14.3 \pm 1.0\kpc$, corresponding to the giant solution because the probability of the dwarf solution is entirely suppressed by the Gaia parallax information. Our values for surface gravity and effective temperature are in perfect agreement with the literature values. The orbit of this star is typical of an eccentric halo  orbit and remains within $15.9^{+1.0}_{-0.9}\kpc$.

\subsection{HE 0134-1519}
\begin{figure*}
\includegraphics[scale=.4]{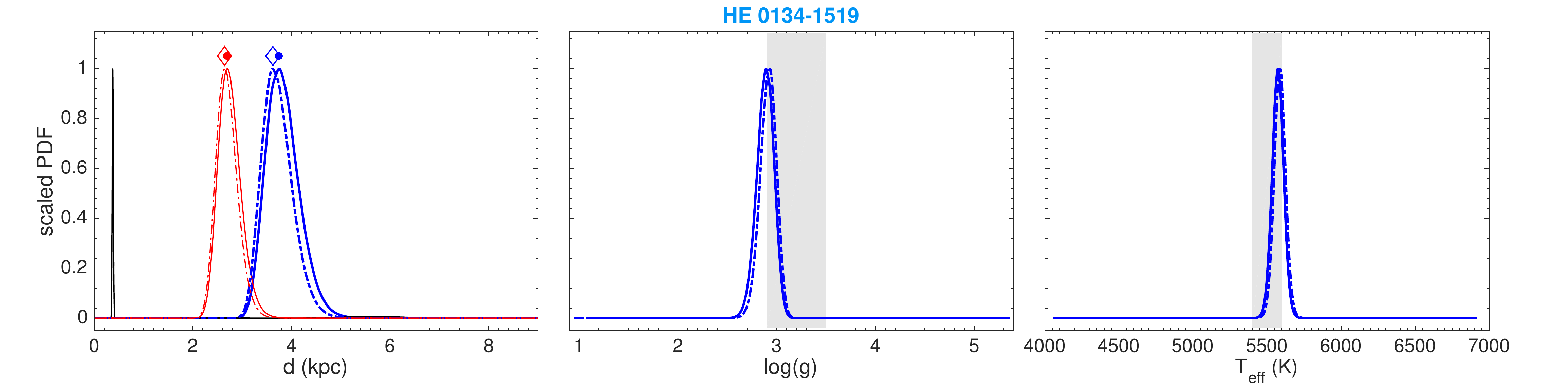}\\ 
\includegraphics[scale=.4]{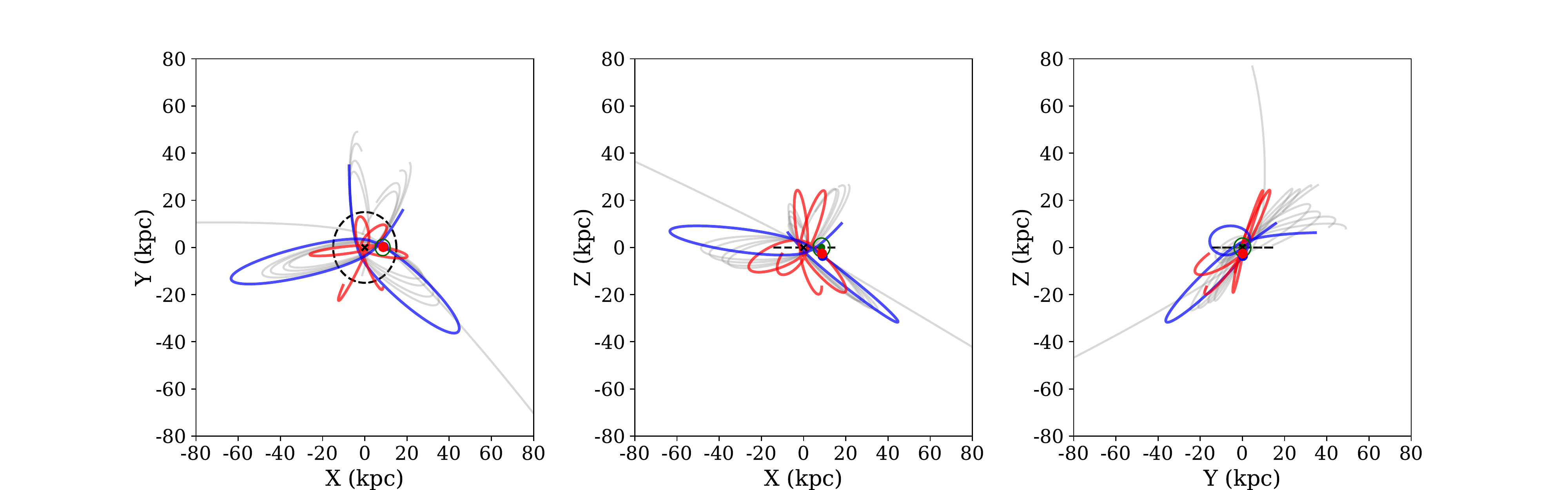} 
\caption{Same as Figure~\ref{SDSS J0023+0307}, but for HE 0134-1519. For this star, the orbit inferred from the product between the astrometric likelihood and MW halo prior is shown with the red line.} 
\label{HE 0134-1519} 
\end{figure*}

Our analysis of HE~0134-1519 \citep{2015ApJ...807..173H} is shown in Figure~\ref{HE 0134-1519}. This is another case for which the astrometric and photometric likelihoods disagree, yielding very different orbits, even though it is clear this star is a giant, in agreement with the literature. Both orbital solutions are indicative of a halo star, but the closer Gaia-only distance solution yields an orbit that remains much closer to the Galactic center  (apocentre of $25.7^{+4.6}_{-1.7} \kpc$ vs. $70.2^{+49.3}_{-20.7}\kpc$).

\subsection{SDSS J014036.21+234458.1}
\begin{figure*}
\includegraphics[scale=.4]{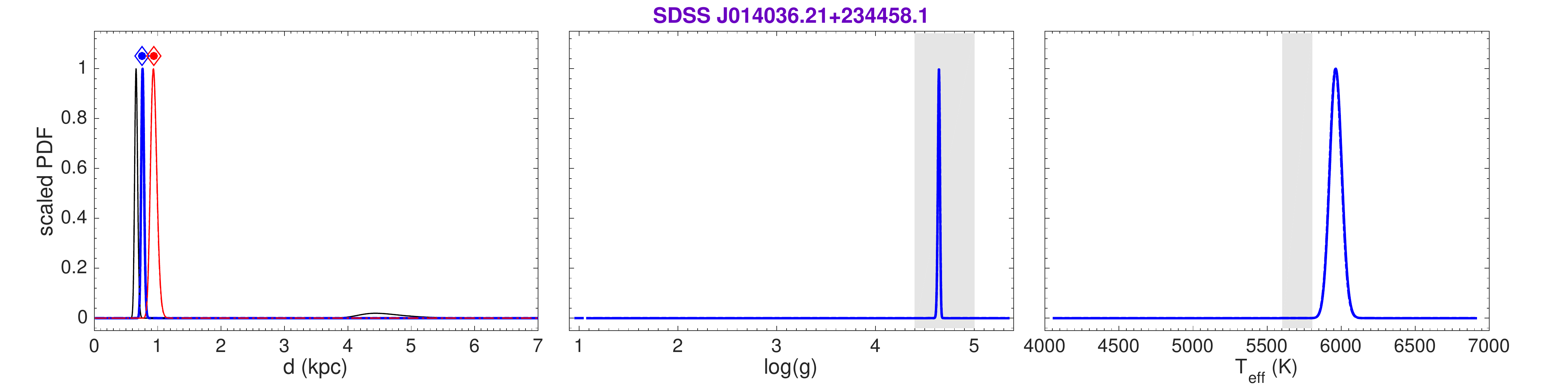}\\ 
\includegraphics[scale=.4]{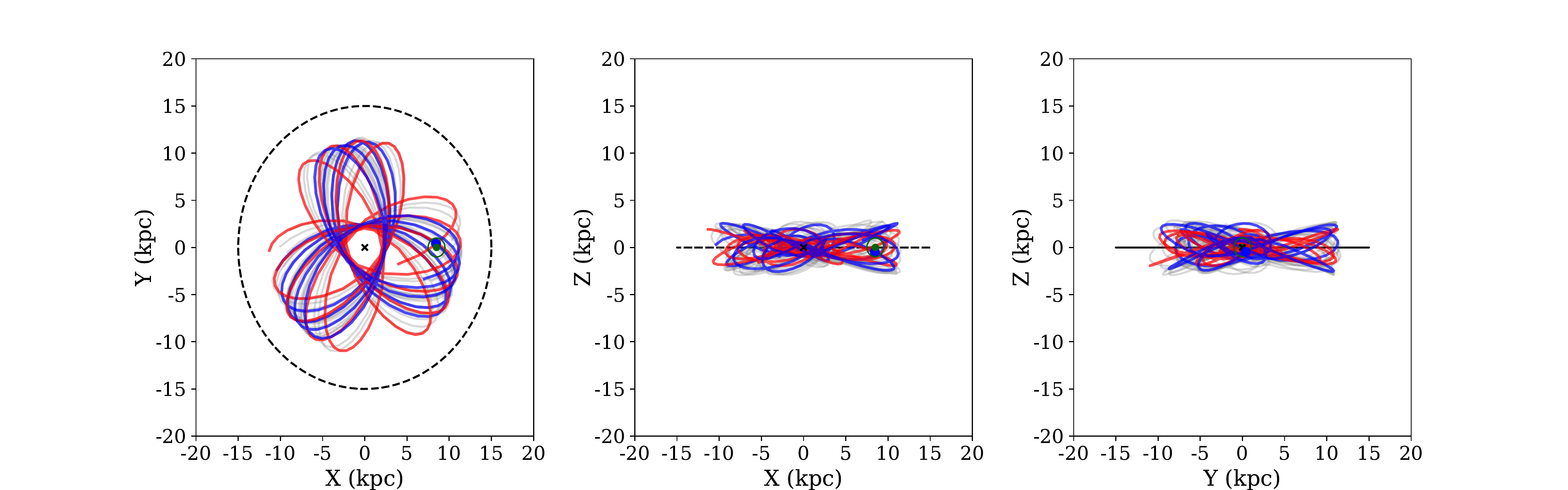} 
\caption{Same as Figure~\ref{SDSS J0023+0307}, but for SDSS J014036.21+234458.1. For this star, the orbit inferred from the product between the astrometric likelihood and MW halo prior is shown with the red line.} 
\label{SDSS J014036.21+234458.1} 
\end{figure*}

For the dwarf star SDSS~J014036.21+234458.1 \citep[][Figure~\ref{SDSS J014036.21+234458.1}]{2013ApJ...762...26Y}, the astrometric and photometric distances are technically in disagreement, but the distance inferences are so similar that it does not impact our results. We infer a distance of $0.76 \pm 0.02\kpc$ and an orbit that brings SDSS~J014036.21+234458.1 close to the MW plane ($|Z| < 2.5 \kpc$).

\subsection{BD+44 493}
\begin{figure*}
\includegraphics[scale=.4]{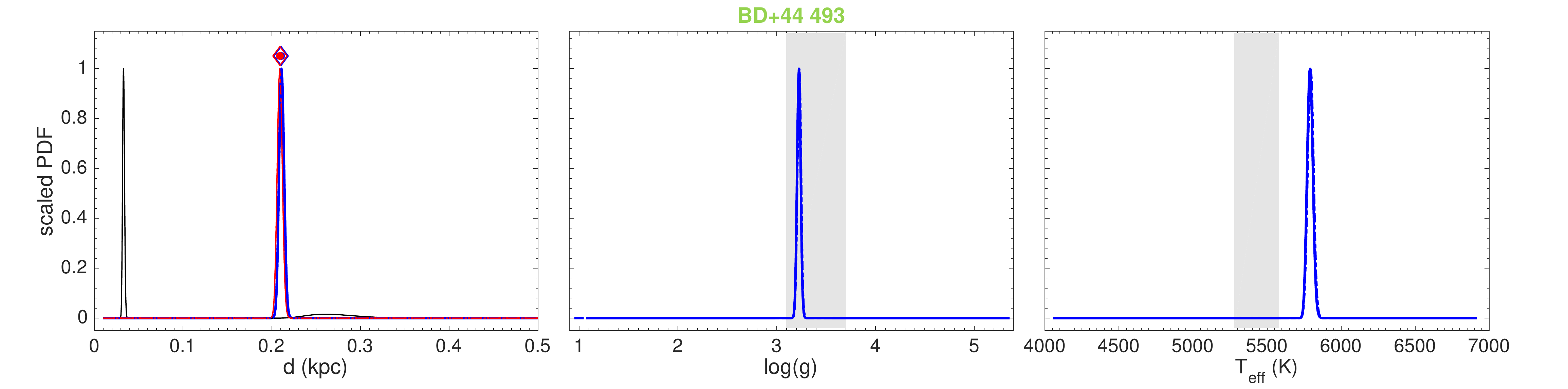}\\ 
\includegraphics[scale=.4]{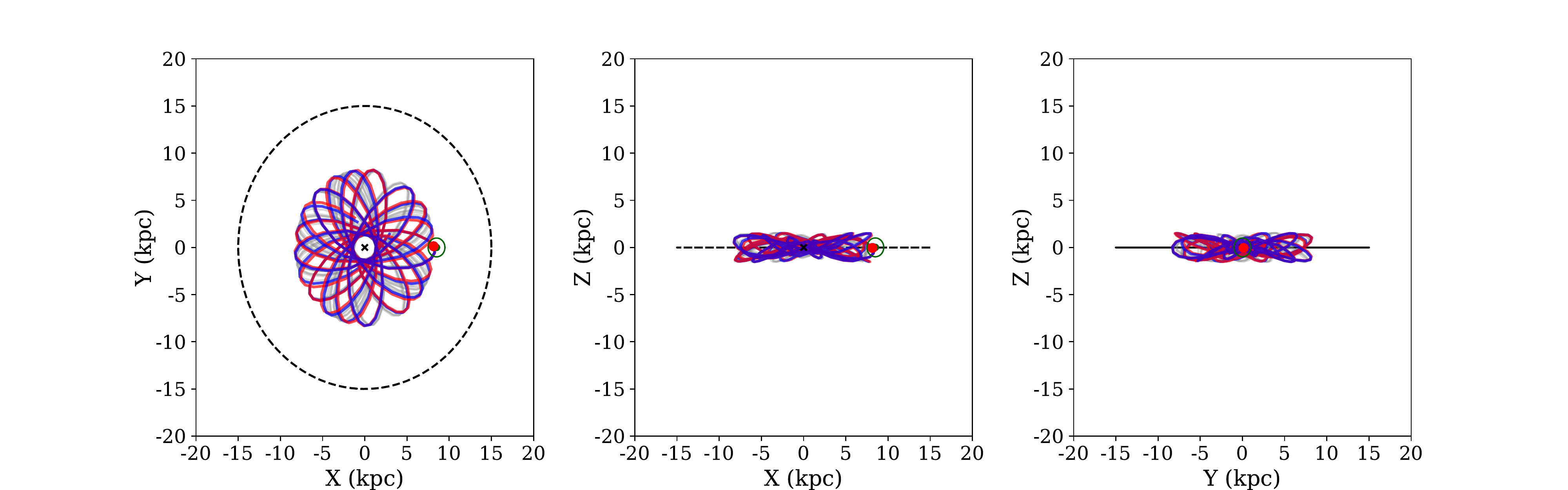} 
\caption{Same as Figure~\ref{SDSS J0023+0307}, but for BD+44 493. For this star, the orbit inferred from the product between the astrometric likelihood and MW halo prior is shown with the red line.} 
\label{BD+44 493} 
\end{figure*}

Our results for BD+44 493 \citep[][Figure~\ref{BD+44 493}]{2013ApJ...773...33I} are strongly constrained by the exquisite Gaia parallax, yielding a distance of $0.211 \pm 0.003\kpc$. Just like with SDSS~J0023+0307, this star is eccentric and stays extremely close to the MW plane ($|Z|<1.5\kpc$). It has an apocentre at the Solar circle.

\subsection{HE 0233-0343}
\begin{figure*}
\includegraphics[scale=.4]{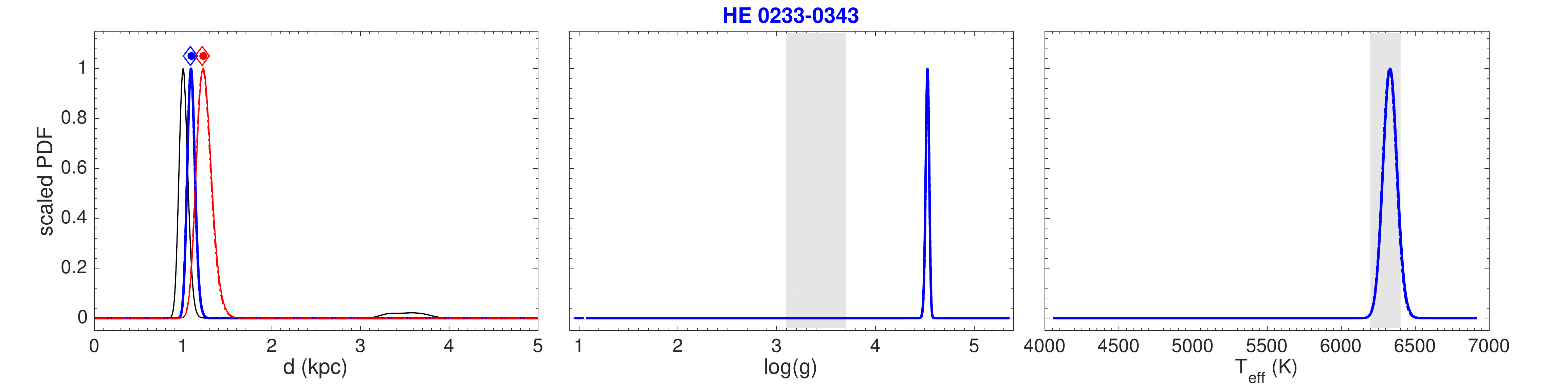}\\ 
\includegraphics[scale=.4]{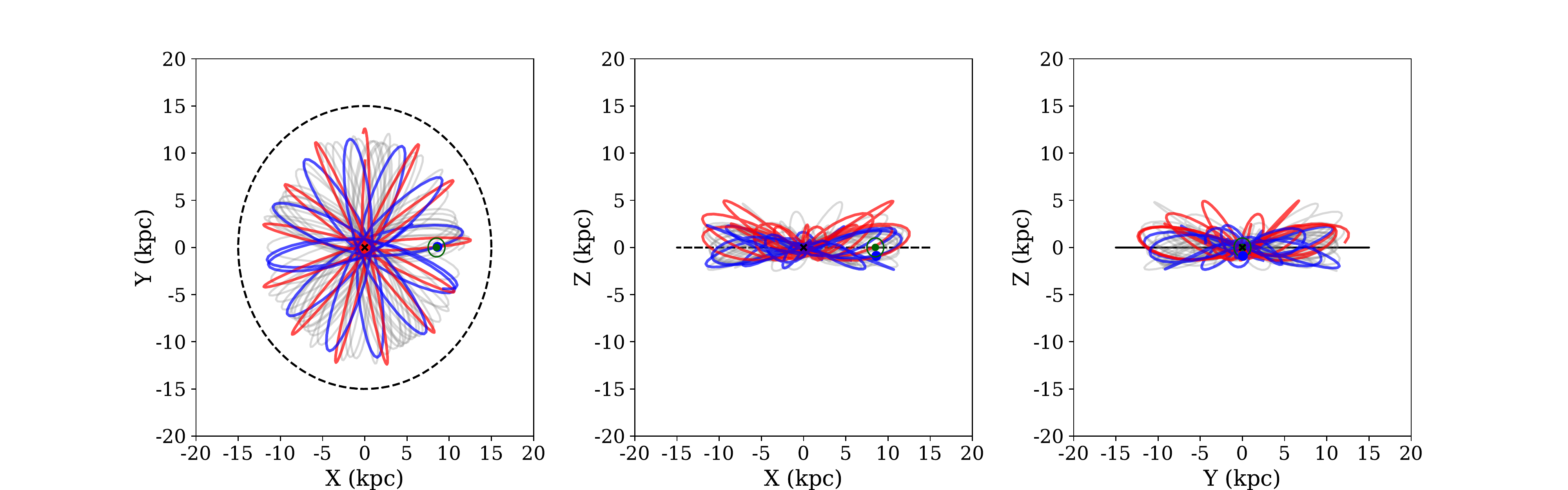} 
\caption{Same as Figure~\ref{SDSS J0023+0307}, but for HE 0233-0343. For this star, the orbit inferred from the product between the astrometric likelihood and MW halo prior is shown with the red line.} 
\label{HE 0233-0343} 
\end{figure*}

The combined astrometric and photometric analysis of HE~0233-0343 \citep[][Figure~\ref{HE 0233-0343}]{2015ApJ...807..173H} yields an accurate distance of $1.09 \pm 0.04\kpc$. Despite this, our $\log(g)$ inference is incompatible with the literature value, but the very accurate Gaia parallax lends support to our inference. Like the previous star, HE~0233-0343 remains confined to the region of the MW disc, with $|Z|<2.6\kpc$ and an apocentre of $11.9^{+0.5}_{-0.4}\kpc$.

\subsection{BPS CS 22963-0004}
\begin{figure*}
\includegraphics[scale=.4]{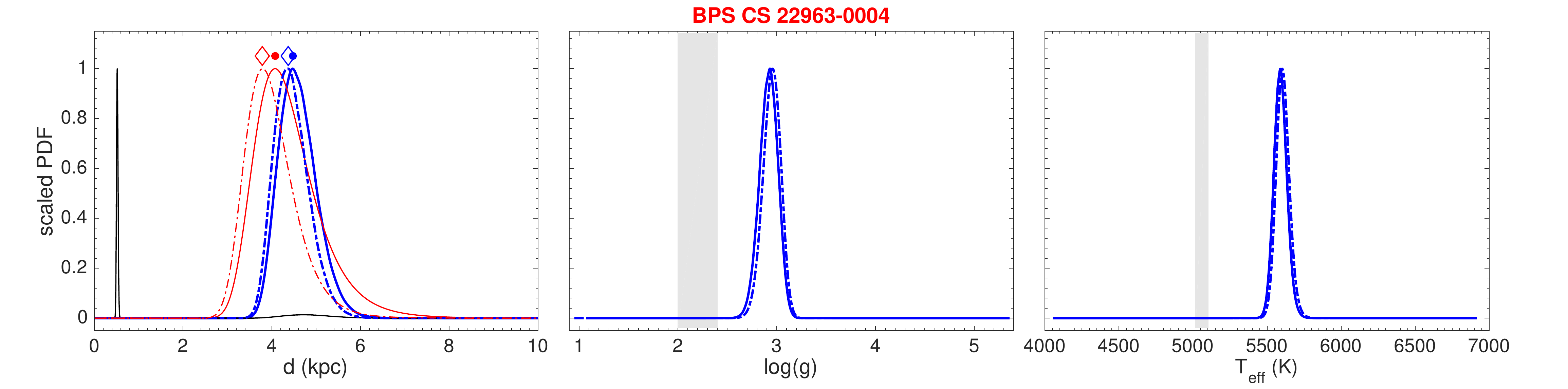}\\ 
\includegraphics[scale=.4]{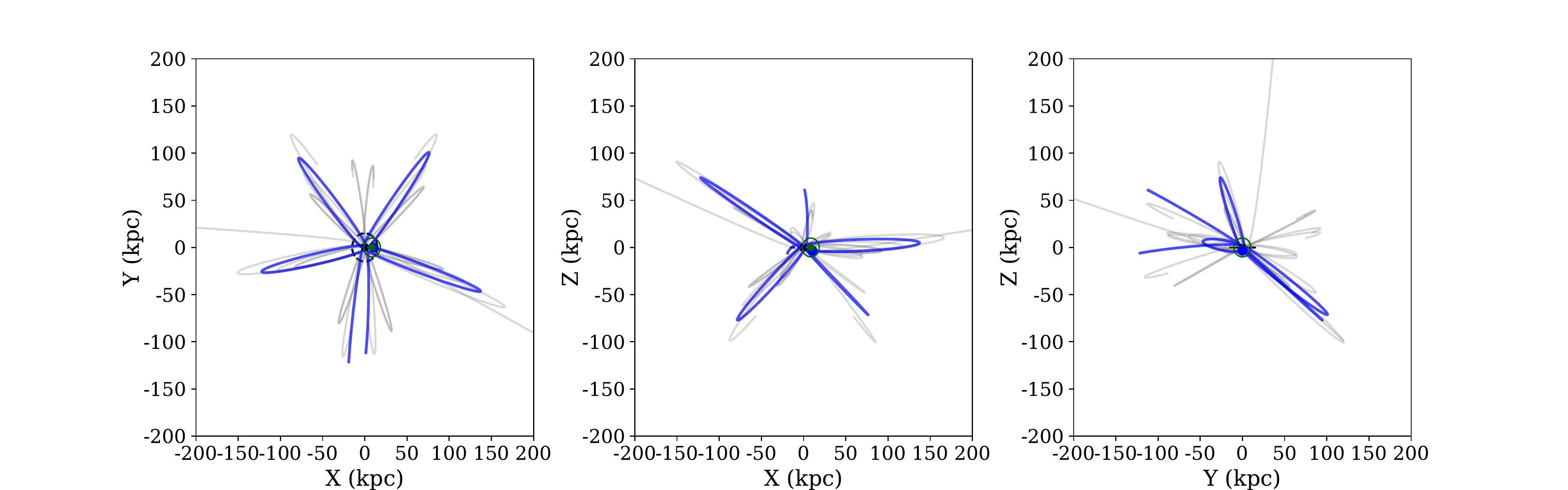} 
\caption{Same as Figure~\ref{SDSS J0023+0307}, but for BPS CS 22963-0004.} 
\label{BPS CS 22963-0004} 
\end{figure*}

For this UMP studied by \citet{Roederer2014}, we infer a distance of $4.5 \pm 0.4\kpc$ (Figure~\ref{BPS CS 22963-0004}). Our astrophysical parameter inference disagrees with the literature values but the MESA isochrones strongly constrain our temperature inference. The difference could hint at systematics in these isochrones or the \citet{Roederer2014} analysis. Despite the currently proximity to this star, its orbit brings it very far into the MW halo, with $r_\mathrm{apo}=155.8^{+183.4}_{-55.0}\kpc$.

\subsection{SDSS J030444.98+391021.1}
\begin{figure*}
\includegraphics[scale=.4]{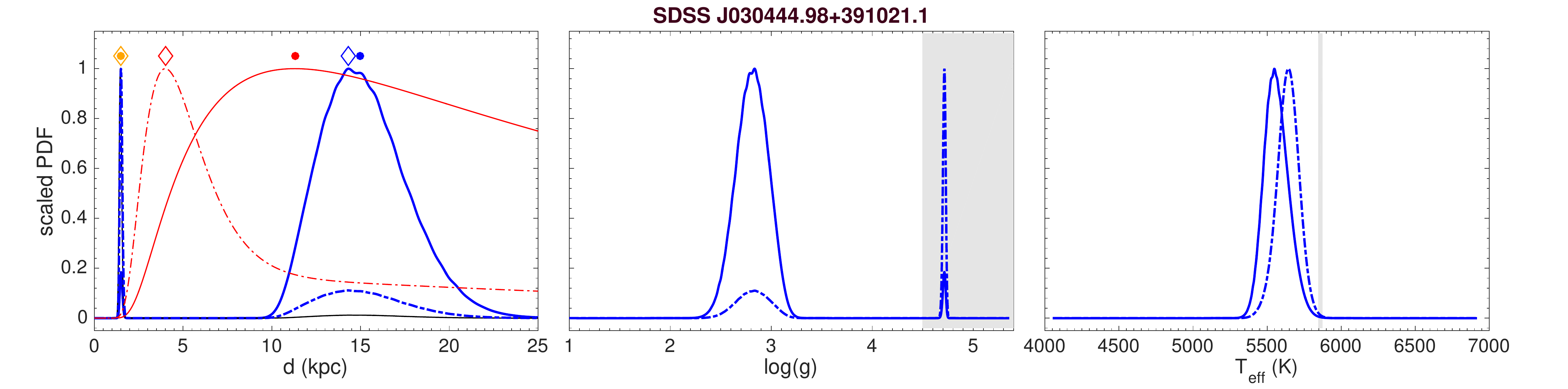}\\ 
\includegraphics[scale=.4]{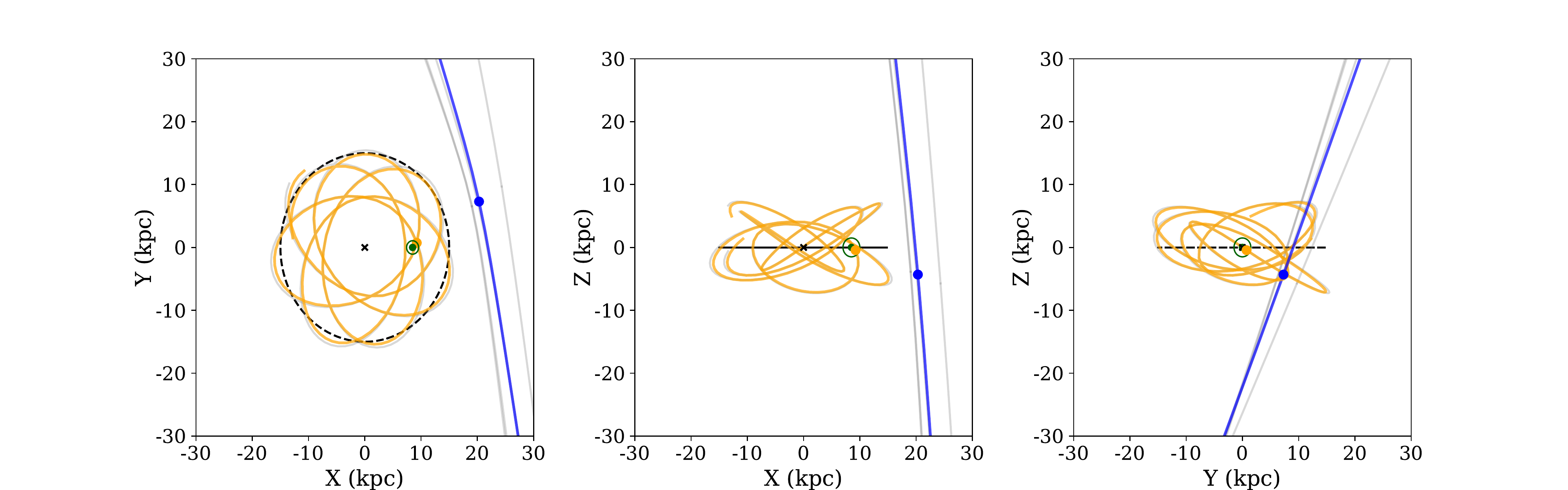} 
\caption{Same as Figure~\ref{SDSS J0023+0307}, but for SDSS J030444.98+391021.1.} 
\label{SDSS J030444.98+391021.1} 
\end{figure*}

The Gaia parallax of SDSS~J030444.98+391021.1 \citep[][Figure~\ref{SDSS J030444.98+391021.1}]{Aguado2017b} is very uncertain ($\varpi= 0.0752 \pm 0.1929\mas$) but, in case of the halo prior, strongly suppresses the dwarf solution (1\% of the PDF). However, the orbital analysis shows that the favoured giant scenario implies that this star is not bound to the MW. According to this, we repeat the analysis with the disc$+$halo prior finding that the inferred distances are not significantly changed but the fractional probability of the peaks is. With this prior, the dwarf solution represents 21$\%$ of the PDF. Taking into account the orbital analysis, the dwarf solution appears to be the more realistic distance estimate ($1.51\pm0.07\kpc$). This solution is also compatible with the $\log(g)$ of \citet{Aguado2017b}, contrary to the result from the giant solution. We note that a slightly larger distance for the dwarf solution would be entirely compatible with the Gaia parallax and we think that the low likelihood of the dwarf solution could be driven further down than it should by a systematic in the models we use. With our favoured close-by distance, this star has the orbit of an inner halo object.

\subsection{SMSS J031300.36-670839.3}
\begin{figure*}
\includegraphics[scale=.4]{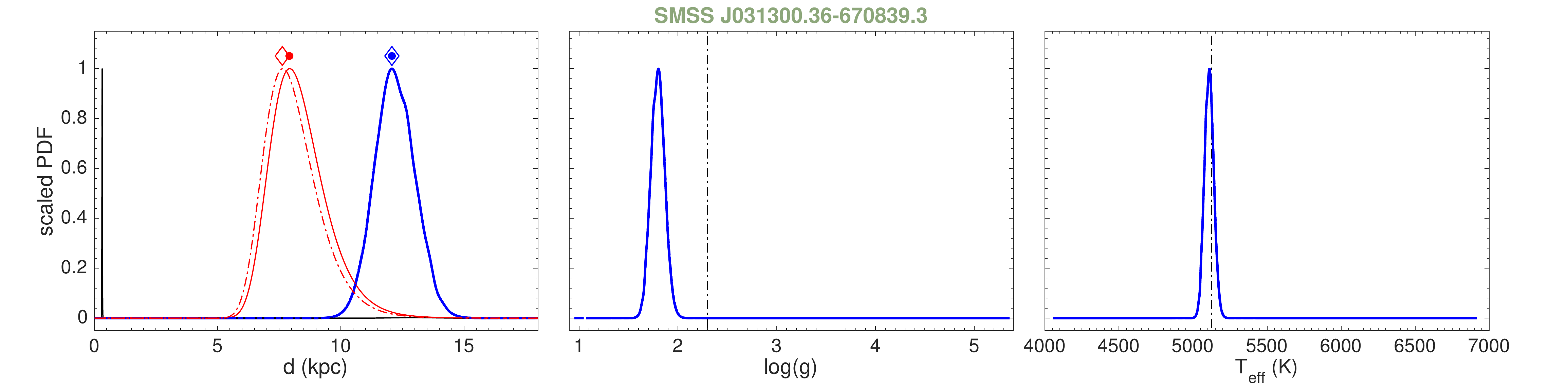}\\ 
\includegraphics[scale=.4]{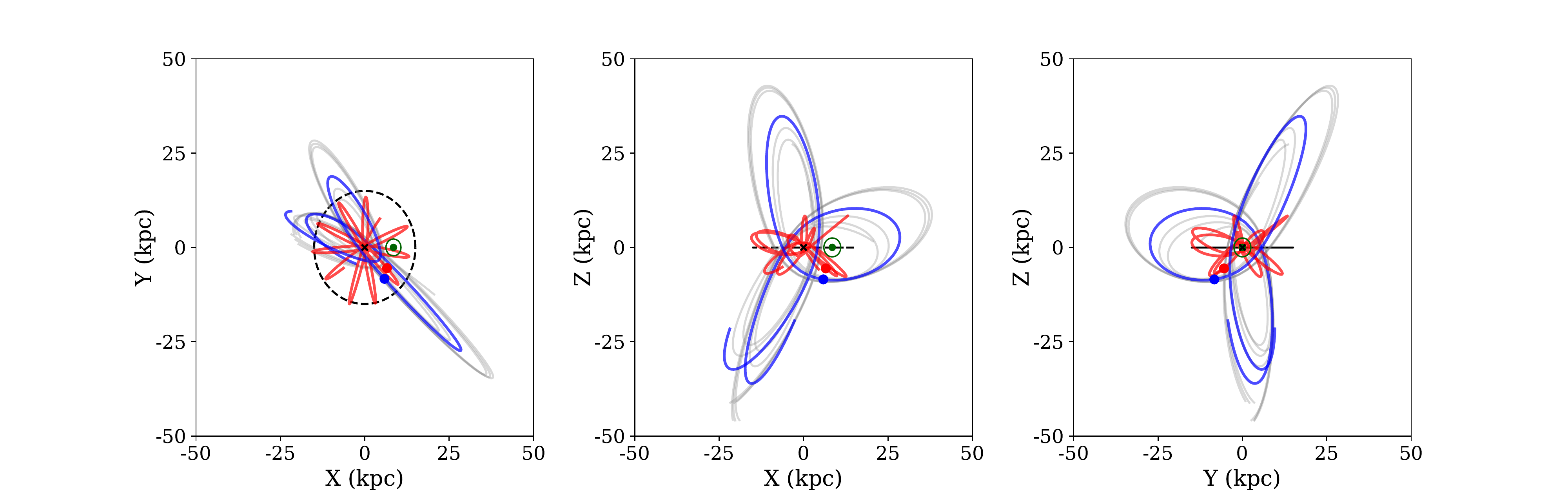} 
\caption{Same as Figure~\ref{SDSS J0023+0307}, but for SMSS J031300.36-670839.3. For this star, the orbit inferred from the product between the astrometric likelihood and MW halo prior is shown with the red line.} 
\label{SMSS J031300.36-670839.3} 
\end{figure*}

For this star with the lowest iron-abundance \citep[$\FeH<-6.53$, ][Figure~\ref{SMSS J031300.36-670839.3}]{Nordlander2017}, we infer a distance of $12.0 \pm 0.8\kpc$  corresponding to the giant solution ($\log(g)=1.8\pm0.1$). The literature $\log(g)$ is however in better agreement with the Gaia-only distance that is a little closer. The orbital analysis implies that this star has a fairly eccentric orbit and that, using the Gaia-only distance, it is compatible with an inner halo object. With the final posterior, we infer an outer halo orbit.

\subsection{HE 0330+0148}\label{HE 0330+0148Sec}
\begin{figure*}
\includegraphics[scale=.4]{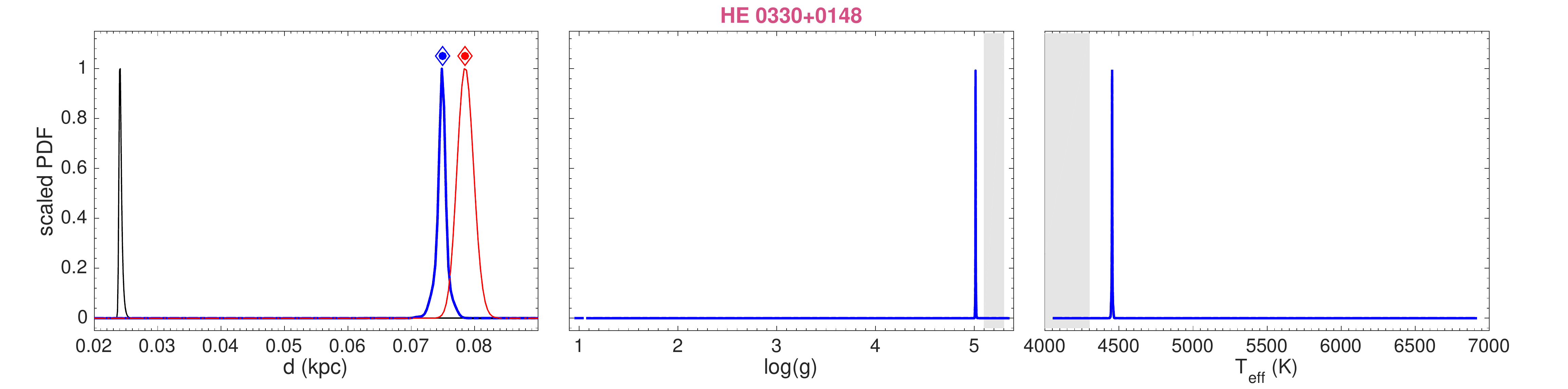}\\ 
\includegraphics[scale=.4]{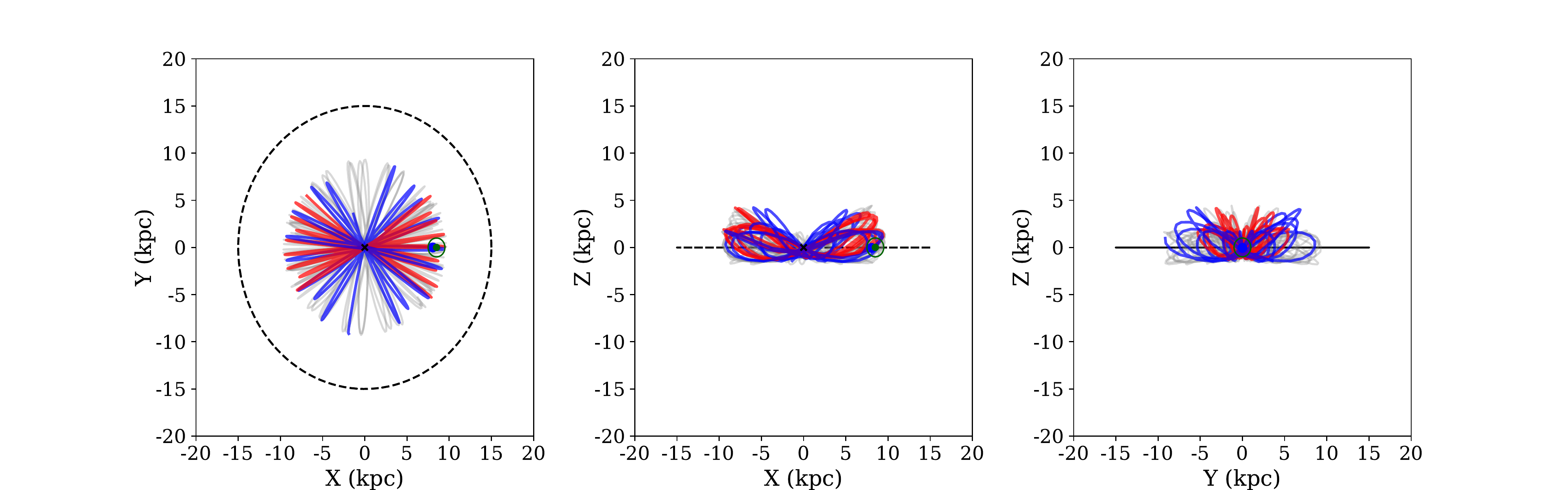} 
\caption{Same as Figure~\ref{SDSS J0023+0307}, but for HE 0330+0148. For this star, the orbit inferred from the product between the astrometric likelihood and MW halo prior is shown with the red line.} 
\label{HE 0330+0148} 
\end{figure*}

As we can see in Figure~\ref{CMD}, the analysis fails for this carbon-enhanced star (also known as G77-61) and its location in the colour-magnitude diagram does not coincide with the isochrone models. The strong carbon bands dominate in the spectrum \citep{Dahn77} , where the Gaia DR2 $BP$ filter is sensitive, leading to an abnormal value of ($BP - RP$)  colour and, as a consequence, this star lays  outside the isochrone range. This could also explain the strong disagreement between the photometric-only and astrometric-only distance likelihood functions (see Figure~\ref{HE 0330+0148}). We don't think that the binarity can affects the photometry because the companion is most likely an unseen white dwarf with a period of ~250 days \citep{Dearborn86}, which means that the Gaia DR2 magnitudes correspond to the magnitude of the star itself and not that of the binary system. In this case, we favour the Gaia-only inference with $78\pm 1\pc$. HE~0330+0148 has a very radial orbit and its current position near the Sun is near its apocentre. Its orbit is close to the MW plane ($|Z| < 2.8 \kpc$). 

\subsection{HE 0557-4840}
\begin{figure*}
\includegraphics[scale=.4]{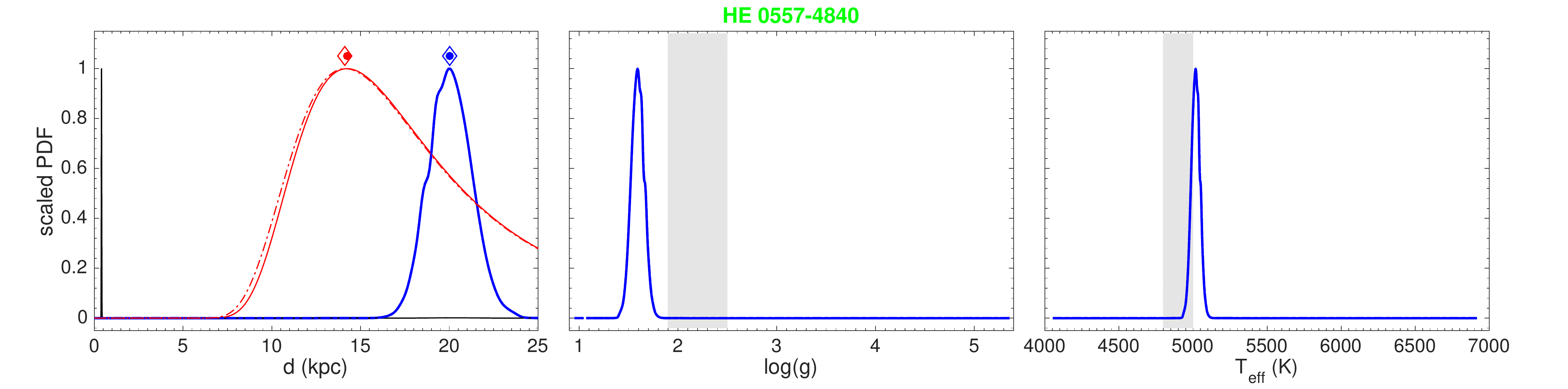}\\ 
\includegraphics[scale=.4]{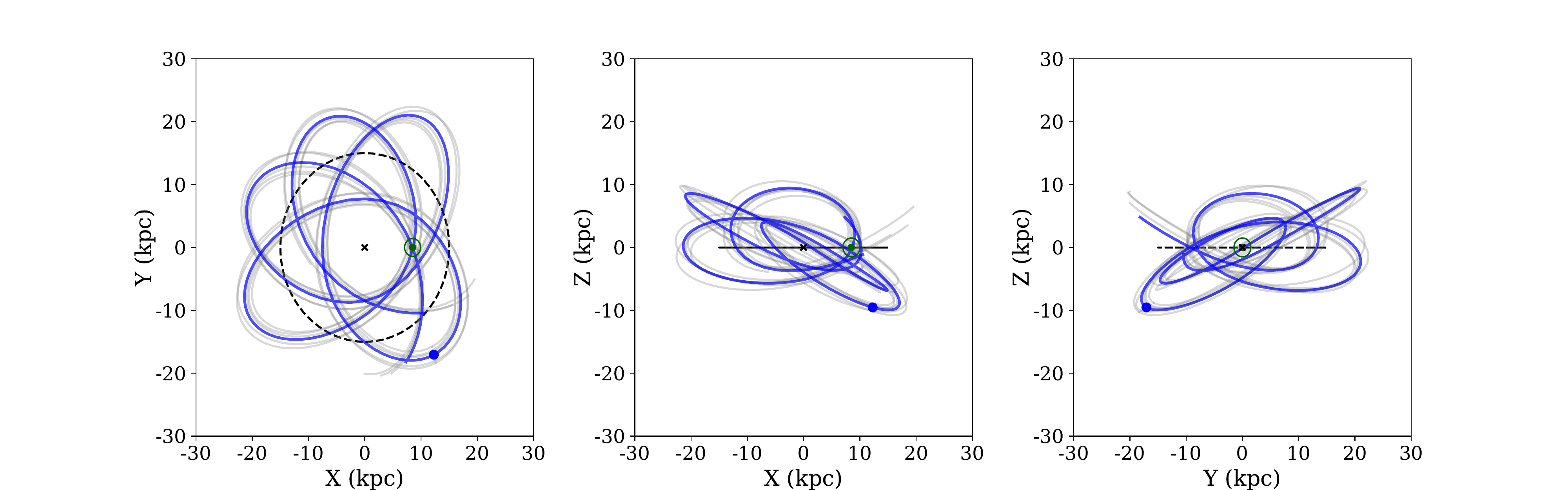} 
\caption{Same as Figure~\ref{SDSS J0023+0307}, but for HE 0557-4840.} 
\label{HE 0557-4840} 
\end{figure*}

The inferred result on HE 0557-4840 \citep[][Figure~\ref{HE 0557-4840}]{2007ApJ...670..774N} shows it is a giant halo star at a distance of $20.0 \pm 1.3\kpc$. Although the peaks of the astrometric and photometric solutions are shifted by $\approx 6\kpc$, these are compatible due to the Gaia parallax that is poorly constrained ($\varpi = 0.0389 \pm 0.0207 \mas$).

\subsection{SDSS J081554.26+472947.5}
\begin{figure*}
\includegraphics[scale=.4]{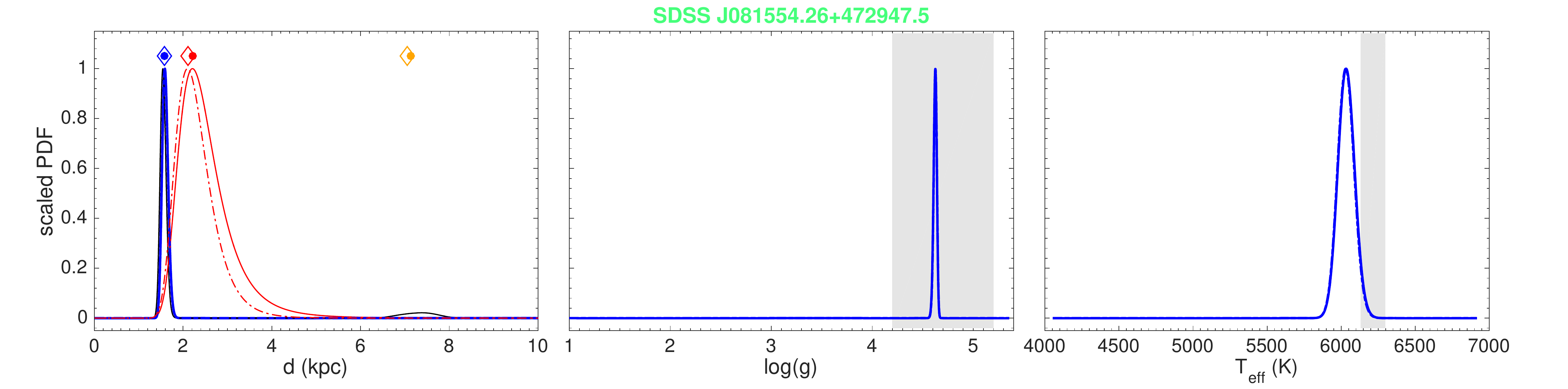}\\ 
\includegraphics[scale=.4]{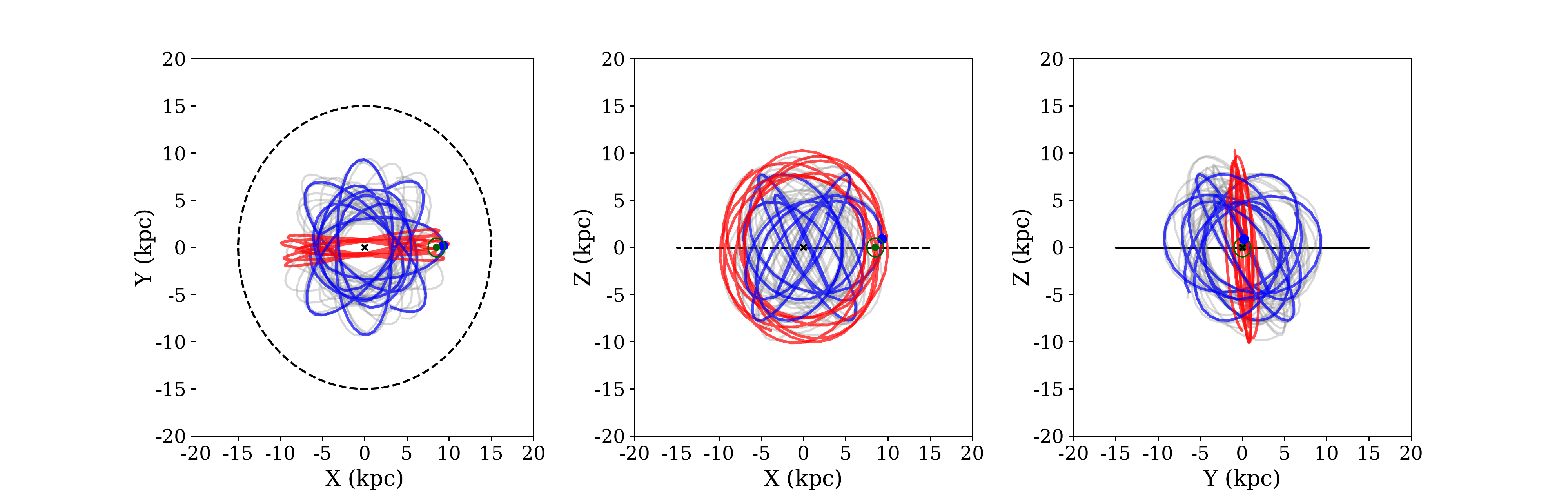} 
\caption{Same as Figure~\ref{SDSS J0023+0307}, but for SDSS J081554.26+472947.5. For this star, the orbit inferred from the product between the astrometric likelihood and MW halo prior is shown with the red line.} 
\label{SDSS J081554.26+472947.5} 
\end{figure*}

Our results on SDSS J081554.26+472947.5 \citep[][Figure~\ref{SDSS J081554.26+472947.5}]{Aguado2018a} show that the star is a dwarf that is located at a distance of $1.59 \pm 0.07\kpc$ and orbits within the inner halo. Our stellar parameter inference is in agreement with the literature values.

\subsection{SDSS J092912.32+023817.0}
\begin{figure*}
\includegraphics[scale=.4]{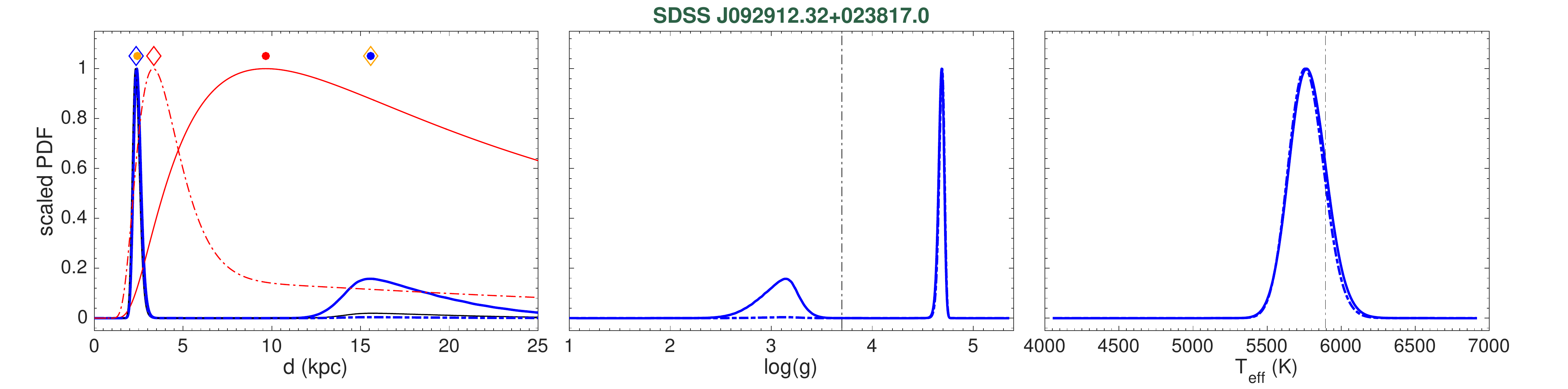}\\ 
\includegraphics[scale=.4]{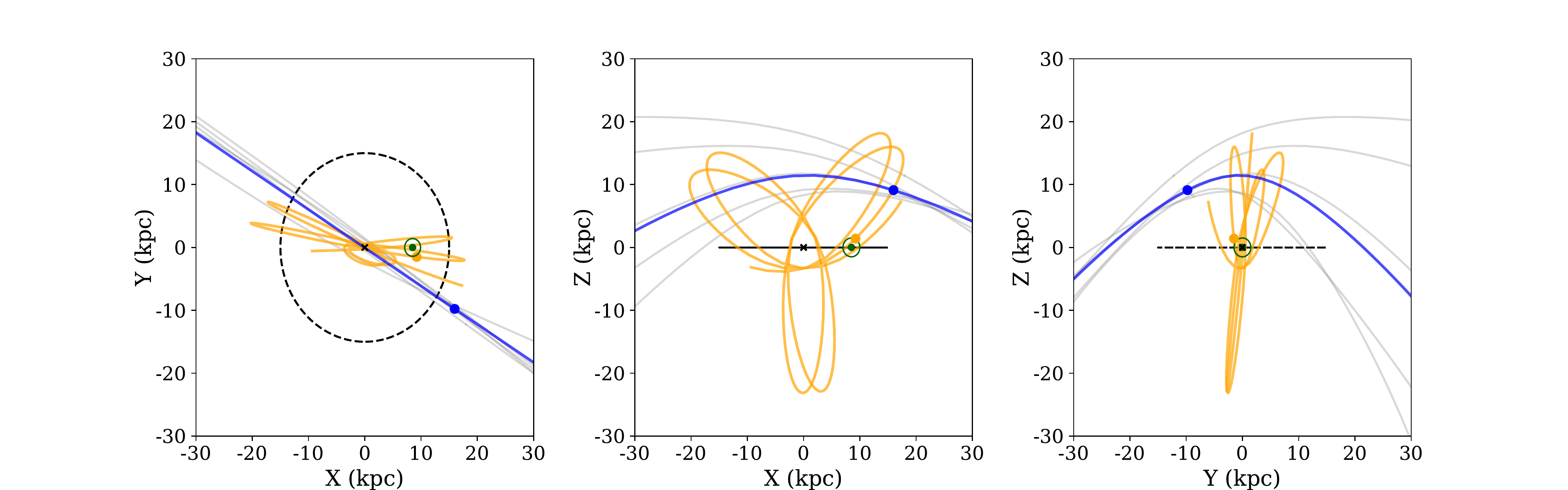} 
\caption{Same as Figure~\ref{SDSS J0023+0307}, but for SDSS J092912.32+023817.0.} 
\label{SDSS J092912.32+023817.0} 
\end{figure*}

The distance PDF for this star \citep[][Figure~\ref{SDSS J092912.32+023817.0}]{2015AA...579A..28B,2016AA...595L...6C} shows two solutions that are not strongly constrained due to the non-informative Gaia parallax ($\varpi= 0.1276 \pm 0.1872 \mas$). Using a MW halo prior, the sub-giant scenario has a greater likelihood (68$\%$ vs. 32$\%$), but it yields an orbit that is not bound to the MW. We therefore reanalyse this star using a disc$+$halo orbit,  finding that the dwarf solution is now preferred (95$\%$ vs. 5$\%$). Hence, this star is located at a distance of $2.4\pm0.2\kpc$ (dwarf solution) and its orbit is perpendicular to the disc with $r_\mathrm{apo,dwarf} = 23.5^{+2.6}_{-1.4} \kpc$.

\subsection{SDSS J094708.27+461010.0}
\begin{figure*}
\includegraphics[scale=.4]{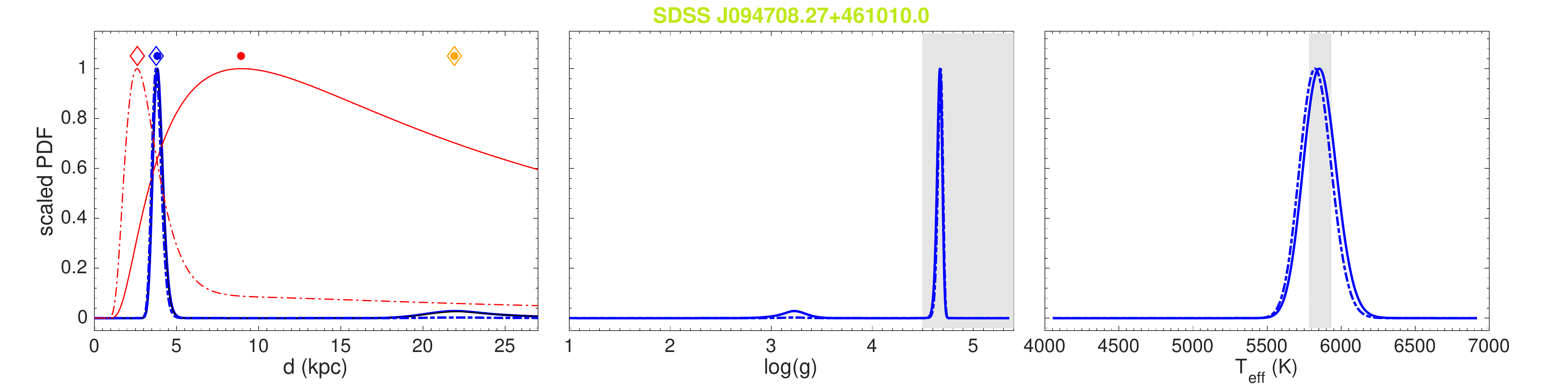}\\ 
\includegraphics[scale=.4]{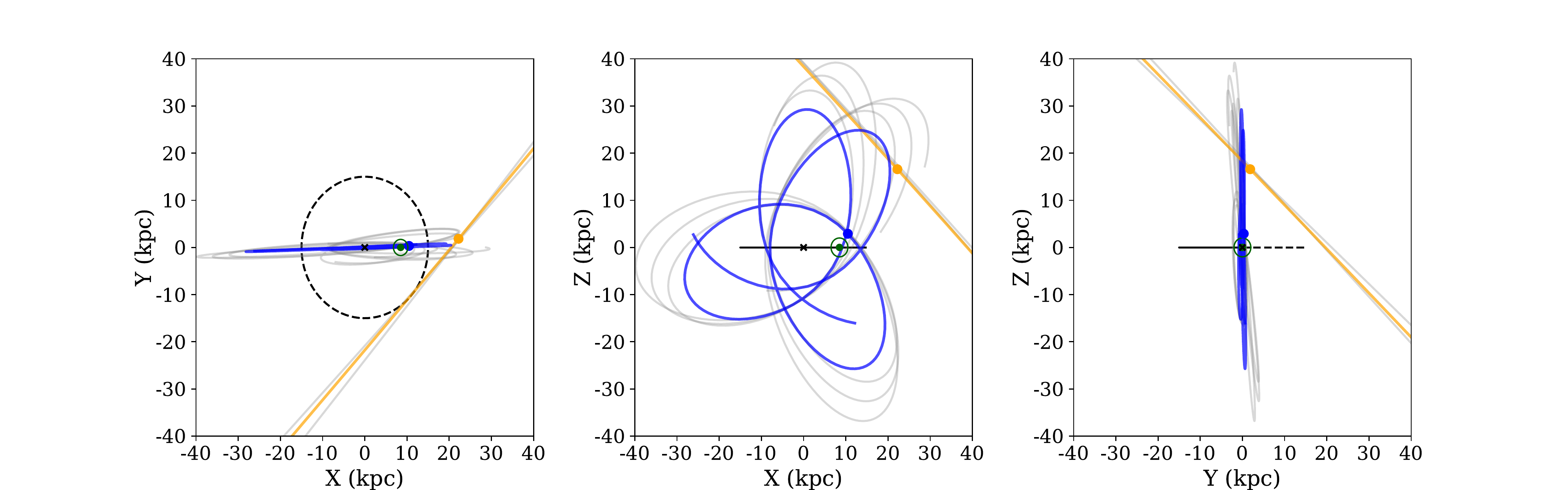} 
\caption{Same as Figure~\ref{SDSS J0023+0307}, but for SDSS J094708.27+461010.0.} 
\label{SDSS J094708.27+461010.0} 
\end{figure*}

The distance to SDSS~J094708.27+461010.0 is not constrained by the Gaia parallax \citep[$\varpi= 0.1989 \pm 0.2299\mas$,][Figure~\ref{SDSS J094708.27+461010.0}]{Aguado2017}. However, for similar reasons to those mentioned above, we favour the dwarf scenario (distance of $3.8 \pm 0.3\kpc$) as a larger distance would mean that this star is not bound to the MW. The orbital analysis shows that its orbital plane is perpendicular to the MW plane. 

\subsection{HE 1012-1540}
\begin{figure*}
\includegraphics[scale=.4]{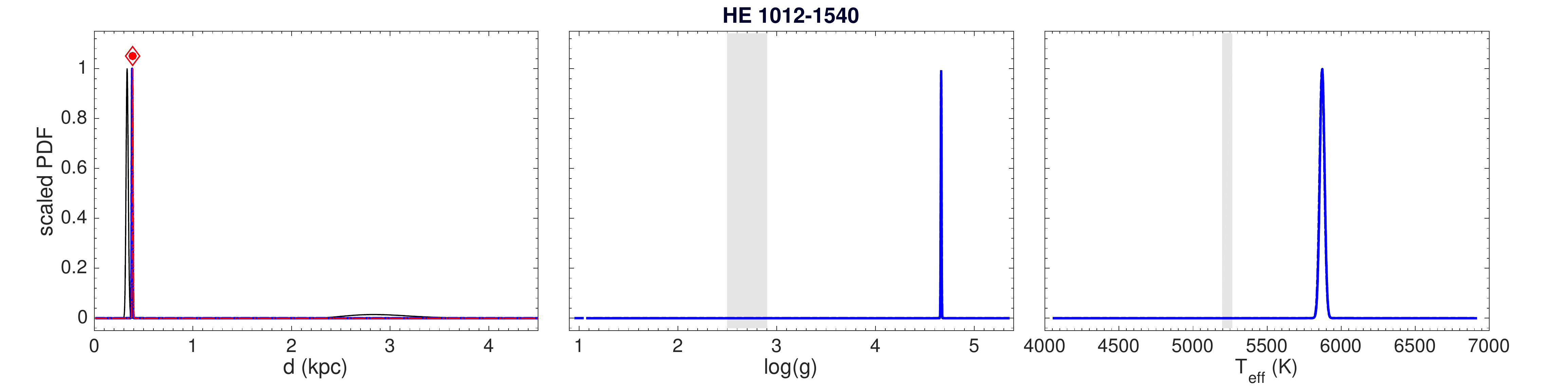}\\ 
\includegraphics[scale=.4]{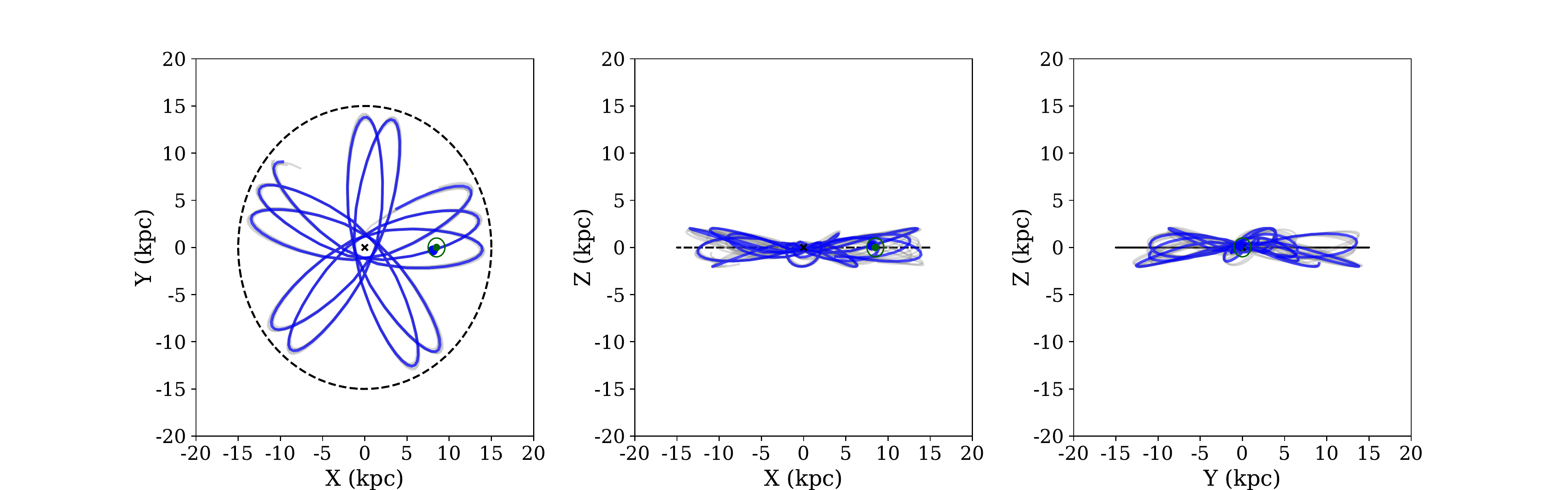} 
\caption{Same as Figure~\ref{SDSS J0023+0307}, but for HE 1012-1540.} 
\label{HE 1012-1540} 
\end{figure*}

For HE~1012-1540 \citep[][Figure~\ref{HE 1012-1540}]{Roederer2014}, the combination of photometric likelihood and the exquisite Gaia parallax leads to a distance of $0.384 \pm 0.004\kpc$ and strongly implies that this is a dwarf star. It is worth noting that the inferred stellar parameters are not in agreement with the literature in which the giant solution is preferred, but the latter seems hardly compatible with the strongly constrained distance. The orbit of this star implies that it remains confined close to the MW plane but has a high eccentricity ($\epsilon=0.83^{+0.00}_{-0.01}$).

\subsection{SDSS J102915+172927}
\begin{figure*}
\includegraphics[scale=.4]{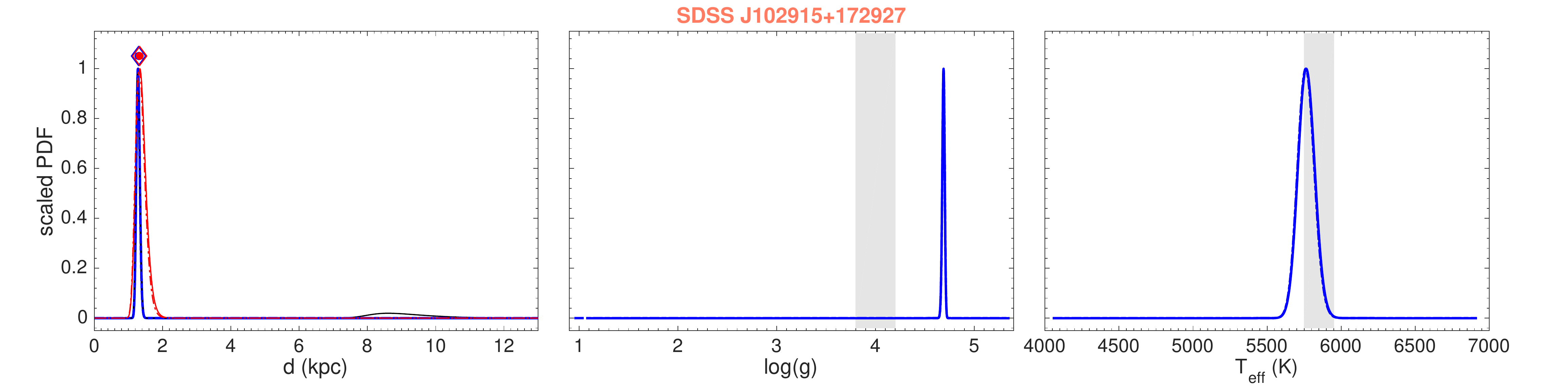}\\ 
\includegraphics[scale=.4]{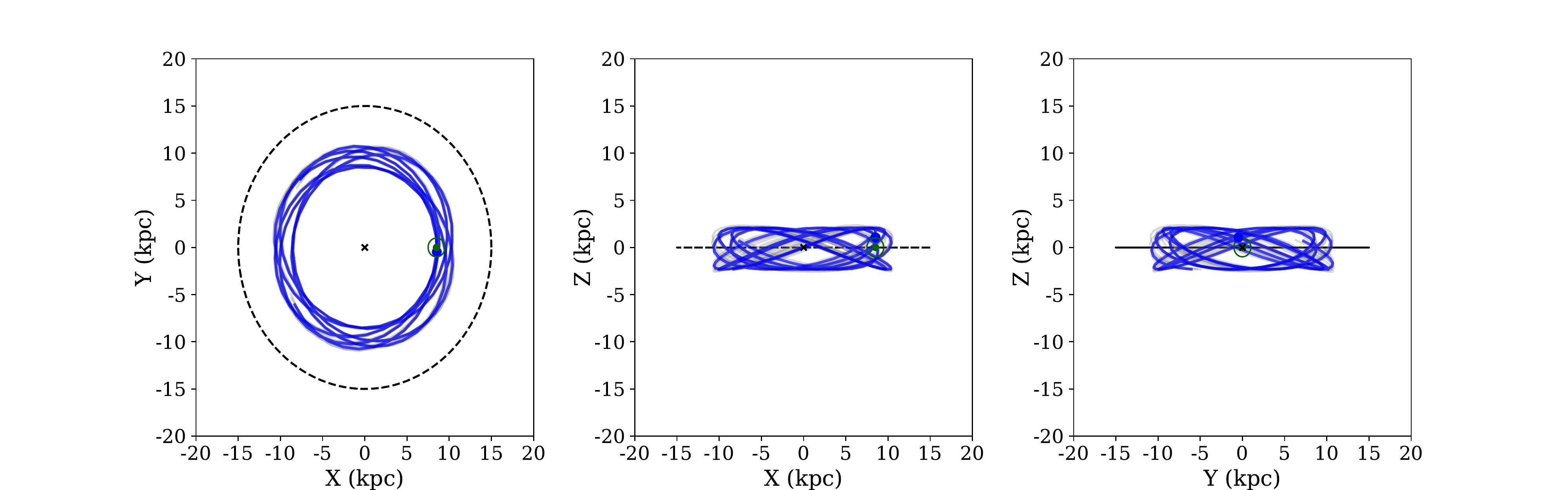} 
\caption{Same as Figure~\ref{SDSS J0023+0307}, but for SDSS J102915+172927.} 
\label{SDSS J102915+172927} 
\end{figure*}

SDSS J102915+172927, which is currently the most metal-poor star known \citep{Caffau2011aa}, is presented in Figure~\ref{SDSS J102915+172927}. The dwarf solution from the photometric likelihood is in agreement with the Gaia parallax and yields a well-constrained distance of $1.28 \pm 0.05\kpc$. We infer a higher surface gravity than in the literature, but our effective temperature inference is compatible. The orbital analysis shows that this star has the orbit of a disc star with an almost circular orbit around the galactic centre ($\epsilon=0.12^{+0.01}_{-0.01}$) that remains close to the MW plane ($|Z|<2.3\kpc$). These orbital properties differ from but supersede those of \citet{Caffau12} that were based on PPMXL Catalogue for proper motions \citep[][$\mu_{\alpha}= -12.8 \pm 3.9 \masyr$ and $\mu_{\delta}= -6.7 \pm 3.9 \masyr$]{PPMXL}.

\subsection{SDSS J103402.70+070116.6} 
\begin{figure*}
\includegraphics[scale=.4]{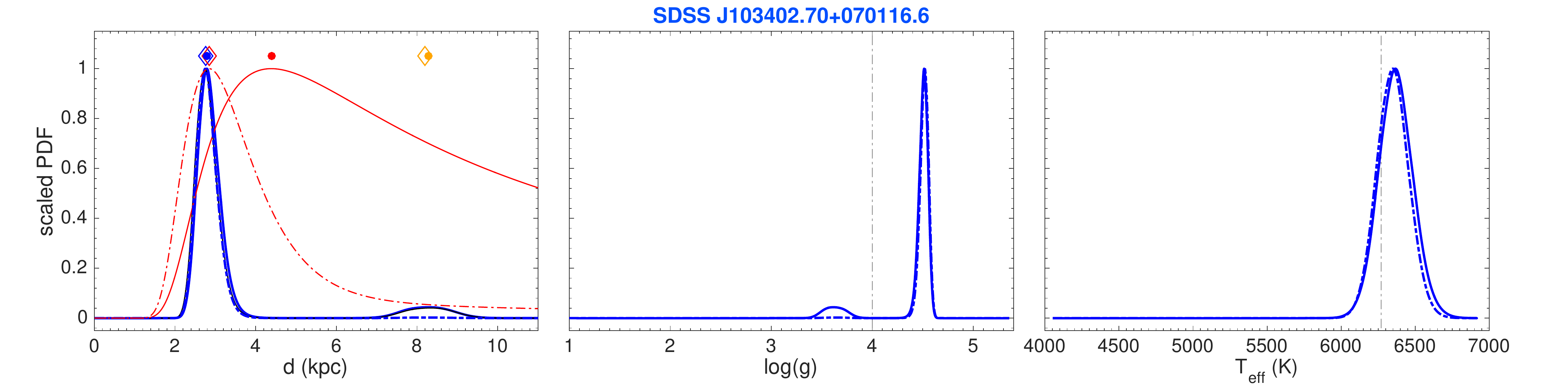}\\ 
\includegraphics[scale=.4]{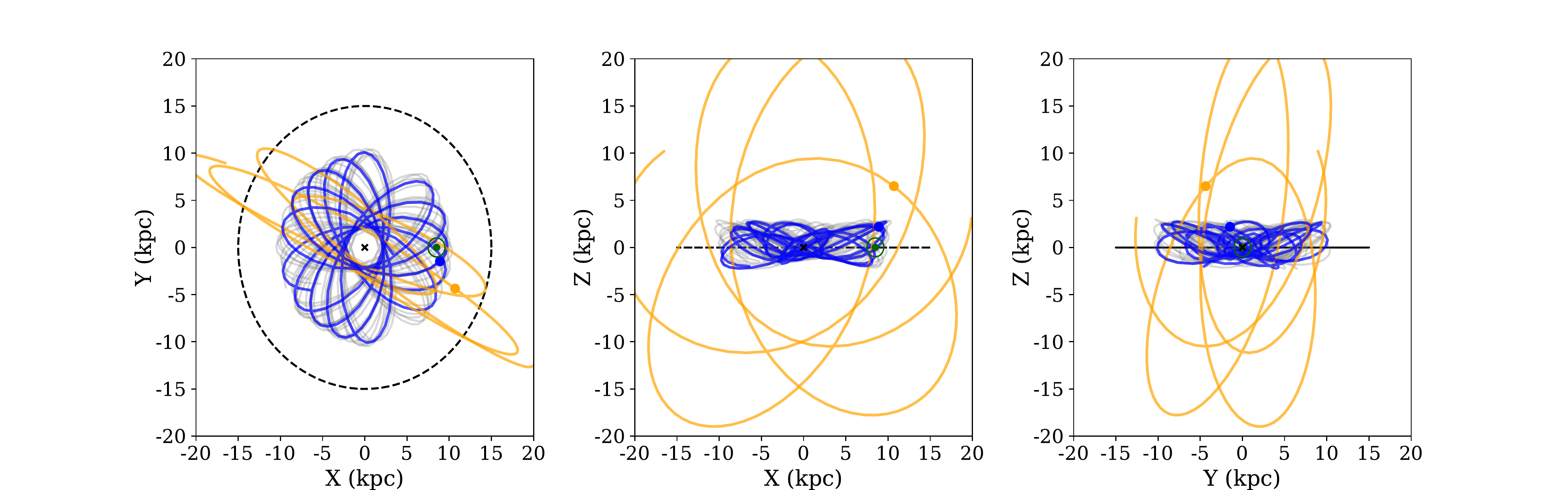} 
\caption{Same as Figure~\ref{SDSS J0023+0307}, but for SDSS J103402.70+070116.6.} 
\label{SDSS J103402.70+070116.6} 
\end{figure*}

Our results for SDSS~J103402.70+070116.6 \citep{Bonifacio2018} are shown in Figure~\ref{SDSS J103402.70+070116.6} and, as we can see, the Gaia parallax does not allow us to break the dwarf/sub-giant degeneracy ($\varpi= 0.2874 \pm 0.1367\mas$). The dwarf solution ($P_\mathrm{dwarf}=89\%$ vs. $P_\mathrm{giant}=11\%$) at $2.79 \pm 0.26\kpc$ implies an eccentric orbit ($\epsilon= 0.63^{+0.03}_{-0.04}$) that remains confined to the Galactic plane ($|Z|< 2.7 \kpc$). On the other hand, the subgiant solution at $8.3\pm0.6\kpc$ brings that star further out in the halo $r_\mathrm{apo}=24.3^{+17.8}_{-0.0} \kpc$. Repeating the analysis with the disc$+$halo prior, the two new solutions are in agreement within the uncertainties with previous results, but now the sub-giant scenario is strongly suppressed (0.6$\%$).

\subsection{SDSS J103556.11+064143.9}
For this star \citep{2015AA...579A..28B}, the Gaia parallax is negative and does not help to constrain the distance ($\varpi=-0.3912  \pm 0.3163 \mas$). Our analysis implies that the dwarf solution at $3.97 \pm 0.34\kpc$ is more likely and this is confirmed by the orbital analysis that yields a large value for the apocentre in case of the giant solution ($r_\mathrm{apo}=147.8^{+25.5}_{-11.8} \kpc$). Just like with SDSS~J103402.70+070116.6 above, the literature $\log(g)$ falls in-between the two solutions we obtain and only the effective temperature inference is compatible with the literature.

\subsection{SDSS J105519.28+232234.0}
\begin{figure*}
\includegraphics[scale=.4]{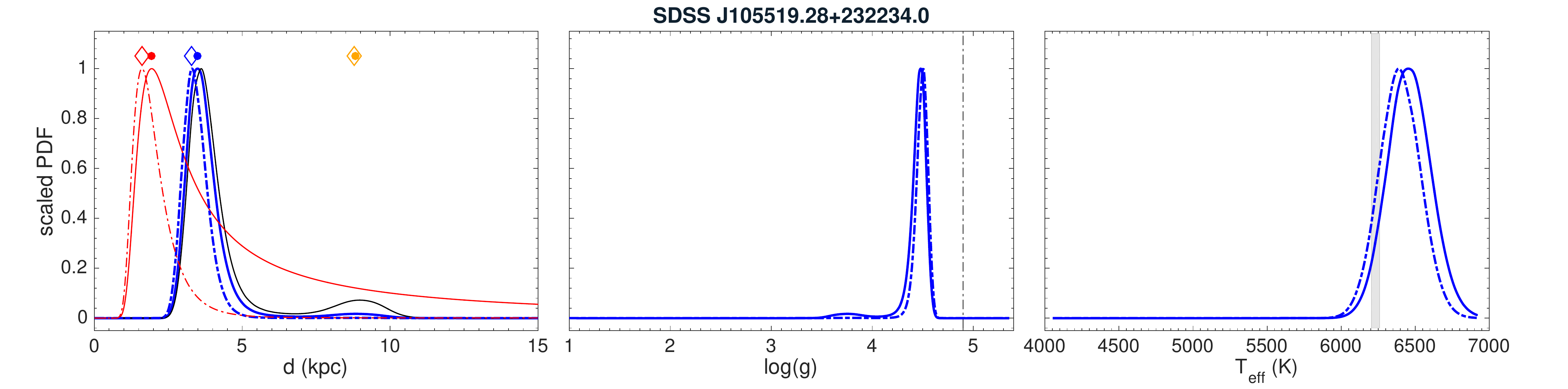}\\ 
\includegraphics[scale=.4]{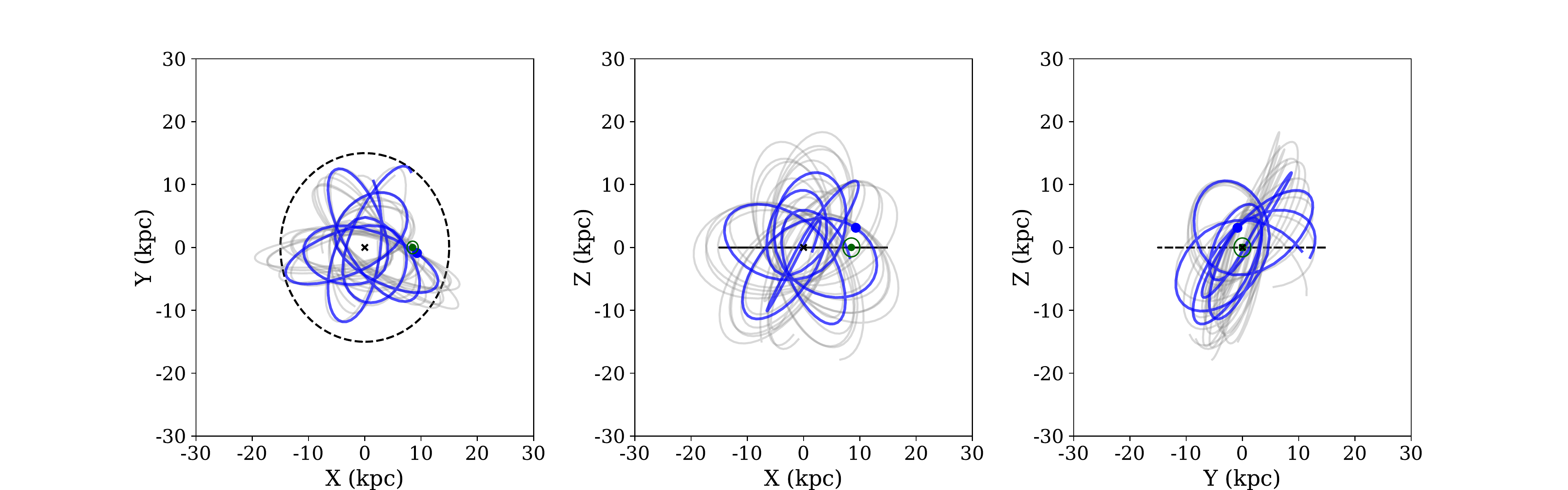} 
\caption{Same as Figure~\ref{SDSS J0023+0307}, but for SDSS J105519.28+232234.0.} 
\label{SDSS J105519.28+232234.0} 
\end{figure*}

The distance PDF for this star \citep[][Figure~\ref{SDSS J105519.28+232234.0}]{Aguado2017b} indicates a strongly preferred distance of $3.49 \pm 0.45\kpc$ corresponding to the dwarf solution, with the effective temperature in agreement with the literature. The inferred orbital parameters indicate an inner halo orbit.

\subsection{SDSS J120441.38+120111.5}
\begin{figure*}
\includegraphics[scale=.4]{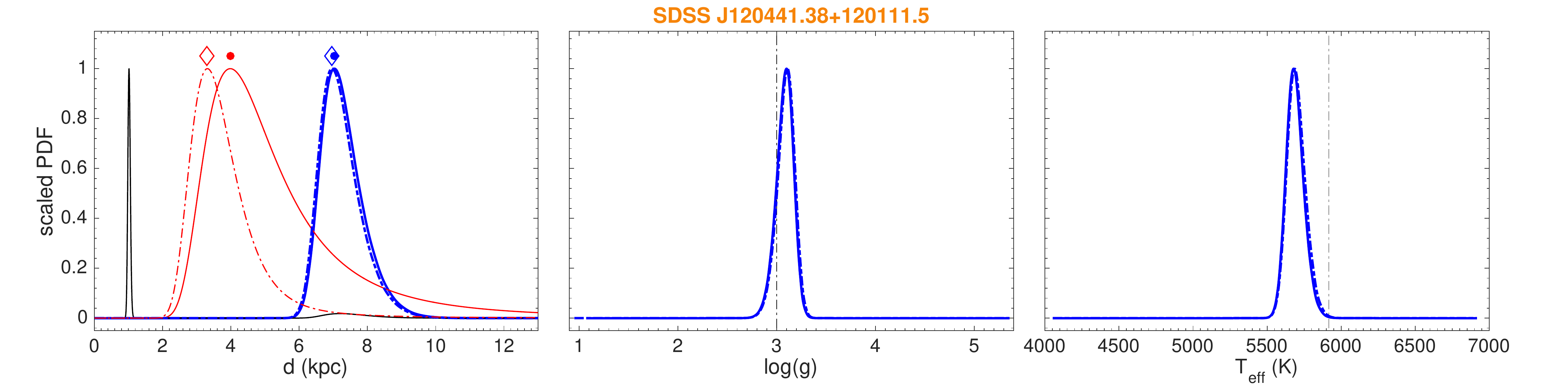}\\ 
\includegraphics[scale=.4]{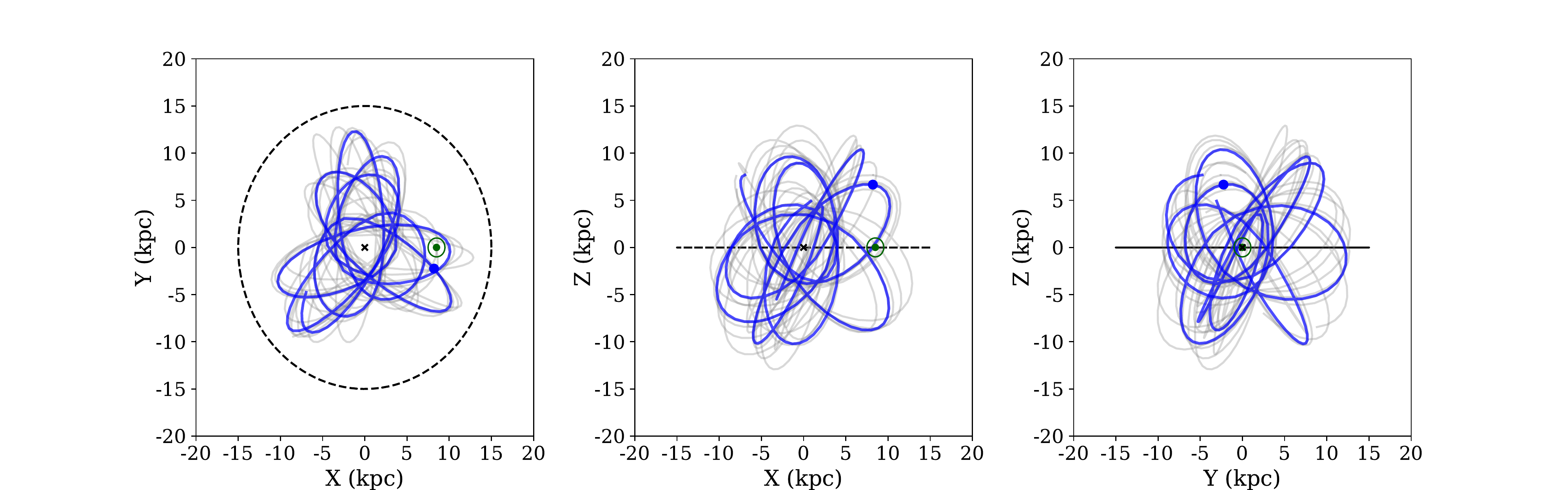} 
\caption{Same as Figure~\ref{SDSS J0023+0307}, but for SDSS J120441.38+120111.5.} 
\label{SDSS J120441.38+120111.5} 
\end{figure*}

The analysis on this star \citep[][Figure~\ref{SDSS J120441.38+120111.5}]{2015ApJ...809..136P} leads to the conclusion that this star is a subgiant located at a distance of $7.03 \pm 0.54\kpc$ from the Sun with an inner halo-like orbit.

\subsection{SDSS J124719.46-034152.4}
\begin{figure*}
\includegraphics[scale=.4]{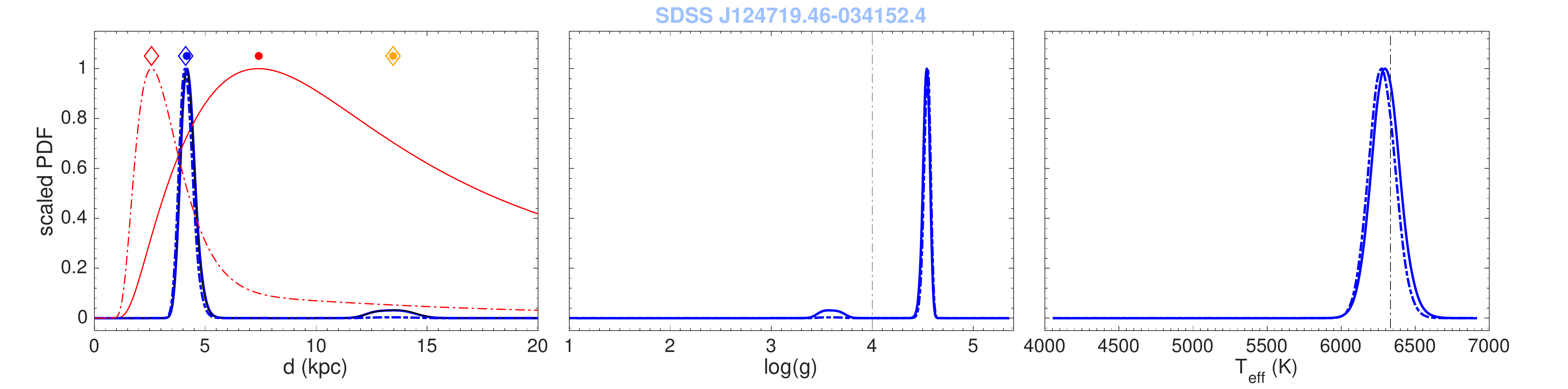}\\ 
\includegraphics[scale=.4]{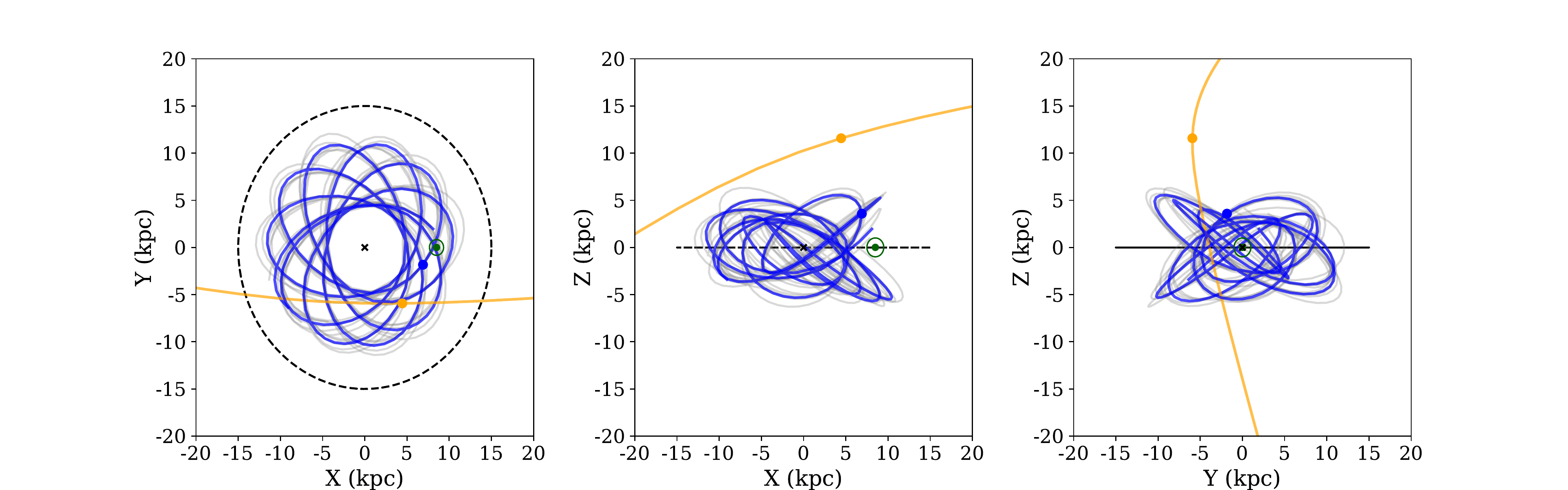} 
\caption{Same as Figure~\ref{SDSS J0023+0307}, but for SDSS J124719.46-034152.4.} 
\label{SDSS J124719.46-034152.4} 
\end{figure*}

The Gaia parallax on this star is poorly constraining ($\varpi=0.3075 \pm 0.2098 \mas$, Figure~\ref{SDSS J124719.46-034152.4}) and, combined with the photometric likelihood, we obtain a favoured distance of $4.17 \pm 0.32\kpc$ corresponding to the dwarf solution that has an inner halo orbit. The far less likely sub-giant solution yields an orbit that is not bound to the MW. For the stellar parameters, the inferred effective temperature is compatible with the literature value  \citep{2013AA...560A..71C}.

\subsection{LAMOST J125346.09+075343.1}
\begin{figure*}
\includegraphics[scale=.4]{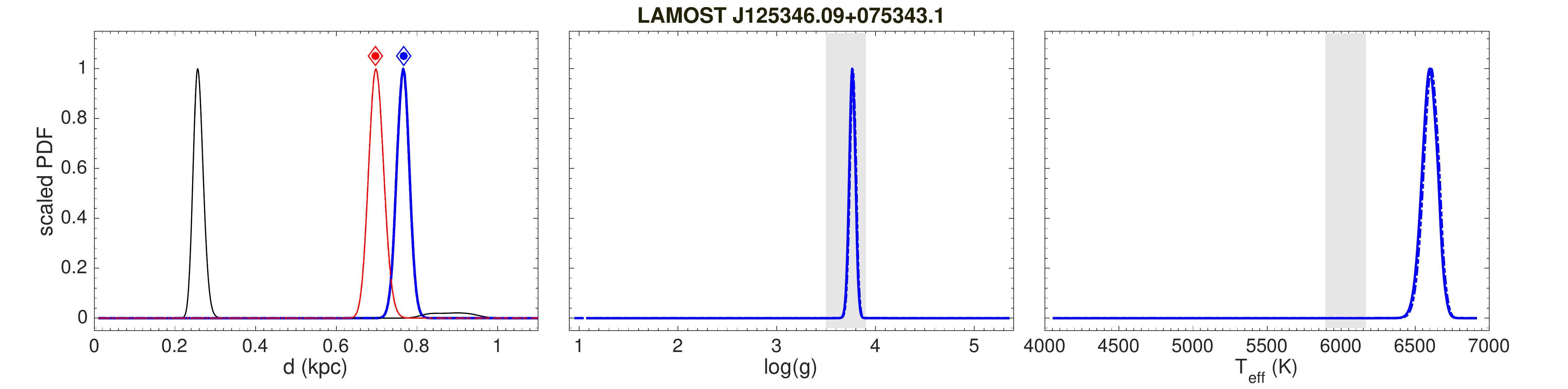}\\ 
\includegraphics[scale=.4]{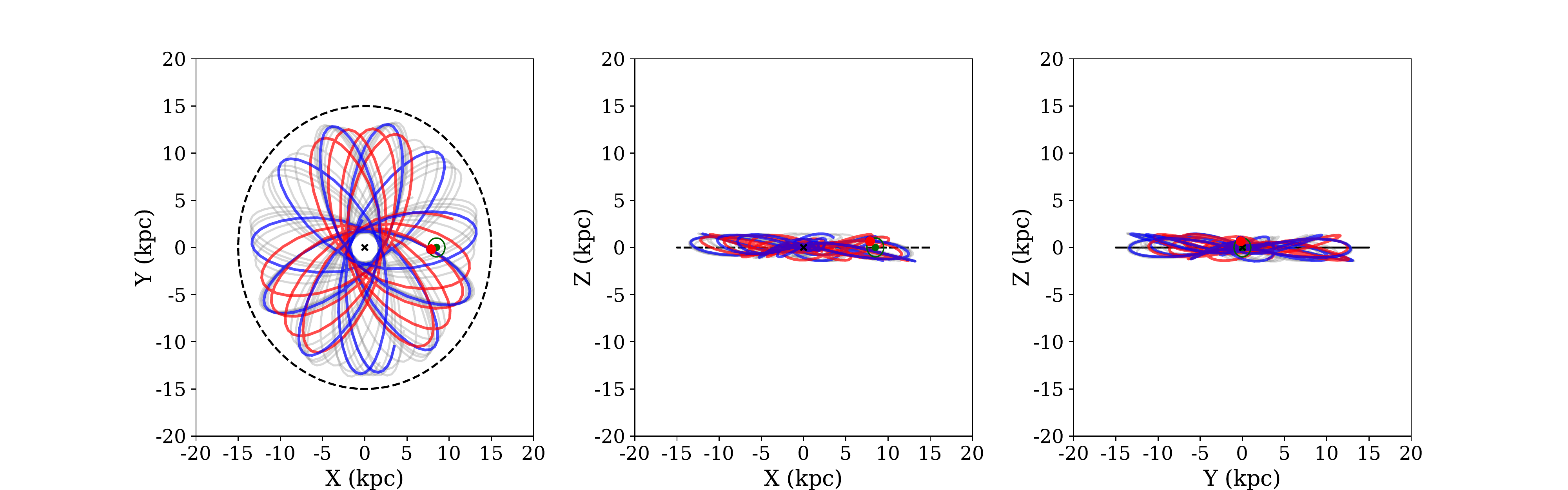} 
\caption{Same as Figure~\ref{SDSS J0023+0307}, but for LAMOST J125346.09+075343.1. For this star, the orbit inferred from the product between the astrometric likelihood and MW halo prior is shown with the red line.} 
\label{LAMOST J125346.09+075343.1} 
\end{figure*}

Figure~\ref{LAMOST J125346.09+075343.1} shows our results for LAMOST~J125346.09+075343.1 \citep{2015PASJ...67...84L} and, as we can see, the Gaia likelihood is not in agreement with the photometric one. Our combined distance analysis favours the sub-giant scenario and a distance of $0.766 \pm 0.016\kpc$, which is close to the Gaia-only inference ($0.698\pm 0.018\kpc$). The surface gravity we infer is compatible with the value in the literature but our analysis implies a hotter star. Both the orbits from Gaia and the distance PDF show that LAMOST~J125346.09+075343.1 remains confined to the MW plane, even though it has a high eccentricity ($\epsilon =0.75^{+0.03}_{-0.02} $). 

\subsection{SDSS J131326.89-001941.4}
\begin{figure*}
\includegraphics[scale=.4]{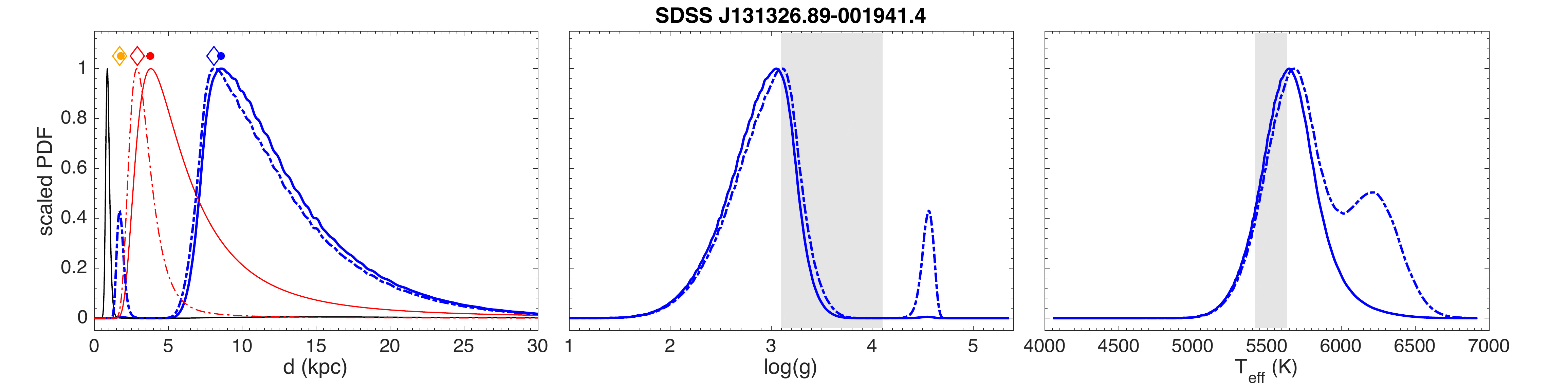}\\ 
\includegraphics[scale=.4]{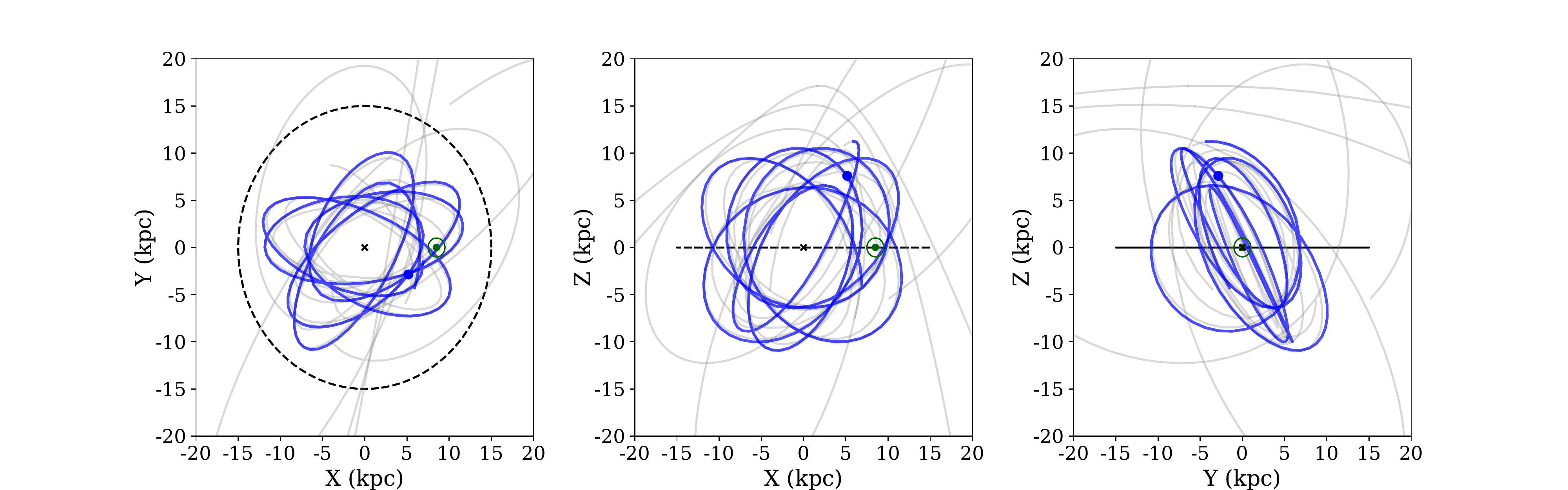} 
\caption{Same as Figure~\ref{SDSS J0023+0307}, but for SDSS J131326.89-001941.4.} 
\label{SDSS J131326.89-001941.4} 
\end{figure*}

The Gaia parallax for this object is poorly constraining ($\varpi=0.2976 \pm 0.0972 \mas$, Figure~\ref{SDSS J131326.89-001941.4}) and we obtain using a pure halo prior that the preferred solution, a giant ($>99\%$ chance), is located at the distance of $8.6 \pm 2.9 \kpc$, with the inferred stellar parameters that are in agreement with the literature \citep{Allende15,FrebelJ1313,Aguado2017b}. From the orbital analysis, this star is classifiable as inner halo.

\subsection{HE 1310-0536}
\begin{figure*}
\includegraphics[scale=.4]{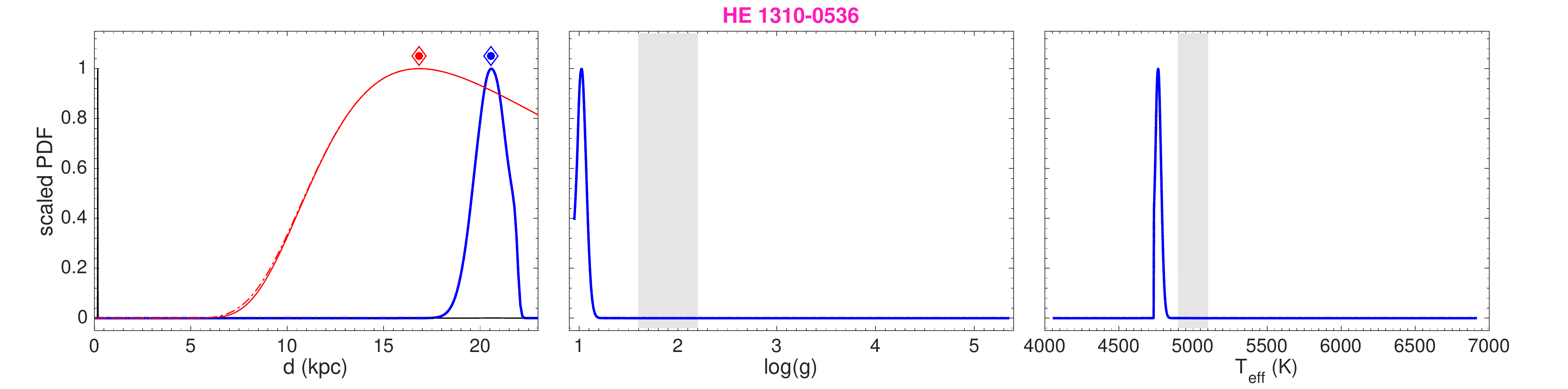}\\ 
\includegraphics[scale=.4]{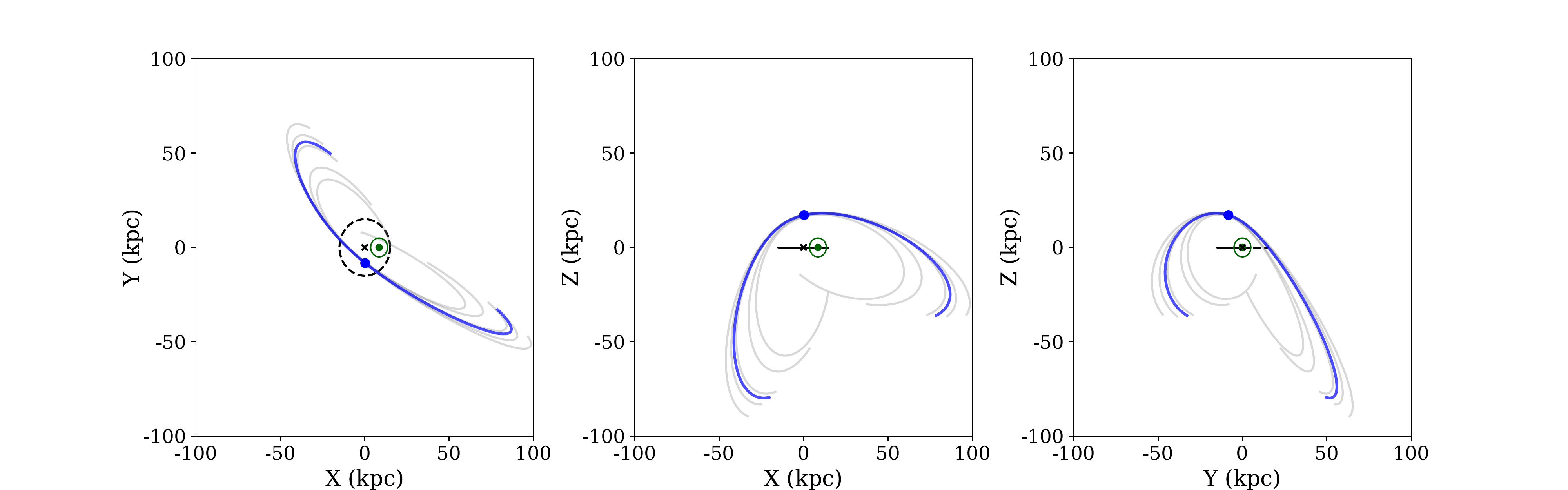} 
\caption{Same as Figure~\ref{SDSS J0023+0307}, but for HE 1310-0536.} 
\label{HE 1310-0536} 
\end{figure*}

The Gaia parallax ($\varpi= 0.0078\pm0.0342 \mas$) rules out the dwarf solution for HE~1310-0536 (Figure~\ref{HE 1310-0536}; \citealt{2015ApJ...807..173H}) and we infer a distance of $20.6 \pm 0.9\kpc$. The inferred stellar parameters are not in agreement with the literature, but this could stem from systematics in the red-giant-branch part of the isochrones we rely on. The orbit of this star clearly brings it in the outer parts of the halo, with $r_\mathrm{apo}=99.7^{+38.3}_{-26.0}\kpc$.

\subsection{HE 1327-2326}
\begin{figure*}
\includegraphics[scale=.4]{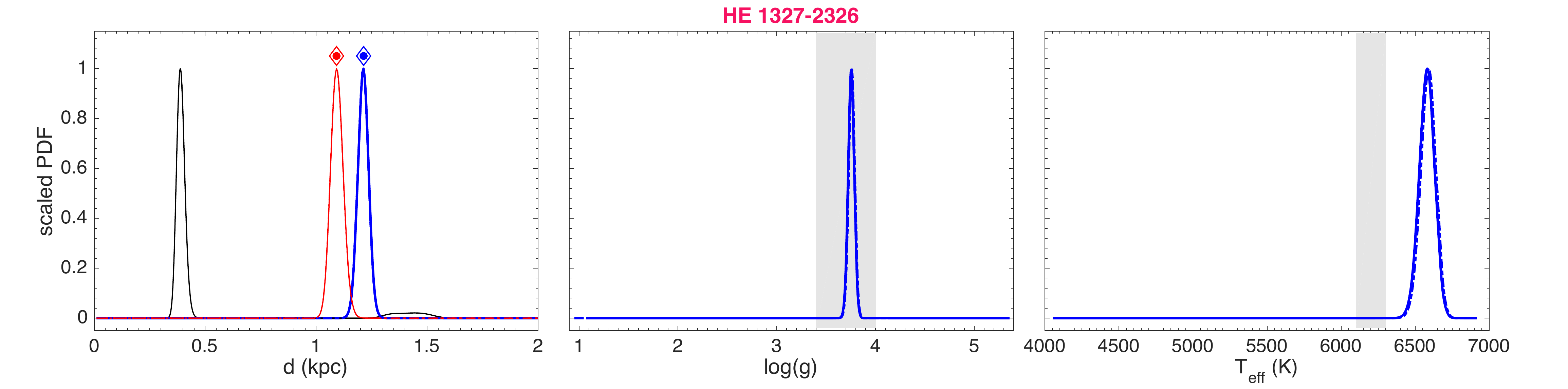}\\ 
\includegraphics[scale=.4]{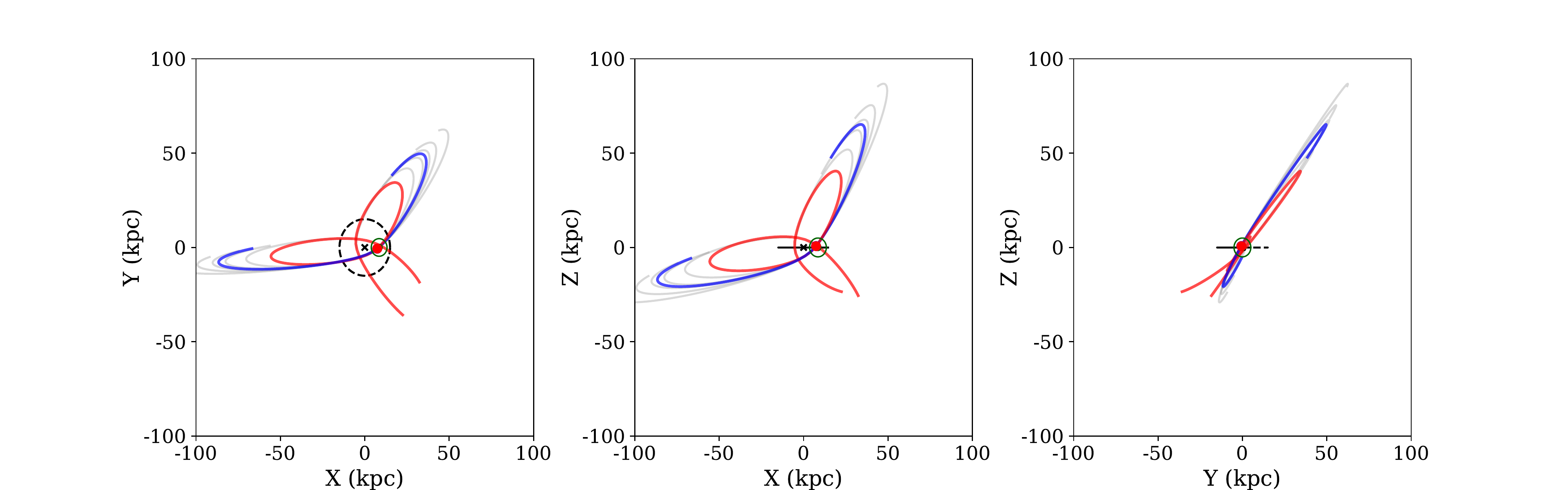} 
\caption{Same as Figure~\ref{SDSS J0023+0307}, but for HE 1327-2326. For this star, the orbit inferred from the product between the astrometric likelihood and MW halo prior is shown with the red line.} 
\label{HE 1327-2326} 
\end{figure*}

The results for HE~1327-2326 are shown in Figure~\ref{HE 1327-2326}, and, despite the fact that  the Gaia and the photometric likelihoods are not in good agreement, the sub-giant scenario is clearly favoured. The distance obtained for the combined analysis is $1.21 \pm 0.02\kpc$ (or $1.09\pm 0.03\kpc$ for the Gaia only analysis) and the inferred effective temperature deviates somewhat from the literature value \citep{0004-637X-684-1-588}. Even though the combined and Gaia-only distances yield significantly different orbits, they both imply halo orbits.

\subsection{HE 1424-0241}
\begin{figure*}
\includegraphics[scale=.4]{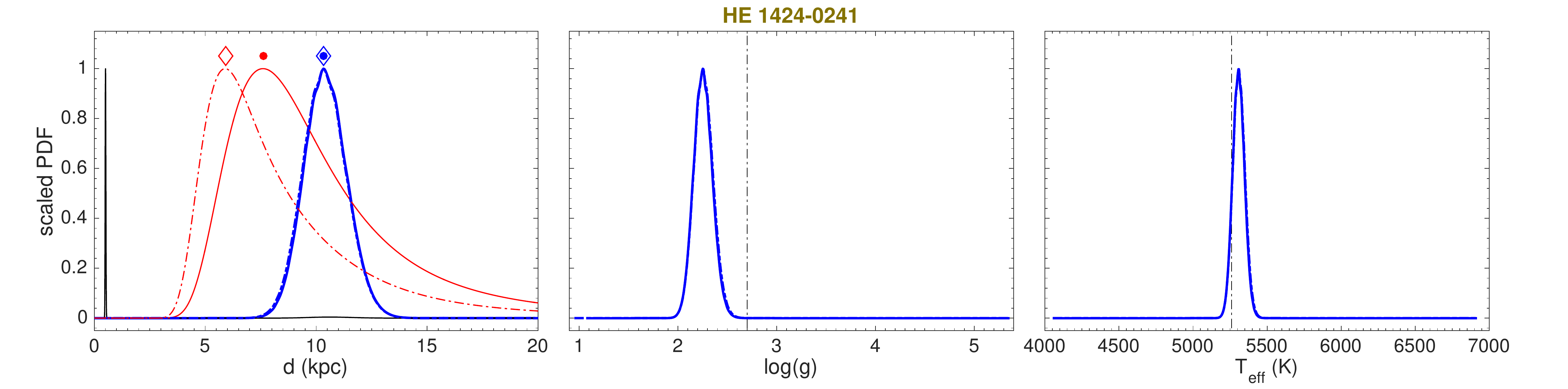}\\ 
\includegraphics[scale=.4]{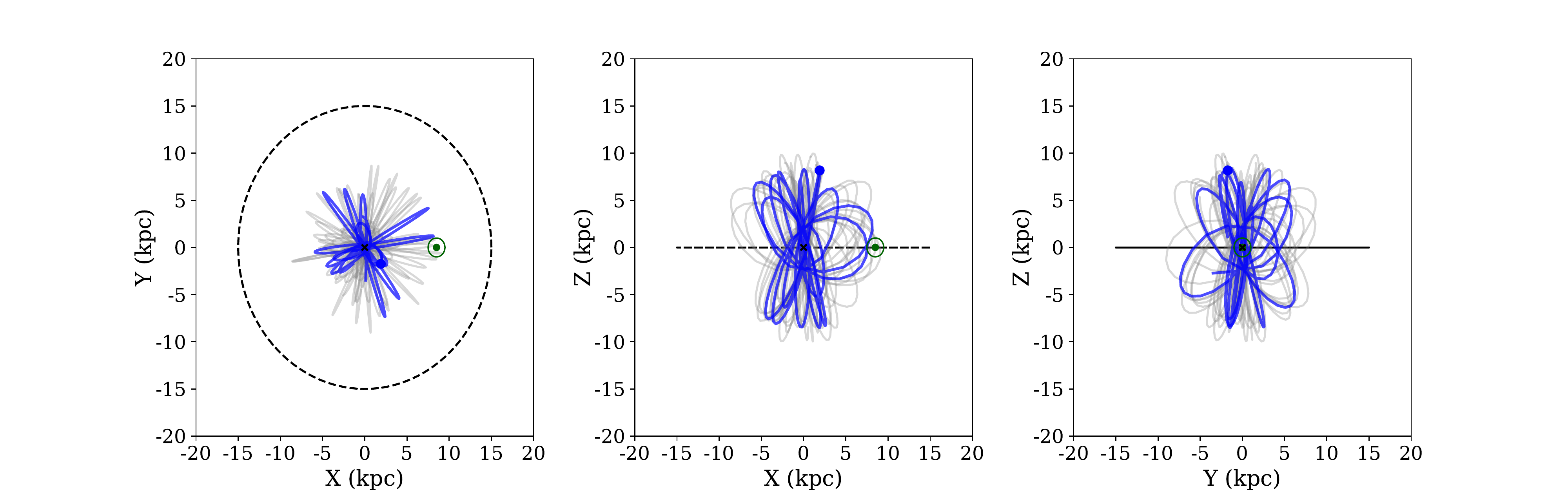} 
\caption{Same as Figure~\ref{SDSS J0023+0307}, but for HE 1424-0241.} 
\label{HE 1424-0241} 
\end{figure*}

This giant star is located at a distance of $10.3 \pm 1.0\kpc$ (Figure~\ref{HE 1424-0241}) and the inferred stellar parameters are in agreement with the literature \citep{2008ApJ...672..320C,2013ApJ...762...25N}. The orbital analysis shows that HE~1424-0241 has an inner-halo orbit with high a eccentricity ($\epsilon=0.81^{+0.06}_{-0.09}$).

\subsection{SDSS J144256.37-001542.7}
\begin{figure*}
\includegraphics[scale=.4]{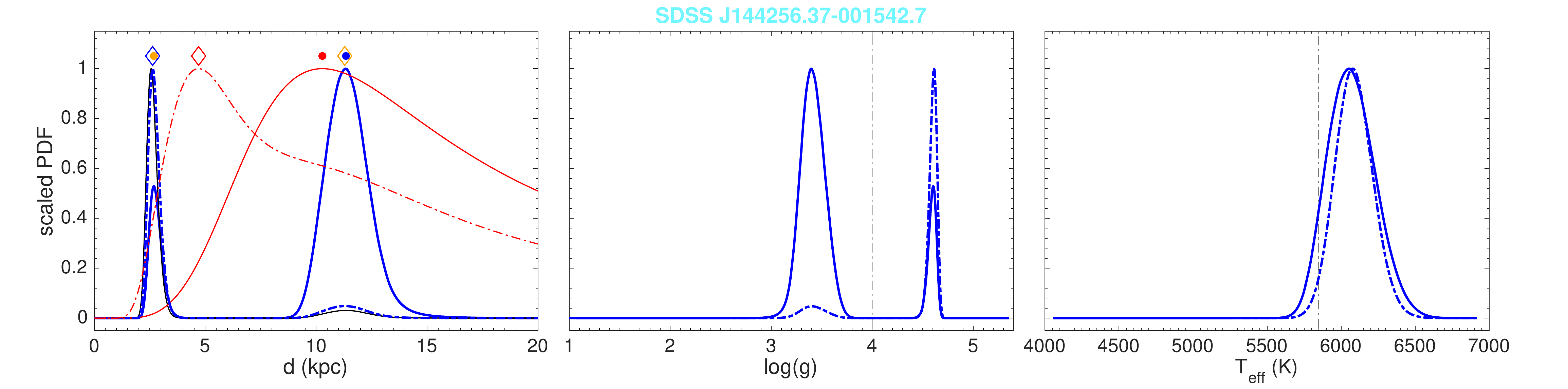}\\ 
\includegraphics[scale=.4]{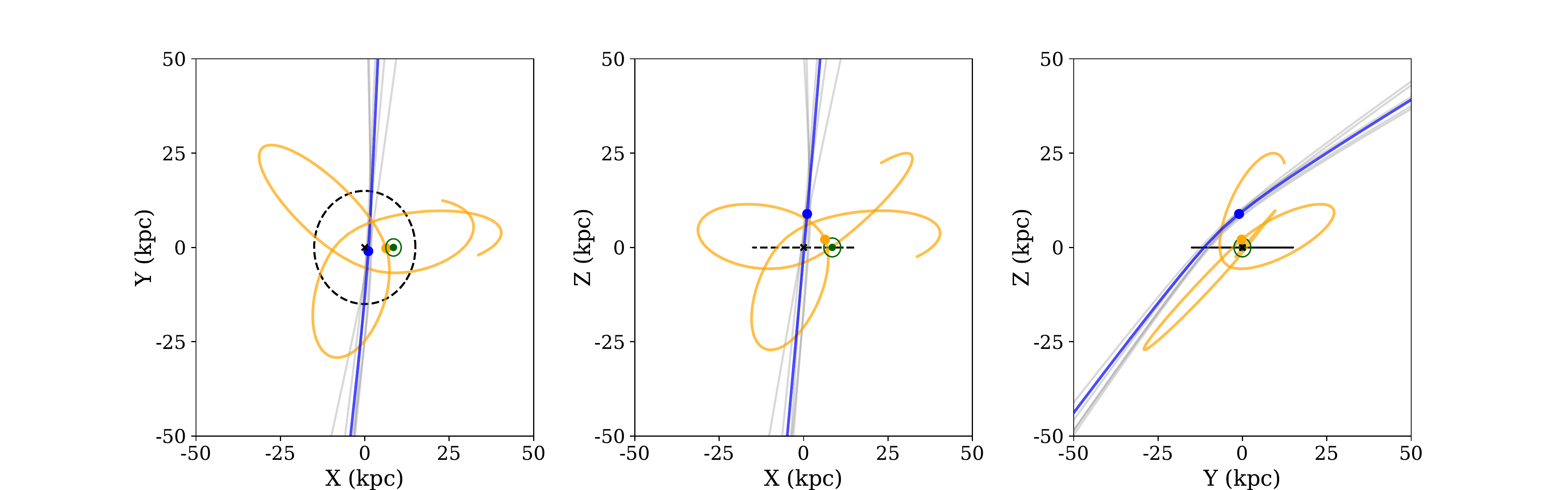} 
\caption{Same as Figure~\ref{SDSS J0023+0307}, but for SDSS J144256.37-001542.7.} 
\label{SDSS J144256.37-001542.7} 
\end{figure*}

The distance PDF for the combined analysis of SDSS~J144256.37-001542.7 (\citealt{2013AA...560A..15C}; Figure~\ref{SDSS J144256.37-001542.7}) still shows two peaks because of the poorly constraining Gaia parallax ($\varpi= -0.3910 \pm 0.2981 \mas$). The giant solution and its distance of $11.3 \pm 1.0\kpc$ is the preferred one with a halo prior ($P_\mathrm{giant}=87\%$ vs. $P_\mathrm{dwarf}=13\%$) but implies an unbound orbit whereas the dwarf solution at $2.68 \pm 0.27\kpc$ yields a more benign halo orbit with $r_\mathrm{apo} = 39.1^{+5.2}_{-2.6}\kpc$. Similar distances are found with a disc$+$halo prior but with the dwarf solution as preferred scenario ($P_\mathrm{dwarf}=84\%$ vs. $P_\mathrm{giant}=16\%$).

\subsection{Pristine 221.8781+9.7844}
\begin{figure*}
\includegraphics[scale=.4]{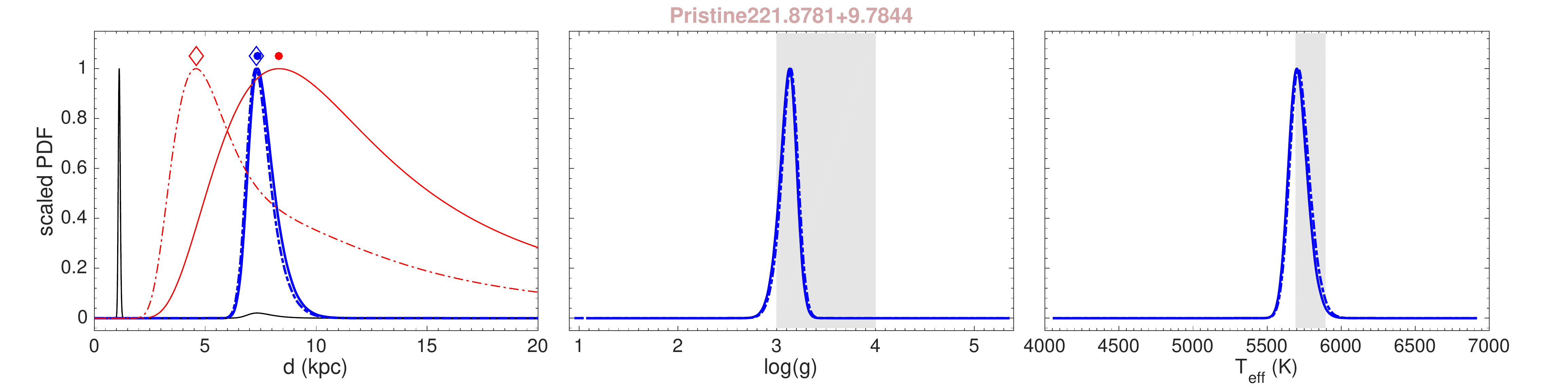}\\ 
\includegraphics[scale=.4]{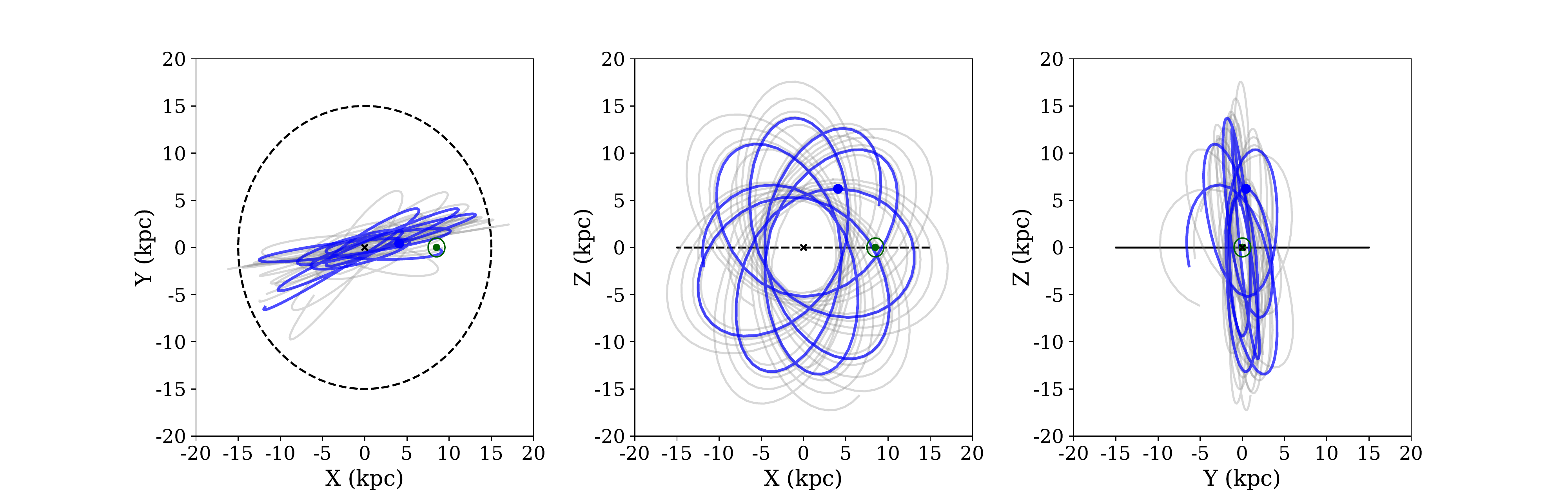} 
\caption{Same as Figure~\ref{SDSS J0023+0307}, but for Pristine221.8781+9.7844.} 
\label{Pristine221.8781+9.7844} 
\end{figure*}

The small Gaia parallax of Pristine~221.8781+9.7844 (\citealt{Pristine221}; $\varpi = 0.1187  \pm 0.0940  \mas$) rules out the dwarf solution. Hence the final picture of a subgiant located at $7.36 \pm 0.55\kpc$ from the Sun. As we can see from Figure~\ref{Pristine221.8781+9.7844}, the inferred stellar parameters agree with the literature and the orbit we infer for Pristine~221.8781+9.7844 indicate that this star has a halo orbit almost perpendicular to the MW plane.

\subsection{SDSS J164234.48+443004.9}
\begin{figure*}
\includegraphics[scale=.4]{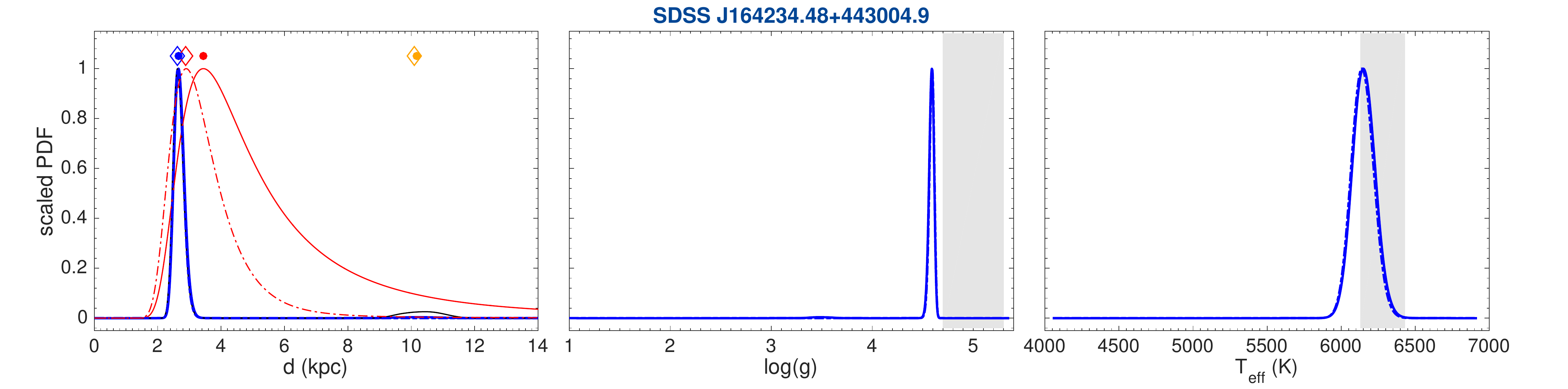}\\ 
\includegraphics[scale=.4]{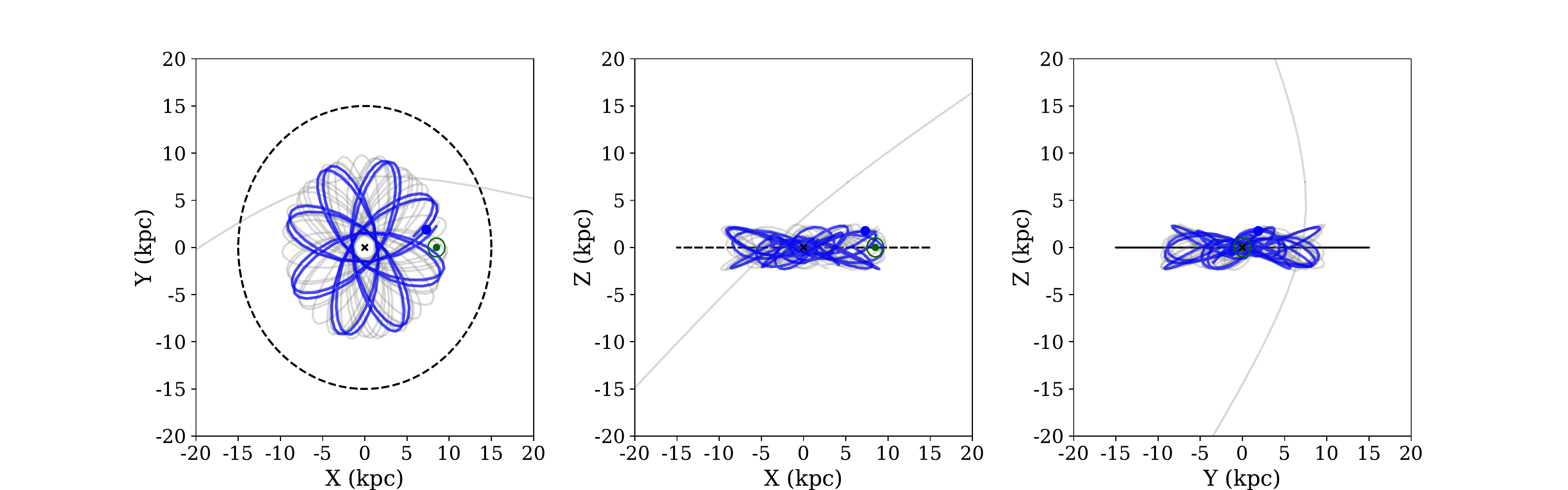} 
\caption{Same as Figure~\ref{SDSS J0023+0307}, but for SDSS J164234.48+443004.9.} 
\label{SDSS J164234.48+443004.9} 
\end{figure*}

SDSS~J164234.48+443004.9 (Figure~\ref{SDSS J164234.48+443004.9}) is a dwarf star located at a distance of $2.66 \pm 0.16\kpc$ (Figure~\ref{SDSS J164234.48+443004.9}). The stellar parameters are compatible with the literature values \citep{Aguado2016}. The orbital analysis suggests that this star remains confined to the MW plane, but has a high eccentricity ($\epsilon= 0.72^{+0.03}_{-0.04}$).

\subsection{SDSS J173403.91+644633.0}
\begin{figure*}
\includegraphics[scale=.4]{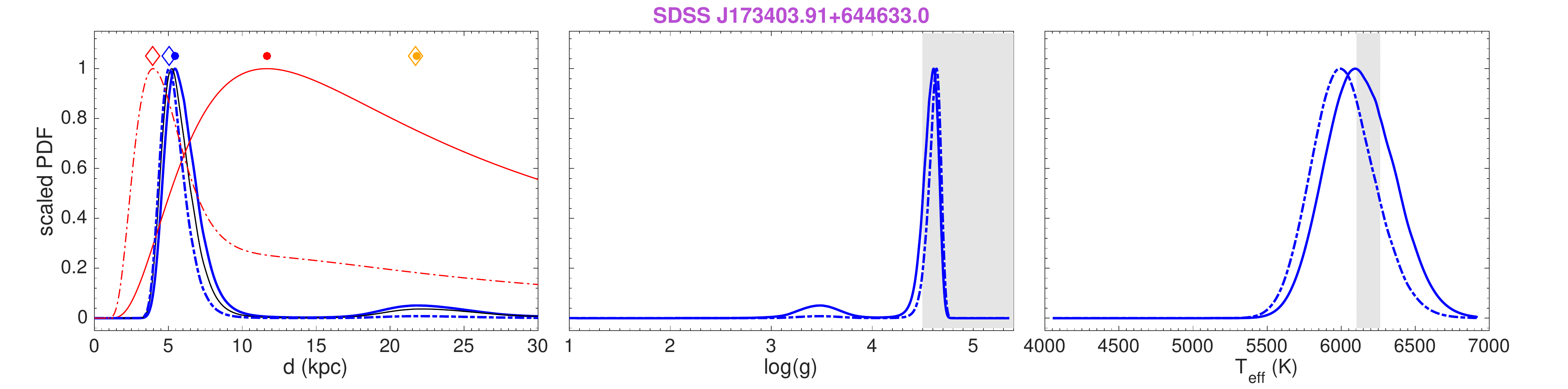}\\ 
\includegraphics[scale=.4]{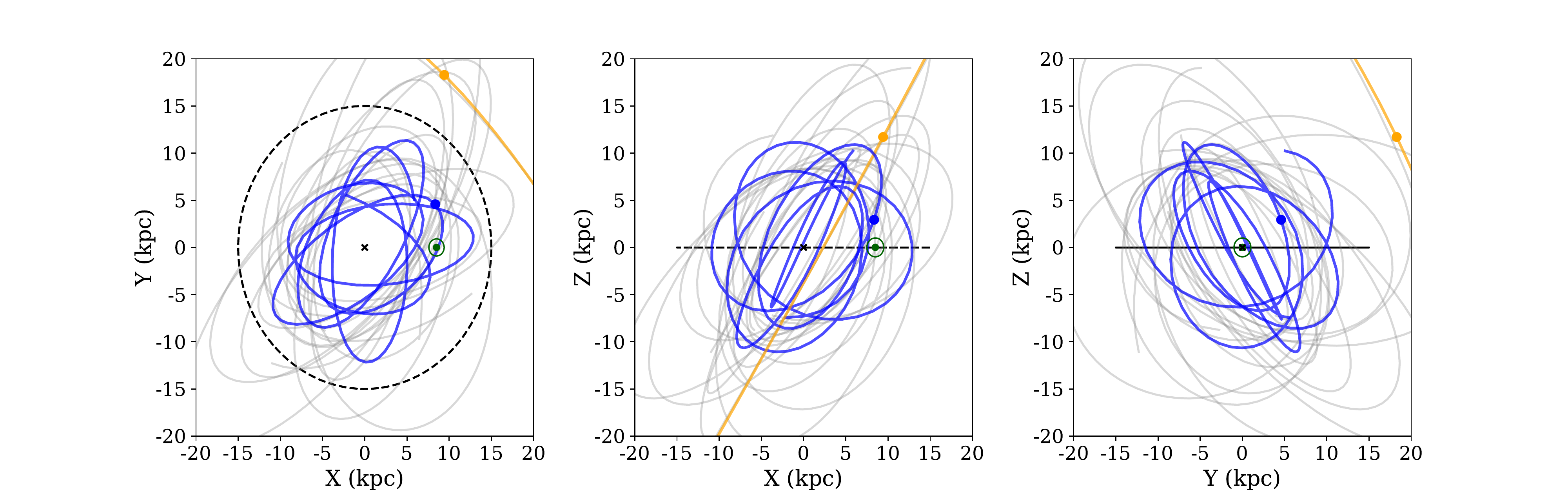} 
\caption{Same as Figure~\ref{SDSS J0023+0307}, but for SDSS J173403.91+644633.0.} 
\label{SDSS J173403.91+644633.0} 
\end{figure*}

SDSS~J173403.91+644633.0 has a non-informative Gaia parallax ($\varpi= -0.1052  \pm 0.2702 \mas$) that does not break the dwarf/giant degeneracy (Figure~\ref{SDSS J173403.91+644633.0}). The dwarf solution with a distance of $5.46 \pm 1.02\kpc$ is nevertheless strongly preferred by the photometric analysis and yields a more realistic inner halo orbit that remains bound to the MW, contrary to the giant solution. The inferred stellar parameters are in agreement with those from the \citet{Aguado2017} analysis.

\subsection{SDSS J174259.67+253135.8}
\begin{figure*}
\includegraphics[scale=.4]{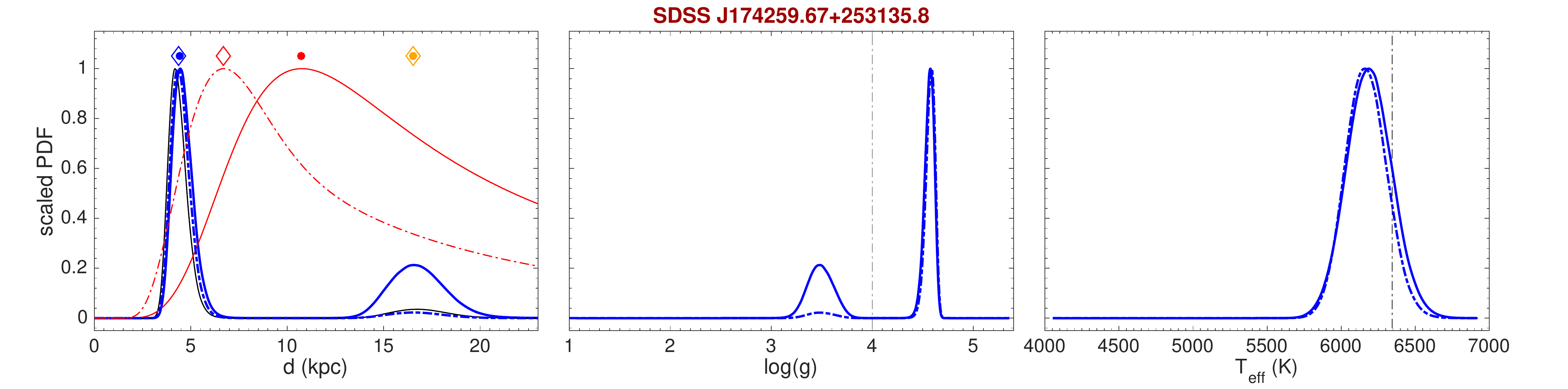}\\ 
\includegraphics[scale=.4]{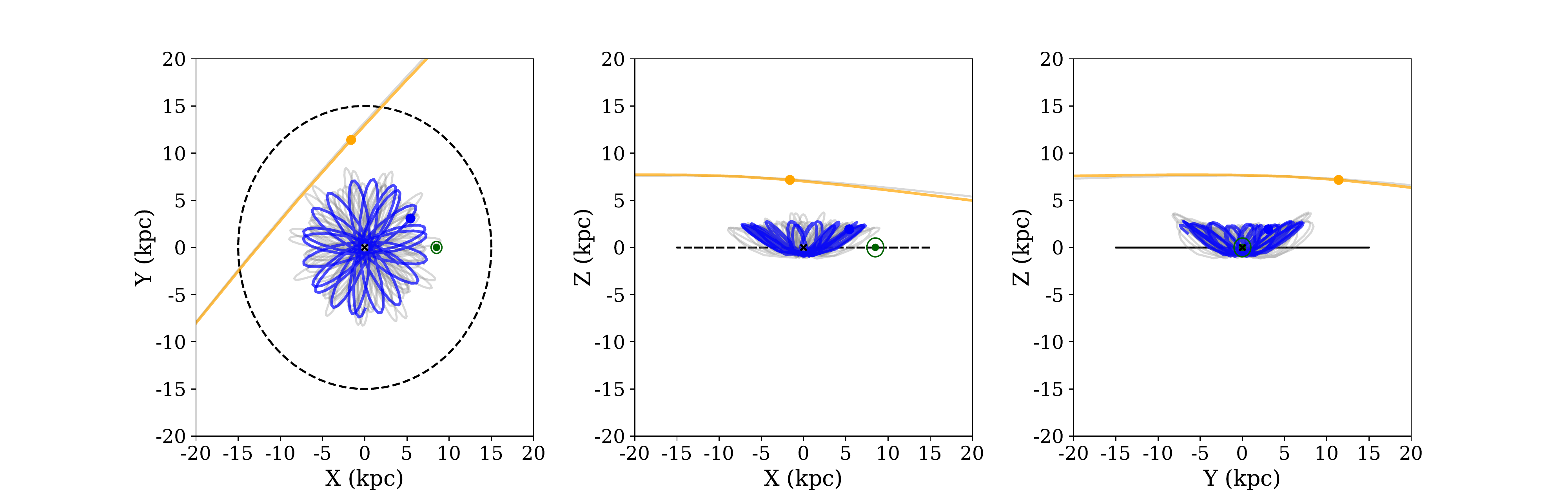} 
\caption{Same as Figure~\ref{SDSS J0023+0307}, but for SDSS J174259.67+253135.8.} 
\label{SDSS J174259.67+253135.8} 
\end{figure*}

Similarly to the previous star, the Gaia parallax of SDSS~J174259.67+253135.8 ($\varpi= -0.1628 \pm 0.1870\mas$) does not allow us to discriminate between the dwarf and giant solutions but the giant solution implies an orbit with a very large apocentre beyond $700\kpc$ and we therefore favour the dwarf solution at $4.46 \pm 0.52 \kpc$ (Figure~\ref{SDSS J174259.67+253135.8}). With this distance, SDSS~J174259.67+253135.8 \citep{2015AA...579A..28B} is on an eccentric orbit that remains close to the MW plane ($|Z|<2.7 \kpc$).

\subsection{2MASS J18082002-5104378}
\begin{figure*}
\includegraphics[scale=.4]{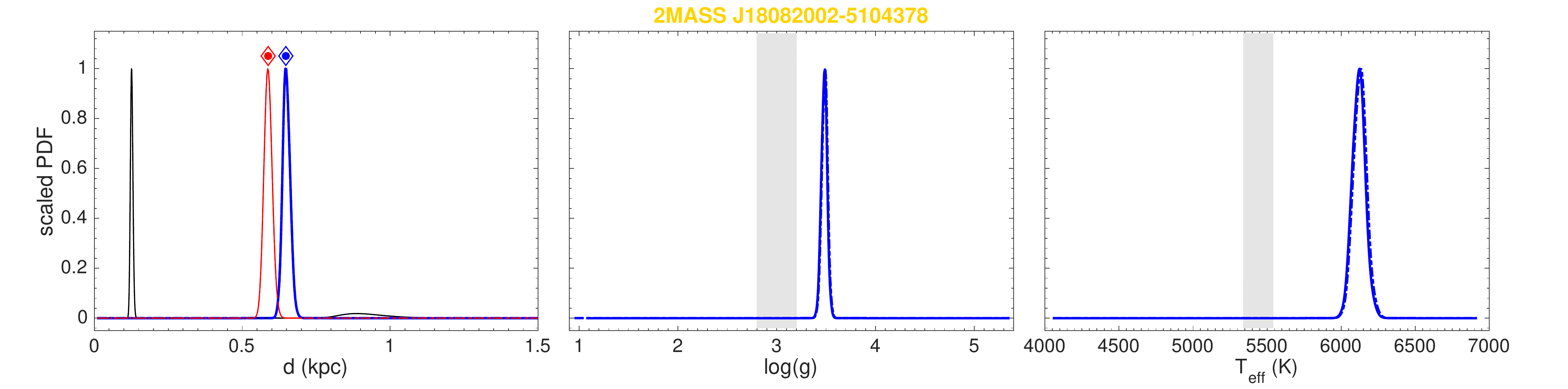}\\ 
\includegraphics[scale=.4]{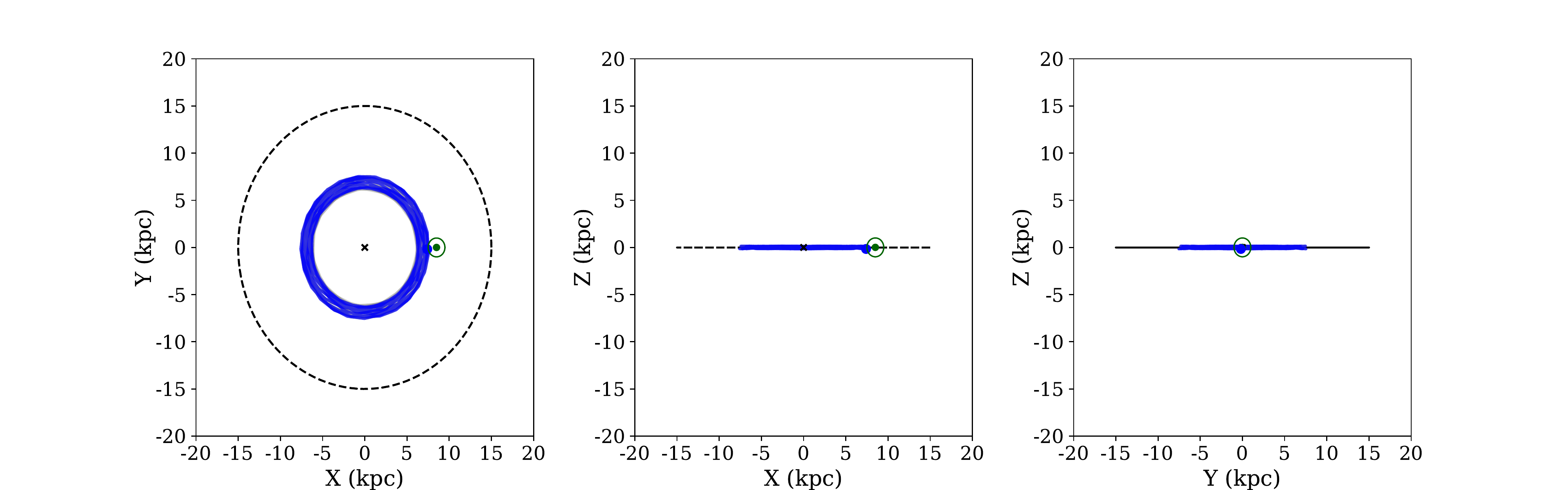} 
\caption{Same as Figure~\ref{SDSS J0023+0307}, but for 2MASS J18082002-5104378.} 
\label{2MASS J18082002-5104378} 
\end{figure*}

\citet{Schlaufman18} show that this star is in a binary system. The orbital parameters they derive show that this binary system has a very low eccentric orbit and is confined to the MW plane ($|Z| < 0.13 \kpc$). From our distance analysis, the photometric likelihood is not in agreement with the exquisite Gaia parallax ($\varpi=1.6775\pm0.0397 \mas$), but we derive a similar overall solution at a distance of $0.647 \pm 0.012\kpc$ and stellar parameters in agreement with the literature values \citep[][Figure~\ref{2MASS J18082002-5104378}]{2016AA...585L...5M}. In agreement with the work from \citet{Schlaufman18}, we derive that the orbit is very close to the MW plane and even confined inside the thin disc ($|Z|< 0.166 \kpc$) with a very low eccentricity of $\epsilon = 0.090^{+0.006}_{-0.005}$.

\subsection{BPS CS 22891-0200}
\begin{figure*}
\includegraphics[scale=.4]{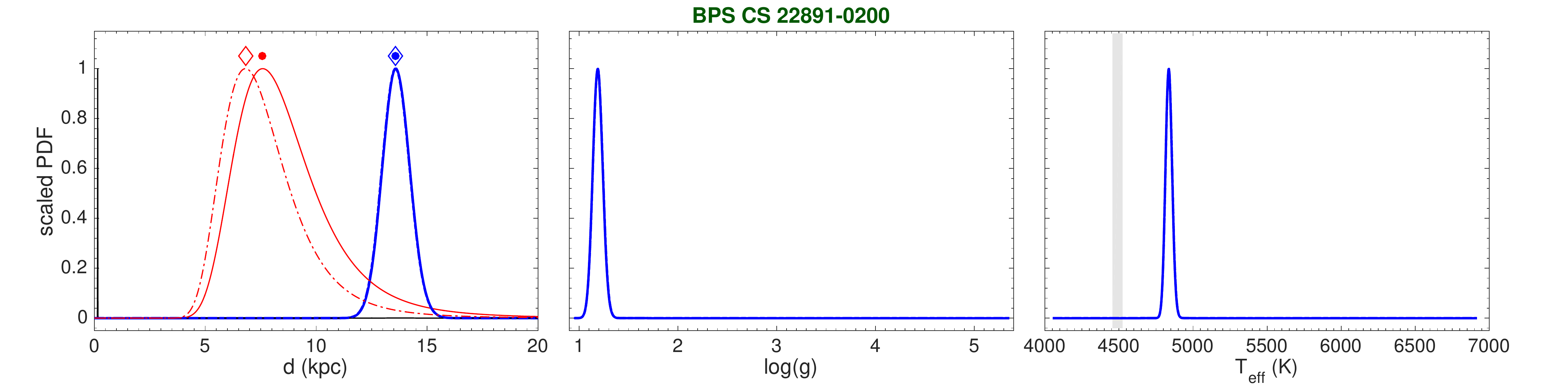}\\ 
\includegraphics[scale=.4]{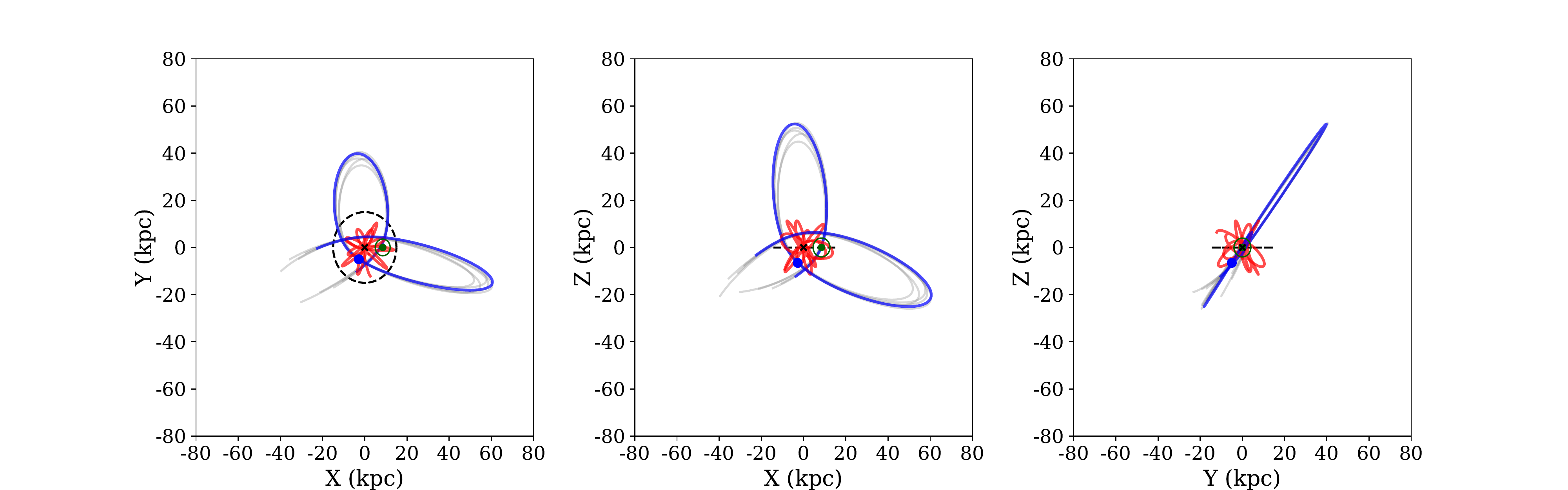} 
\caption{Same as Figure~\ref{SDSS J0023+0307}, but for BPS CS 22891-0200. The literature value for surface gravity is out of range in the plot. For this star, the orbit inferred from the product between the astrometric likelihood and MW halo prior is shown with the red line.} 
\label{BPS CS 22891-0200} 
\end{figure*}

The  PDF of BPS~CS 22891-0200 (\citealt{Roederer2014}; Figure~\ref{BPS CS 22891-0200}) shows that is a giant star near the tip (see also Figure~\ref{CMD}), located at the distance of $14.7 \pm 0.5\kpc$. Our inferred stellar parameters do not match the values from the literature. The orbit of BPS~CS 22891-0200 brings it far out into the halo of the MW ($r_\mathrm{apo}=64.0^{+18.9}_{-11.1}\kpc$).

\subsection{BPS CS 22885-0096}
\begin{figure*}
\includegraphics[scale=.4]{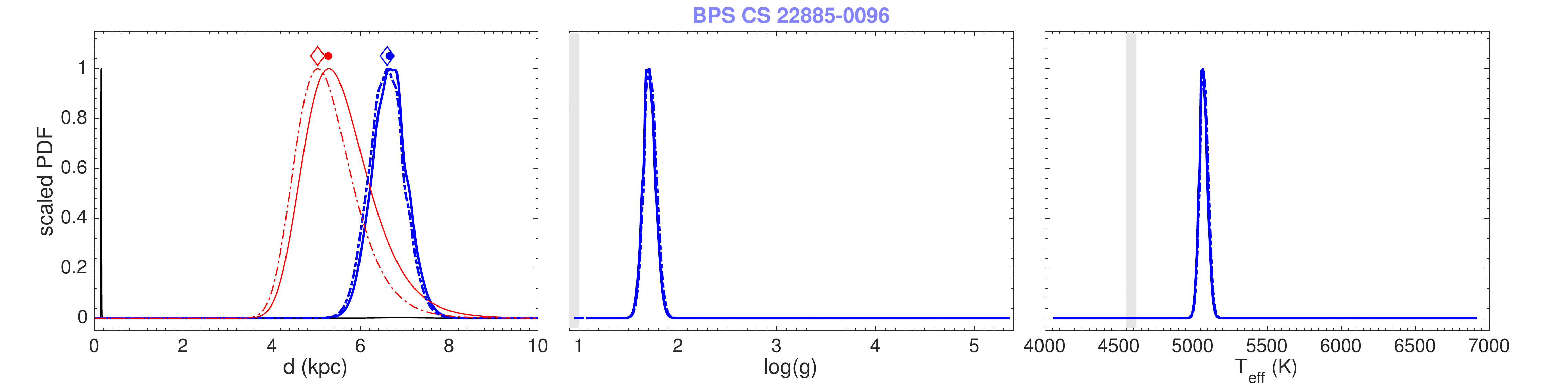}\\ 
\includegraphics[scale=.4]{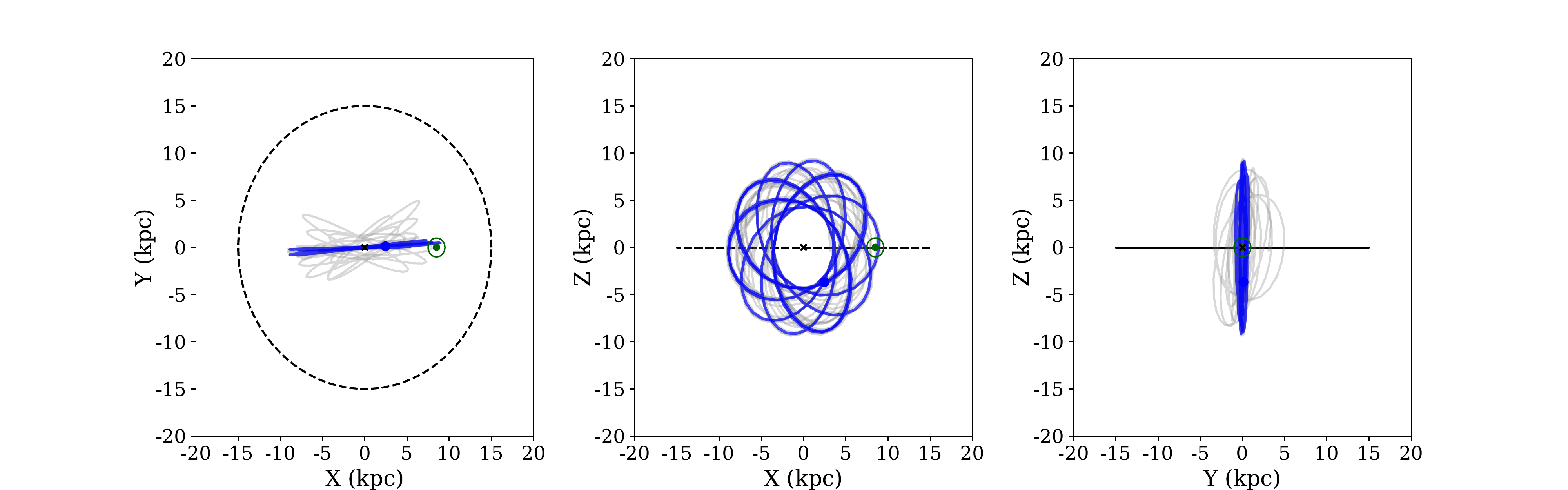} 
\caption{Same as Figure~\ref{SDSS J0023+0307}, but for BPS CS 22885-0096.} 
\label{BPS CS 22885-0096} 
\end{figure*}

Figure~\ref{BPS CS 22885-0096} presents our results for BPS~CS 22885-0096 \citep{Roederer2014}, indicating that it is a giant at a distance of $6.65 \pm 0.22\kpc$, even though the stellar parameters we infer differ from the literature values. The orbit of this star is confined to a very narrow plane that is perpendicular to the MW plane.

\subsection{BPS CS 22950-0046}
\begin{figure*}
\includegraphics[scale=.4]{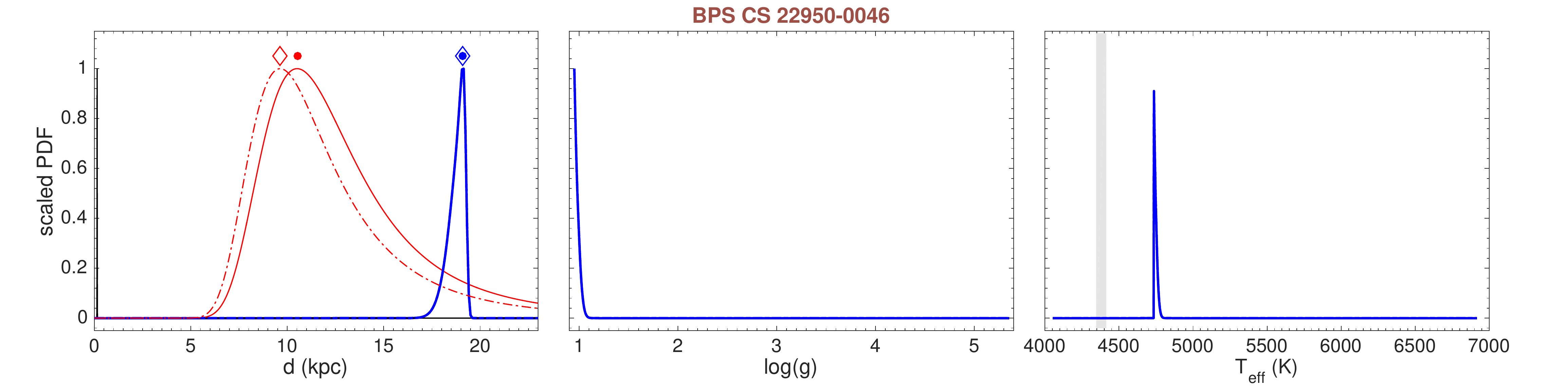}\\ 
\includegraphics[scale=.4]{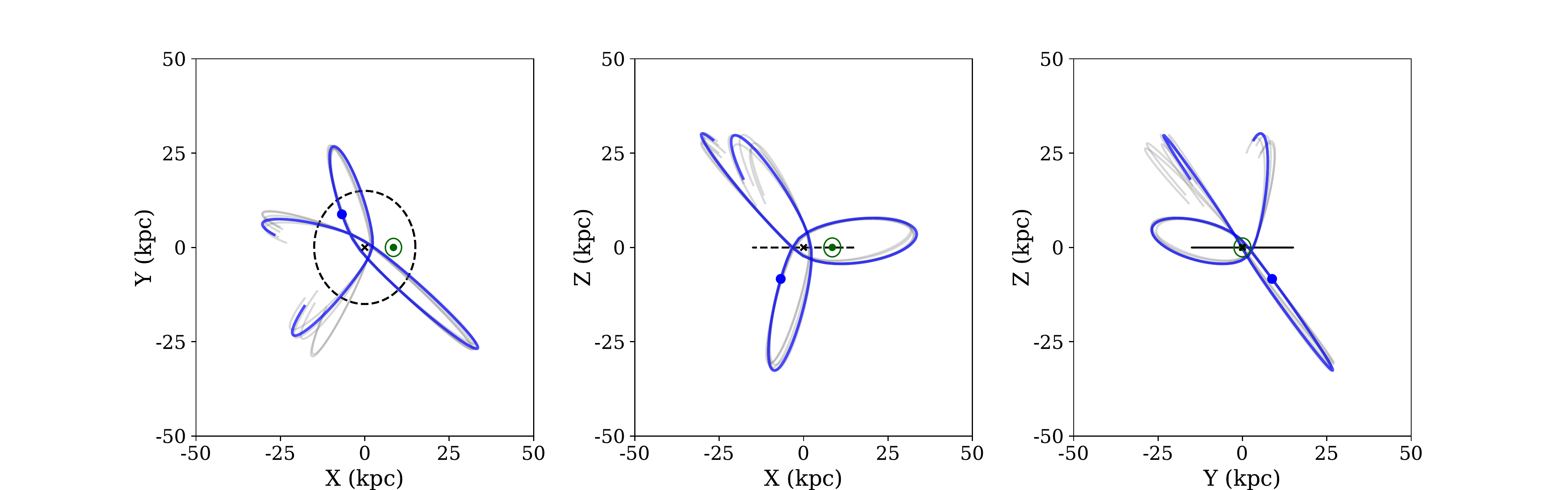} 
\caption{Same as Figure~\ref{SDSS J0023+0307}, but for BPS CS 22950-0046. The literature value for surface gravity is out of range in the plot.} 
\label{BPS CS 22950-0046} 
\end{figure*}

The Gaia parallax for this star ($\varpi=0.0587  \pm 0.0270 \mas$) clearly rules out the dwarf solution (Figure~\ref{BPS CS 22950-0046}). As the plots show, this halo giant star is at a distance of $19.1 \pm 0.3 \kpc$ and the inferred stellar parameters are not in agreement with the literature \citep{Roederer2014}.

\subsection{BPS CS 30336-0049}
\begin{figure*}
\includegraphics[scale=.4]{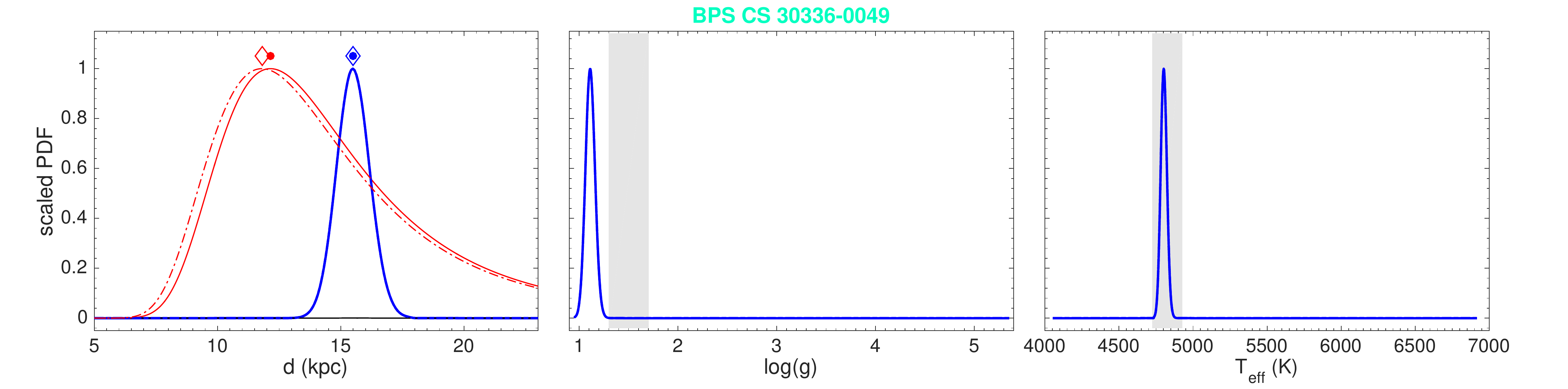}\\ 
\includegraphics[scale=.4]{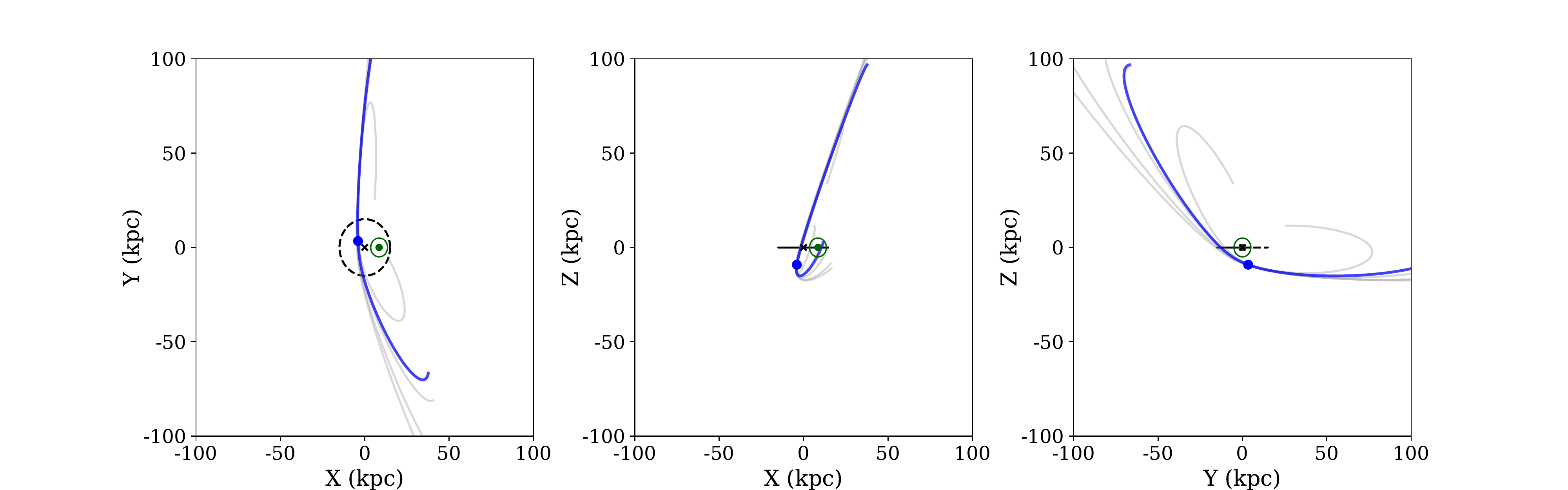} 
\caption{Same as Figure~\ref{SDSS J0023+0307}, but for BPS CS 30336-0049.} 
\label{BPS CS 30336-0049} 
\end{figure*}

Figure~\ref{BPS CS 30336-0049} shows that BPS~CS 30336-0049 is located at $15.5 \pm 0.7\kpc$ and follows an orbit that brings it far into the MW halo ($r_\mathrm{apo}=122.7^{+51.1}_{-41.4}\kpc$). The inferred effective temperature matches the value from the literature \citep{0004-637X-681-2-1524}, while our constraints on the gravity yields a slightly lower $\log(g)$. 

\subsection{HE 2139-5432}
\begin{figure*}
\includegraphics[scale=.4]{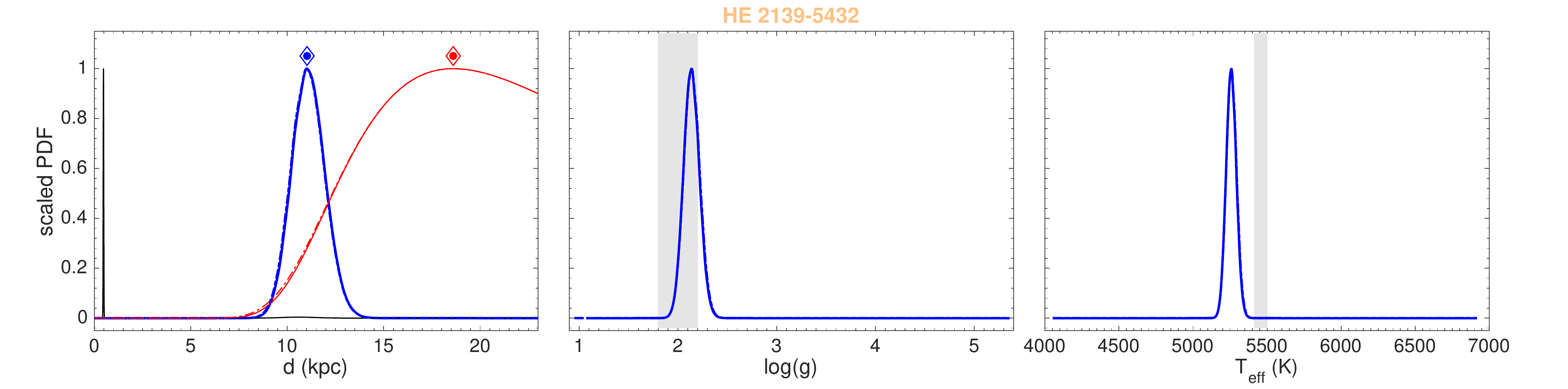}\\ 
\includegraphics[scale=.4]{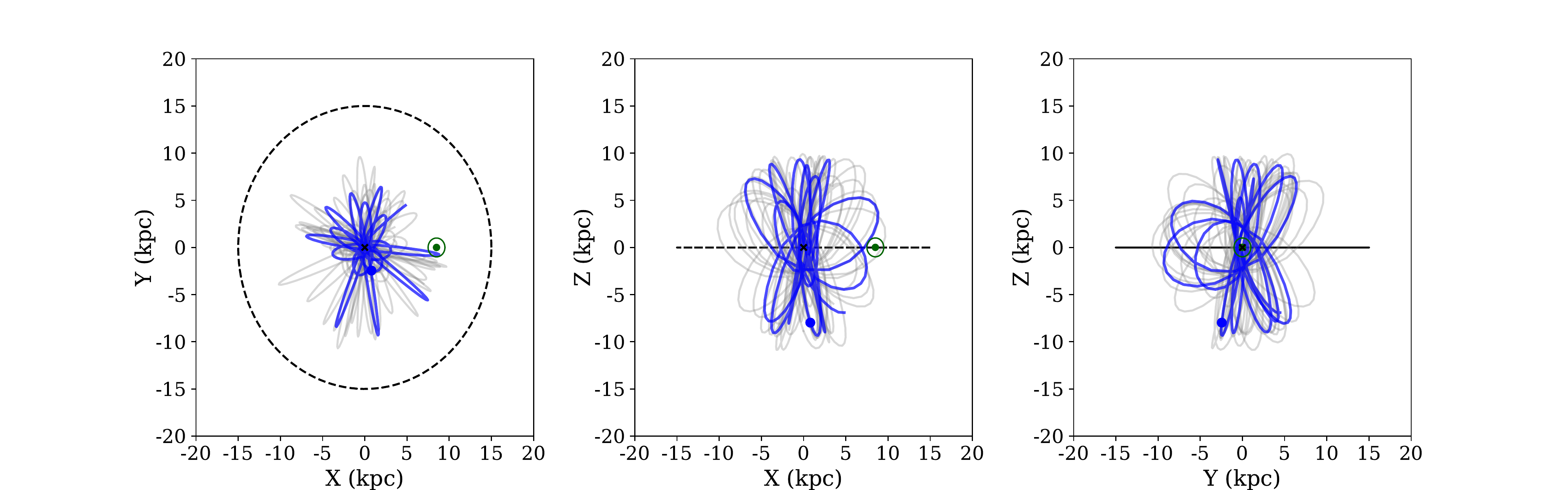} 
\caption{Same as Figure~\ref{SDSS J0023+0307}, but for HE 2139-5432.} 
\label{HE 2139-5432} 
\end{figure*}

Our results on HE 2139-5432 are summarised in Figure~\ref{HE 2139-5432}, and they lead to the conclusion that this star is a giant located at a distance of $11.0 \pm 0.9\kpc$ from the Sun. The inferred surface gravity is in agreement with the literature \citep{2013ApJ...762...25N} but the effective temperature is slightly cooler. The inferred orbit indicates that HE~2139-5432 is an inner halo star with a high eccentricity ($\epsilon= 0.79^{+0.05}_{-0.06} $).

\subsection{HE 2239-5019}
\begin{figure*}
\includegraphics[scale=.4]{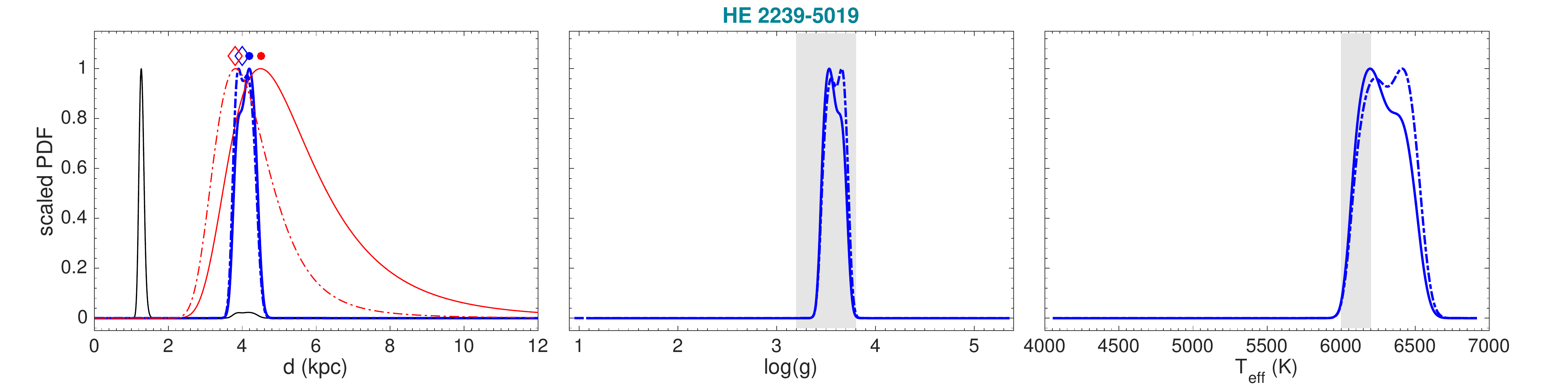}\\ 
\includegraphics[scale=.4]{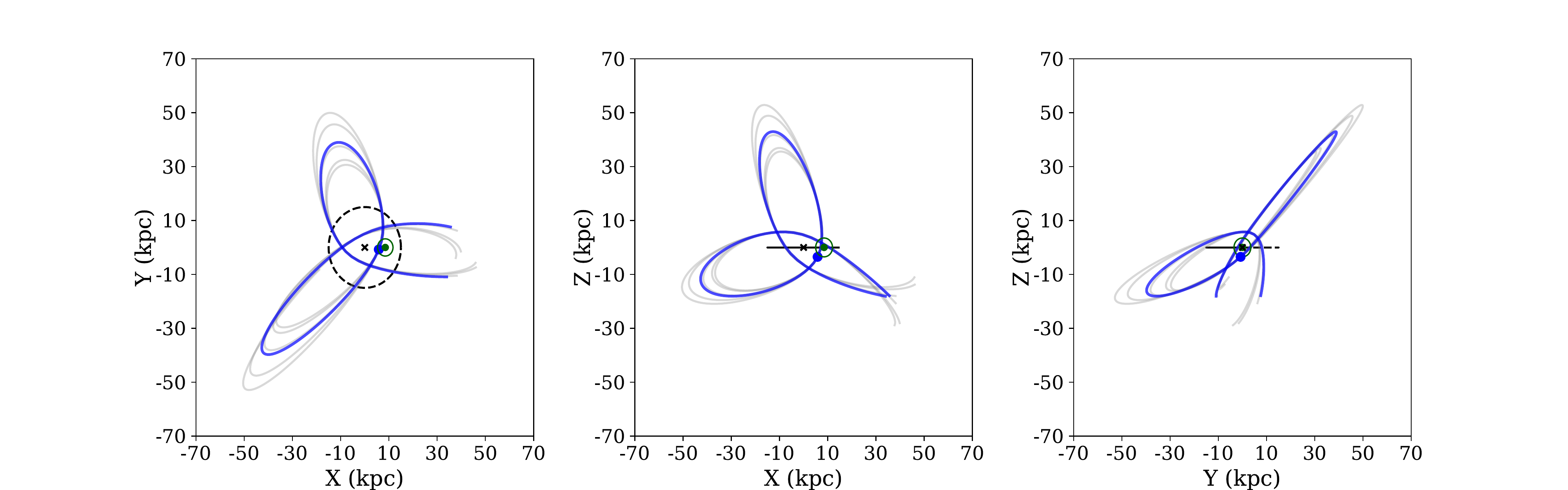} 
\caption{Same as Figure~\ref{SDSS J0023+0307}, but for HE 2239-5019.} 
\label{HE 2239-5019} 
\end{figure*}

For this star, the photometric and the astrometric likelihoods are in agreement, indicating the subgiant scenario at $4.19 \pm 0.28\kpc$ is the valid solution (Figure~\ref{HE 2239-5019}). The orbit of HE~2239-5019 brings it at fairly large distances in the halo, with $r_\mathrm{apo}=52.9^{+16.6}_{-10.4}\kpc$. The inferred surface gravity and effective temperature are compatible with the values from literature \citep{2015ApJ...807..173H}.

\subsection{HE 2323-0256}
\begin{figure*}
\includegraphics[scale=.4]{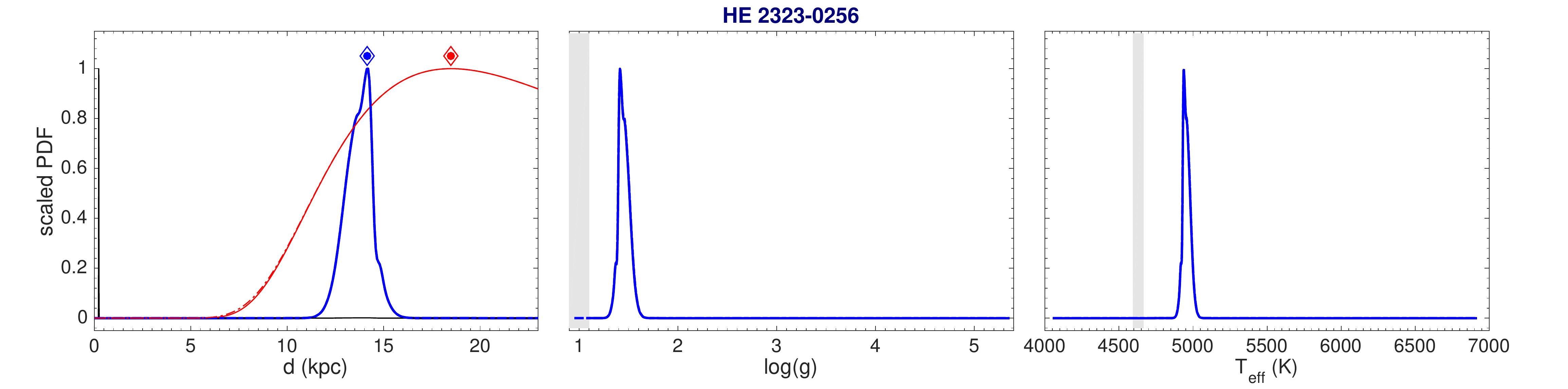}\\ 
\includegraphics[scale=.4]{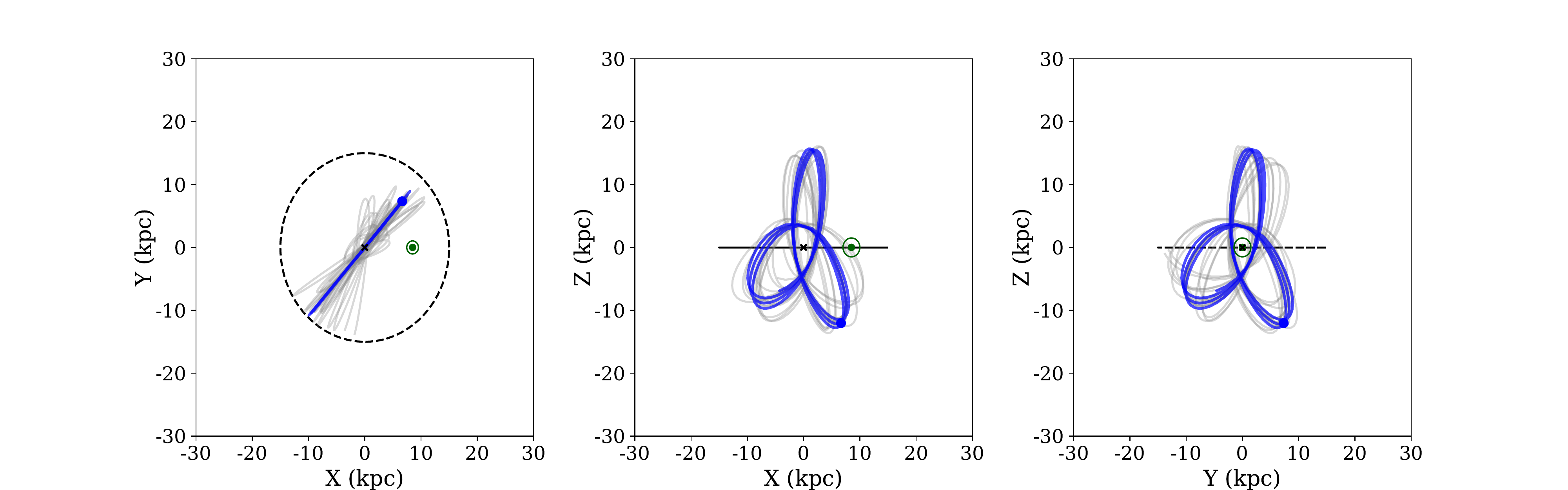} 
\caption{Same as Figure~\ref{SDSS J0023+0307}, but for HE 2323-0256.} 
\label{HE 2323-0256} 
\end{figure*}

Although the Gaia parallax is uncertain ($\varpi=0.0038  \pm 0.0359 \mas$), it helps break the dwarf/giant degeneracy. The final solution is that of a giant at a distance of $14.2 \pm 0.6\kpc$, belonging to the halo (Figure~ \ref{HE 2323-0256}). We obtain higher values for the effective temperature and surface gravity than \citet{Roederer2014}.

% \clearpage
 
\begin{figure*}
\includegraphics[scale=.4]{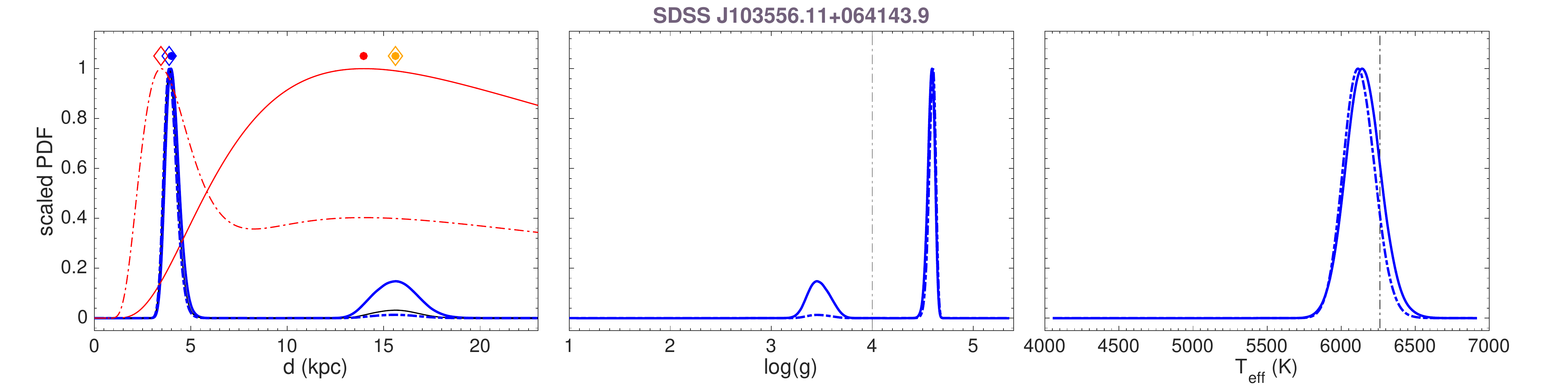}\\ 
\includegraphics[scale=.4]{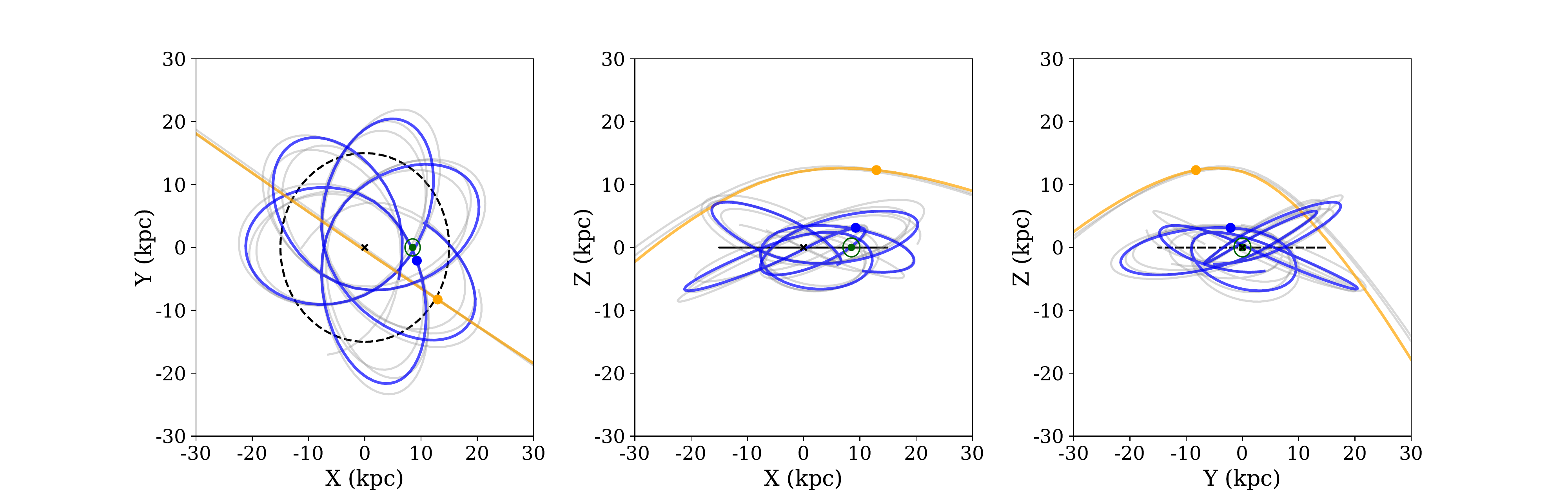} 
\caption{Same as Figure~\ref{SDSS J0023+0307}, but for SDSS J103556.11+064143.9.} 
\label{SDSS J103556.11+064143.9} 
\end{figure*} 
 
 \clearpage
 
 \pagebreak
 \clearpage
 
\section{Comparison with values from the literature}\label{comp}
\begin{figure*}
\includegraphics[scale=.5]{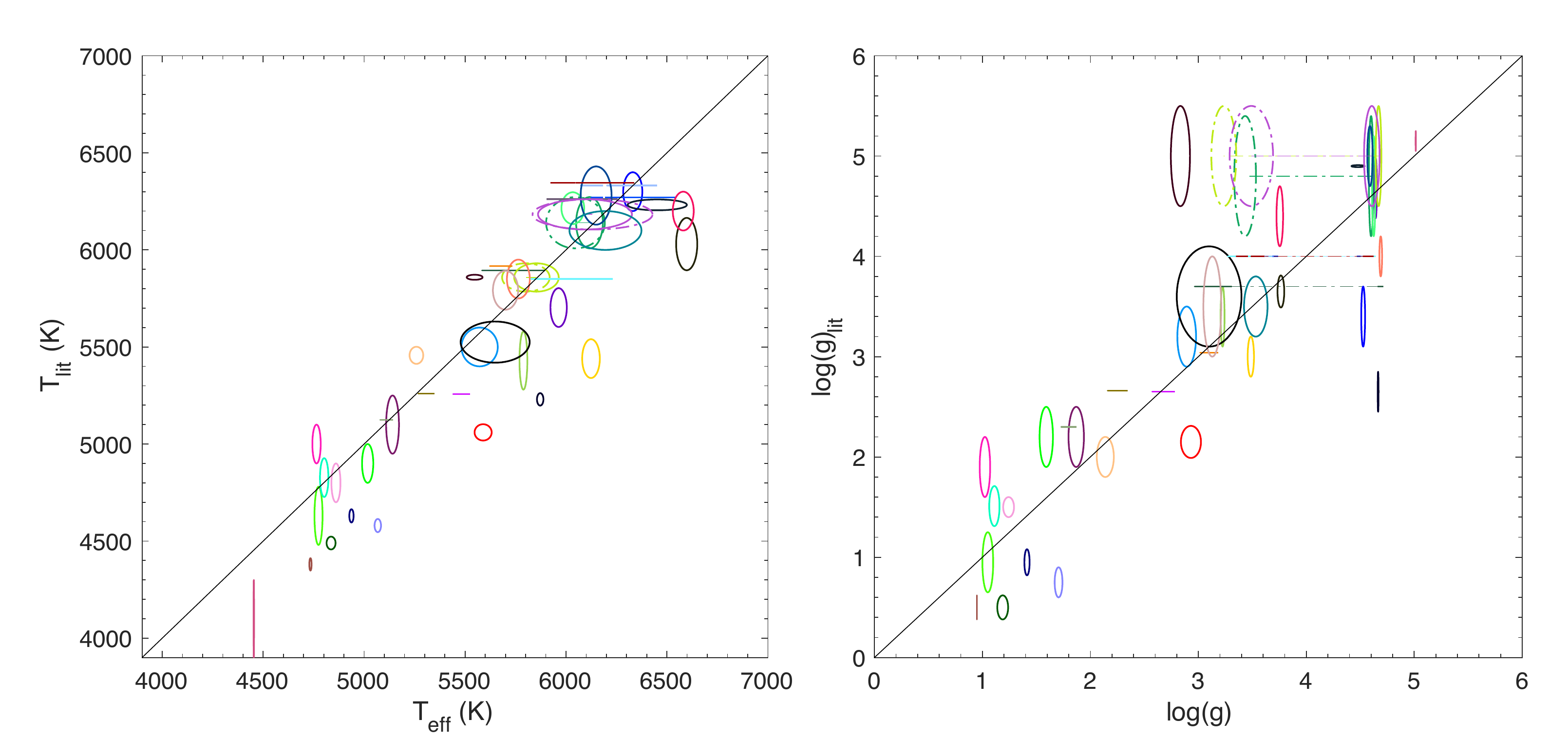}
\caption{Comparison between inferred effective temperature $T_{\mathrm{eff}}$ (left panel), surface gravity $\log(g)$ (right panel) and the values from the  literature.  The ellipses represent the position of the stars within 1 sigma and the black line corresponds to the 1:1 relation. If the dwarf-giant degeneracy is not broken, the two possible solutions are represented and connected by a dot-dashed line of the same colour code. Each colour represents a star and the colour-code is the same as the colour-code for the markers in Figures~1~-~2 and the panel's titles in Figures~\ref{HE 0020-1741}~-~\ref{HE 2323-0256}. Solutions with integrated probability ($\int_{d-3\sigma}^{d+3\sigma} P(r) dr$) lower than $5\%$ are not shown and solutions with integrated probability in the range $[5\%, 50\%]$ are shown with dot-dashed ellipses.} 
\label{comparison} 
\end{figure*}

A global comparison between the stellar parameters inferred in this work and the values found in the literature is reported in the two panels of  Figure~\ref{comparison}. As we can see, we find a broad agreement for the effective temperature (left panel) and the surface gravity (right panel).
Possible systematics are involved both in our method (e.g. $T_{\mathrm{eff}}-\log(g)$ relation in the MESA/MIST isochrones) and the multiple spectroscopic methods used by different authors (e.g. grid based models, synthetic spectra, data-driven analysis etc.).

\end{document}